%% file: main.tex
\documentclass[longtitle,preprint,5p,times,twocolumn]{elsarticle}

\usepackage{amssymb}

\usepackage{caption}
\usepackage{xspace}
\usepackage{longtable}
\usepackage{pdflscape}
\usepackage{multirow}
\usepackage{ragged2e}
\usepackage{booktabs}
\usepackage{hyperref}
\usepackage{lineno}

\newcommand{\jpsi}{\mbox{$J/\psi$}}
\newcommand{\mrwell}{\mbox{$\mu$RWELL}\xspace}
\newcommand{\PbWOiv}{\mbox{PbWO$_4$}\xspace}
\newcommand{\geant}{\mbox{\sc Geant4}\xspace}

\journal{Nuclear Instruments and Methods A}

\begin{document}

\begin{frontmatter}

\title{Design of the ECCE Detector for the Electron Ion Collider}

\def\theaffn{\arabic{affn}} 
\input{authors/authors-alphabetical}
\input{authors/affiliations}

\begin{abstract}
The EIC Comprehensive Chromodynamics Experiment (ECCE) detector has been designed to address the full scope of the proposed Electron Ion Collider (EIC) physics program as presented by the National Academy of Science and provide a deeper understanding of the quark-gluon structure of matter.   To accomplish this, the ECCE detector offers nearly acceptance and energy coverage along with excellent tracking and particle identification.
The ECCE detector was designed to be built within the budget envelope set out by the EIC project while simultaneously managing cost and schedule risks.
This detector concept has been selected to be the basis for the EIC project detector.
\end{abstract}


\end{frontmatter}


\setcounter{tocdepth}{1}
\tableofcontents 

\input{sections/introduction}
\input{sections/detector}
\input{sections/summary}
\input{sections/acknowledgements}

\bibliographystyle{elsarticle-num} 
\bibliography{refs.bib,refs-ecce.bib}

\end{document}

%% file: authors/authors-alphabetical.tex
%
%
%
%

\author[MoreheadState]{J.~K.~Adkins}
\author[RIKEN,RBRC]{Y.~Akiba}
\author[UKansas]{A.~Albataineh}
\author[ODU]{M.~Amaryan}
\author[Oslo]{I.~C.~Arsene}
\author[MSU]{C. Ayerbe Gayoso}
\author[Sungkyunkwan]{J.~Bae}
\author[UVA]{X.~Bai}
\author[BNL,JLab]{M.D.~Baker}
\author[York]{M.~Bashkanov}
% ...

\cortext[cor1]{Deceased}

%% file: authors/affiliations.tex
%

\affiliation[AANL]{organization={A. Alikhanyan National Laboratory},
	 city={Yerevan},
	 country={Armenia}} 
 
\affiliation[AcademiaSinica]{organization={Institute of Physics, Academia Sinica},
	 city={Taipei},
	 country={Taiwan}} 
 
\affiliation[AUGIE]{organization={Augustana University},
	 city={Sioux Falls},
	 state={SD},
	 country={USA}} 
	 
\affiliation[BGU]{organizatoin={Ben-Gurion University of the Negev}, 
      city={Beer-Sheva},
      country={Israel}}

\affiliation[BNL]{organization={Brookhaven National Laboratory},
	 city={Upton},
	 state={NY},
	 country={USA}} 
 
\affiliation[BrunelUniversity]{organization={Brunel University London},
	 city={Uxbridge},
	 country={UK}} 
 
\affiliation[CanisiusCollege]{organization={Canisius College},
	 city={Buffalo},
	 state={NY},
	 country={USA}} 
 
\affiliation[CCNU]{organization={Central China Normal University},
	 city={Wuhan},
	 country={China}} 
 
\affiliation[Charles]{organization={Charles University},
	 city={Prague},
	 country={Czech Republic}} 
 
\affiliation[CIAE]{organization={China Institute of Atomic Energy, Fangshan},
	 city={Beijing},
	 country={China}} 
 
\affiliation[CNU]{organization={Christopher Newport University},
	 city={Newport News},
	 state={VA},
	 country={USA}} 
 
\affiliation[Columbia]{organization={Columbia University},
	 city={New York},
	 state={NY},
	 country={USA}} 
 
\affiliation[CUA]{organization={Catholic University of America},
	 city={Washington DC},
	 country={USA}} 
 
\affiliation[CzechTechUniv]{organization={Czech Technical University},
	 city={Prague},
	 country={Czech Republic}} 
 
\affiliation[Duquesne]{organization={Duquesne University},
	 city={Pittsburgh},
	 state={PA},
	 country={USA}}

 
\affiliation[FIU]{organization={Florida International University},
	 city={Miami},
	 state={FL},
	 country={USA}} 
 
\affiliation[GeorgiaState]{organization={Georgia State University},
	 city={Atlanta},
	 state={GA},
	 country={USA}} 
 
\affiliation[Glasgow]{organization={University of Glasgow},
	 city={Glasgow},
	 country={UK}} 
 
\affiliation[GSI]{organization={GSI Helmholtzzentrum fuer Schwerionenforschung GmbH},
	 city={Darmstadt},
	 country={Germany}} 
 
\affiliation[GWU]{organization={The George Washington University},
	 city={Washington, DC},
	 country={USA}}

 
\affiliation[HUJI]{organization={Hebrew University},
	 city={Jerusalem},
	 country={Isreal}} 
 
\affiliation[IJCLabOrsay]{organization={Universite Paris-Saclay, CNRS/IN2P3, IJCLab},
	 city={Orsay},
	 country={France}} 
	 
\affiliation[CEA]{organization={IRFU, CEA, Universite Paris-Saclay},
     cite= {Gif-sur-Yvette},
     country={France}
}

\affiliation[IMP]{organization={Chinese Academy of Sciences},
	 city={Lanzhou},
	 country={China}} 
 
\affiliation[IowaState]{organization={Iowa State University},
	 city={Iowa City},
	 state={IA},
	 country={USA}} 
 
\affiliation[JazanUniversity]{organization={Jazan University},
	 city={Jazan},
	 country={Sadui Arabia}} 
 
\affiliation[JLab]{organization={Thomas Jefferson National Accelerator Facility},
	 city={Newport News},
	 state={VA},
	 country={USA}} 

 
\affiliation[KobeUniversity]{organization={Kobe University},
	 city={Kobe},
	 country={Japan}} 
 
\affiliation[Kyungpook]{organization={Kyungpook National University},
	 city={Daegu},
	 country={Republic of Korea}} 
 
\affiliation[LANL]{organization={Los Alamos National Laboratory},
	 city={Los Alamos},
	 state={NM},
	 country={USA}} 

 
\affiliation[LehighUniversity]{organization={Lehigh University},
	 city={Bethlehem},
	 state={PA},
	 country={USA}} 
 
\affiliation[LLNL]{organization={Lawrence Livermore National Laboratory},
	 city={Livermore},
	 state={CA},
	 country={USA}} 
 
\affiliation[MoreheadState]{organization={Morehead State University},
	 city={Morehead},
	 state={KY},
	 }
 
\affiliation[MIT]{organization={Massachusetts Institute of Technology},
	 city={Cambridge},
	 state={MA},
	 country={USA}} 
 
\affiliation[MSU]{organization={Mississippi State University},
	 city={Mississippi State},
	 state={MS},
	 country={USA}} 
 
\affiliation[NCKU]{organization={National Cheng Kung University},
	 city={Tainan},
	 country={Taiwan}} 
 
\affiliation[NCU]{organization={National Central University},
	 city={Chungli},
	 country={Taiwan}} 
 
\affiliation[Nihon]{organization={Nihon University},
	 city={Tokyo},
	 country={Japan}} 
 
\affiliation[NMSU]{organization={New Mexico State University},
	 city={Las Cruces},
	 state={NM},
	 country={USA}} 
 
\affiliation[NRNUMEPhI]{organization={National Research Nuclear University MEPhI},
	 city={Moscow},
	 country={Russian Federation}} 
 
\affiliation[NRCN]{organization={Nuclear Research Center - Negev},
	 city={Beer-Sheva},
	 country={Isreal}} 
 
\affiliation[NTHU]{organization={National Tsing Hua University},
	 city={Hsinchu},
	 country={Taiwan}} 
 
\affiliation[NTU]{organization={National Taiwan University},
	 city={Taipei},
	 country={Taiwan}} 
 
\affiliation[ODU]{organization={Old Dominion University},
	 city={Norfolk},
	 state={VA},
	 country={USA}} 
 
\affiliation[Ohio]{organization={Ohio University},
	 city={Athens},
	 state={OH},
	 country={USA}} 
 
\affiliation[ORNL]{organization={Oak Ridge National Laboratory},
	 city={Oak Ridge},
	 state={TN},
	 country={USA}} 
	 
\affiliation[Oxford]{organization={University of Oxford},
    city={Oxford},
    country={UK}}

\affiliation[PNNL]{organization={Pacific Northwest National Laboratory},
	 city={Richland},
	 state={WA},
	 country={USA}} 
 
\affiliation[Pusan]{organization={Pusan National University},
	 city={Busan},
	 country={Republic of Korea}} 
 
\affiliation[Rice]{organization={Rice University},
	 city={Houston},
	 state={TX},
	 country={USA}} 
 
\affiliation[RIKEN]{organization={RIKEN Nishina Center},
	 city={Wako},
	 state={Saitama},
	 country={Japan}} 
 
\affiliation[Rutgers]{organization={The State University of New Jersey},
	 city={Piscataway},
	 state={NJ},
	 country={USA}}

\affiliation[CFNS]{organization={Center for Frontiers in Nuclear Science},
	 city={Stony Brook},
	 state={NY},
	 country={USA}} 
 
\affiliation[StonyBrook]{organization={Stony Brook University},
	 city={Stony Brook},
	 state={NY},
	 country={USA}} 
 
\affiliation[RBRC]{organization={RIKEN BNL Research Center},
	 city={Upton},
	 state={NY},
	 country={USA}} 
	 
\affiliation[SDU]{organizaton={Shandong University},
     city={Qingdao},
     state={Shandong},
     country={China}}
     
\affiliation[Seoul]{organization={Seoul National University},
	 city={Seoul},
	 country={Republic of Korea}} 
 
\affiliation[Sejong]{organization={Sejong University},
	 city={Seoul},
	 country={Republic of Korea}} 
 
\affiliation[Shinshu]{organization={Shinshu University},
         city={Matsumoto},
	 state={Nagano},
	 country={Japan}} 
 
\affiliation[Sungkyunkwan]{organization={Sungkyunkwan University},
	 city={Suwon},
	 country={Republic of Korea}} 
 
\affiliation[TAU]{organization={Tel Aviv University},
	 city={Tel Aviv},
	 country={Israel}} 

\affiliation[USTC]{organization={University of Science and Technology of China},
     city={Hefei},
     country={China}}

\affiliation[Tsinghua]{organization={Tsinghua University},
	 city={Beijing},
	 country={China}} 
 
\affiliation[Tsukuba]{organization={Tsukuba University of Technology},
	 city={Tsukuba},
	 state={Ibaraki},
	 country={Japan}} 
 
\affiliation[CUBoulder]{organization={University of Colorado Boulder},
	 city={Boulder},
	 state={CO},
	 country={USA}} 
 
\affiliation[UConn]{organization={University of Connecticut},
	 city={Storrs},
	 state={CT},
	 country={USA}} 
 
\affiliation[UNGeorgia]{organization={University of North Georgia},
     cite={Dahlonega}, 
     state={GA},
     country={USA}}
     
\affiliation[UH]{organization={University of Houston},
	 city={Houston},
	 state={TX},
	 country={USA}} 
 
\affiliation[UIUC]{organization={University of Illinois}, 
	 city={Urbana},
	 state={IL},
	 country={USA}} 
 
\affiliation[UKansas]{organization={Unviersity of Kansas},
	 city={Lawrence},
	 state={KS},
	 country={USA}} 
 
\affiliation[UKY]{organization={University of Kentucky},
	 city={Lexington},
	 state={KY},
	 country={USA}} 
 
\affiliation[Ljubljana]{organization={University of Ljubljana, Ljubljana, Slovenia},
	 city={Ljubljana},
	 country={Slovenia}} 
 
\affiliation[NewHampshire]{organization={University of New Hampshire},
	 city={Durham},
	 state={NH},
	 country={USA}} 
 
\affiliation[Oslo]{organization={University of Oslo},
	 city={Oslo},
	 country={Norway}} 
 
\affiliation[Regina]{organization={ University of Regina},
	 city={Regina},
	 state={SK},
	 country={Canada}} 
 
\affiliation[USeoul]{organization={University of Seoul},
	 city={Seoul},
	 country={Republic of Korea}} 
 
\affiliation[UTsukuba]{organization={University of Tsukuba},
	 city={Tsukuba},
	 country={Japan}} 
	 
\affiliation[UoT]{organization={University of Texas},
    city={Austin},
    state={Texas},
    country={USA}}
 
\affiliation[UTK]{organization={University of Tennessee},
	 city={Knoxville},
	 state={TN},
	 country={USA}} 
 
\affiliation[UVA]{organization={University of Virginia},
	 city={Charlottesville},
	 state={VA},
	 country={USA}} 
 
\affiliation[Vanderbilt]{organization={Vanderbilt University},
	 city={Nashville},
	 state={TN},
	 country={USA}} 
 
 
\affiliation[VirginiaUnion]{organization={Virginia Union University},
	 city={Richmond},
	 state={VA},
	 country={USA}} 
 
\affiliation[WayneState]{organization={Wayne State University},
	 city={Detroit},
	 state={MI},
	 country={USA}} 
 
\affiliation[WI]{organization={Weizmann Institute of Science},
	 city={Rehovot},
	 country={Israel}} 
 
\affiliation[WandM]{organization={The College of William and Mary},
	 city={Williamsburg},
	 state={VA},
	 country={USA}} 
 
\affiliation[Yamagata]{organization={Yamagata University},
	 city={Yamagata},
	 country={Japan}} 
 
\affiliation[Yarmouk]{organization={Yarmouk University},
	 city={Irbid},
	 country={Jordan}} 
 
\affiliation[Yonsei]{organization={Yonsei University},
	 city={Seoul},
	 country={Republic of Korea}} 
 
\affiliation[York]{organization={University of York},
	 city={York},
	 country={UK}}


%% file: sections/introduction.tex
\section{Introduction}
\label{introduction}

The physics program at the Electron-Ion Collider (EIC) -- planned for construction at Brookhaven National Laboratory (BNL), in close partnership with the Thomas Jefferson National Accelerator Facility (TJNAF) -- will be the culmination of decades of research into the quark and gluon substructure of hadrons and nuclei. It will provide scientific opportunities well into the next three decades.  The EIC will address a broad set of scientific questions whose impact and context were assessed in a major 2018 report by the National Academies of Science (NAS)~\cite{NAP25171}: 

\begin{figure*}[ht]
    \centering
    \includegraphics[width=0.8\textwidth]{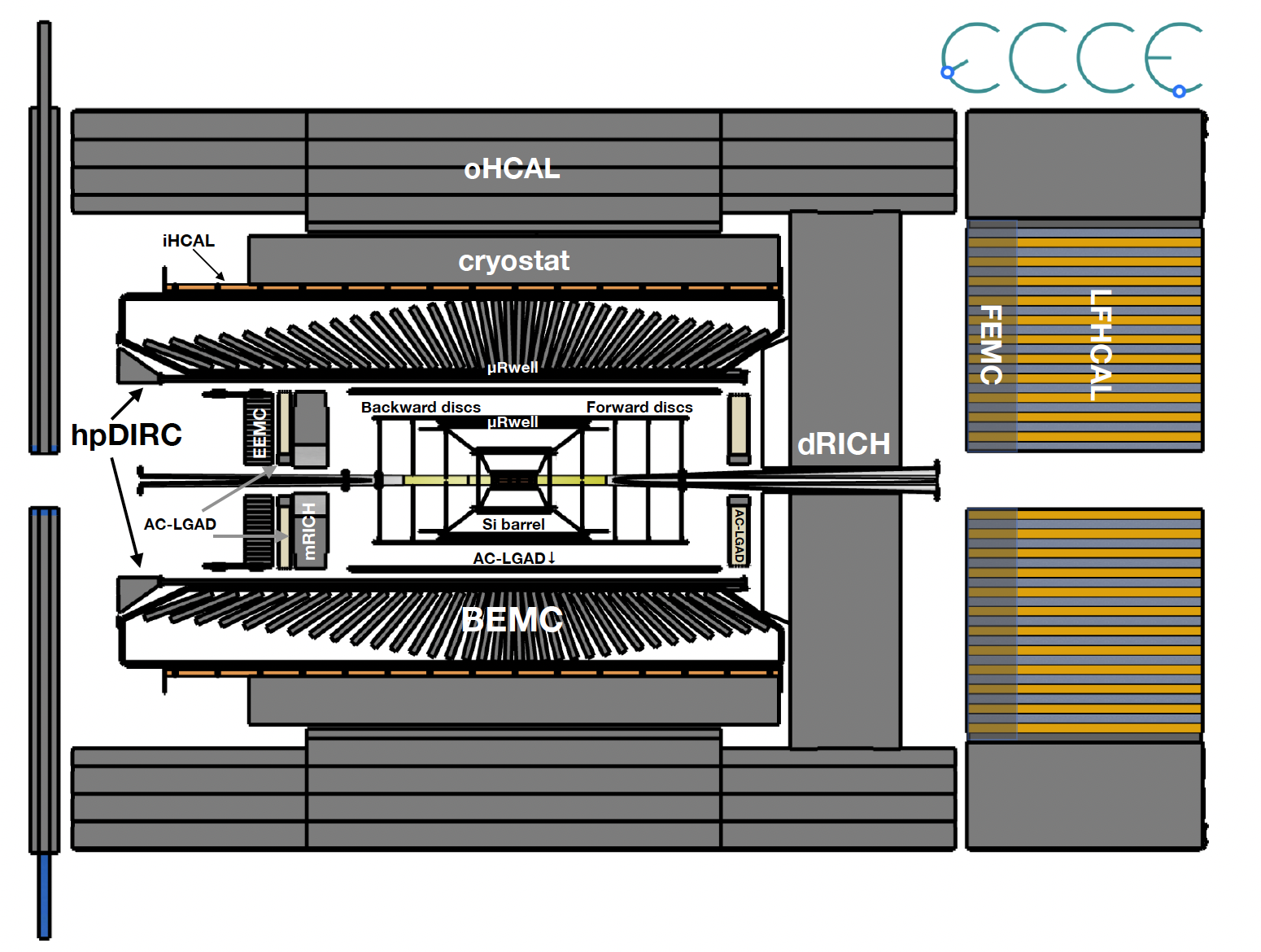}
        \caption{Side view of the full ECCE detector system, oriented with the hadron beam arriving from the left and the electron beam arriving from the right.}
    \label{fig:ecce-sectional-diagram-1.1}
\end{figure*}

The first major opportunity for the EIC will be to complete our understanding of quark and gluons in the nucleon.   Presently, the longitudinal momenta of quarks and gluons in nucleons and nuclei have been measured with great precision at previous facilities -- most notably CEBAF at JLab and the HERA collider at DESY -- but the EIC will allow the full three-dimensional momentum and spatial structure of nucleons and nuclei to be fully elucidated.  including spin, which requires the separation of the intrinsic spin of the constituent particles from their orbital motion.  

The second major opportunity for the EIC is to provide insight into how the mutual interactions of quarks and gluons generate the mass of nucleons and other hadrons.  The nucleon mass is one of the single most important scales in all of physics, as it is the basis for nuclear masses, and thus the mass of essentially all of visible matter in the universe.

The third major opportunity for the EIC is to understand the emergent properties of dense systems of quarks and gluons.   It is well understood that the density of quarks and gluons which carry the smallest $x$, the fraction of the nuclear momentum (or that of its constituent nucleons), can grow so large that their mutual interactions enter a non-linear regime in which elegant, universal features emerge in what may be a new, distinct state of matter characterized by a “saturation momentum scale”.  Probing this state requires high energy beams and large nuclear size (A), and will answer longstanding questions raised by the heavy ion programs at RHIC and the LHC.

To carry out this ambitious physics program, the EIC requires a comprehensive experimental program carefully designed to extract physics from the scattering of electrons off of hadrons and nuclei.  The ideal EIC detector must measure nearly every particle emerging from the interaction point, including its direction, its momentum, as well as its hadron species.  
Each of these aspects of the EIC physics program, and how a single comprehensive detector system could address them, was studied by the EIC scientific community and led to the community-authored “Yellow Report”~\cite{AbdulKhalek:2021gbh}.  
From this report and from the physics requirements of the EIC science program~\cite{NAP25171},
six detector performance requirements for the EIC detector system were identified: 
\begin{enumerate}
\item The outgoing electron must be distinguished  from other produced particles in the event, with a pion rejection of $10^{3}$ — $10^{4}$ even at large angles, in order to characterize the kinematic properties of the initial scattering process.  These include the momentum fraction of the struck target constituent ($x$) and the squared momentum transfer ($Q^{2}$).  
\item A large-acceptance magnetic spectrometer is needed to measure the scattered electron momentum, as well as the momenta of the other charged hadrons and leptons.  
The magnet dimensions and field strength should be matched to the scientific program and the medium-energy scale of the EIC.
This requires a nearly 4$\pi$ angular aperture, and the ability to make precise measurements of the momentum down to low-transverse momentum, $p_t$, to measure its point of origin, and to distinguish prompt particles from charm and bottom hadron decays. 
\item A high-purity hadron particle identification (PID) system, able to provide continuous ($e/\pi$), ($K/\pi$), and ($K$/p) discrimination out to the highest momentum (60 GeV/c), is important for identifying particles containing different light quark flavors with semi-inclusive DIS.  
\item A hermetic electromagnetic calorimeter system, 
with matching hadronic sections, is required to measure neutral particles (particularly photons and neutrons) and, in tandem with the spectrometer, to reconstruct hadronic jets.  These jets carry kinematic information about the struck quark or gluon, as well as its radiative properties via its substructure.  
\item Far-forward detector systems, in the direction of the incoming hadron or nucleus, are needed in order to perform measurements of deeply-virtual Compton scattering and diffractive processes, e.g. by measuring the small deflections of the incoming proton and suppressing incoherent interactions with nuclei.  
\item Far-backward detectors, in the direction of the incoming electron, are needed to reach the very lowest values of $Q^2$.  Also these detectors need to measure luminosity for both absolute cross-section measurements and precision spin-dependent asymmetries. 
\end{enumerate}

As a response to the joint BNL/JLab call for detector proposals, this document presents the EIC Comprehensive Chromodynamics Experiment (ECCE), which has been designed, simulated and extensively studied by the 96 institutes in the newly-formed ECCE proto-collaboration.  The ECCE detector has been designed to address the full scope of the EIC physics program as presented in the EIC white paper~\cite{Accardi:2012qut} and the NAS report~\cite{NAP25171}.  The specific requirements of each of the ECCE detector systems flows, in turn, from the more general detector requirements described in the Yellow Report~\cite{AbdulKhalek:2021gbh}.  Through judicious reuse of existing equipment, ECCE meets the Yellow Report requirements, and can be built within the EIC project budget.

The ECCE concept reuses the BaBar superconducting solenoid as well as the sPHENIX barrel flux return and hadronic calorimeter.  These two pieces of equipment are currently being installed in RHIC Interaction Region 8 (IR8) as part of the sPHENIX detector~\cite{PHENIX:2015siv}.  Engineering studies have confirmed that these two components can be relocated to IR6, the IR where the EIC project currently plans to site the on-project detector.  

Additional details concerning ECCE subsystems, performance, and selected
physics objectives are provided in separate articles within this same
collection.\cite{
ecce-paper-det-2022-02,
Schmidt:2022lsp,
ecce-note-comp-2021-01-nim,
ecce-paper-comp-2022-02,
ecce-paper-phys-2022-01,
ecce-paper-phys-2022-02,
ecce-paper-phys-2022-03}

%% file: sections/detector.tex
\section{ECCE Detector Overview}
\label{sec:detector_overview}

\begin{figure*}[ht]
    \centering
    \includegraphics[width=\textwidth]{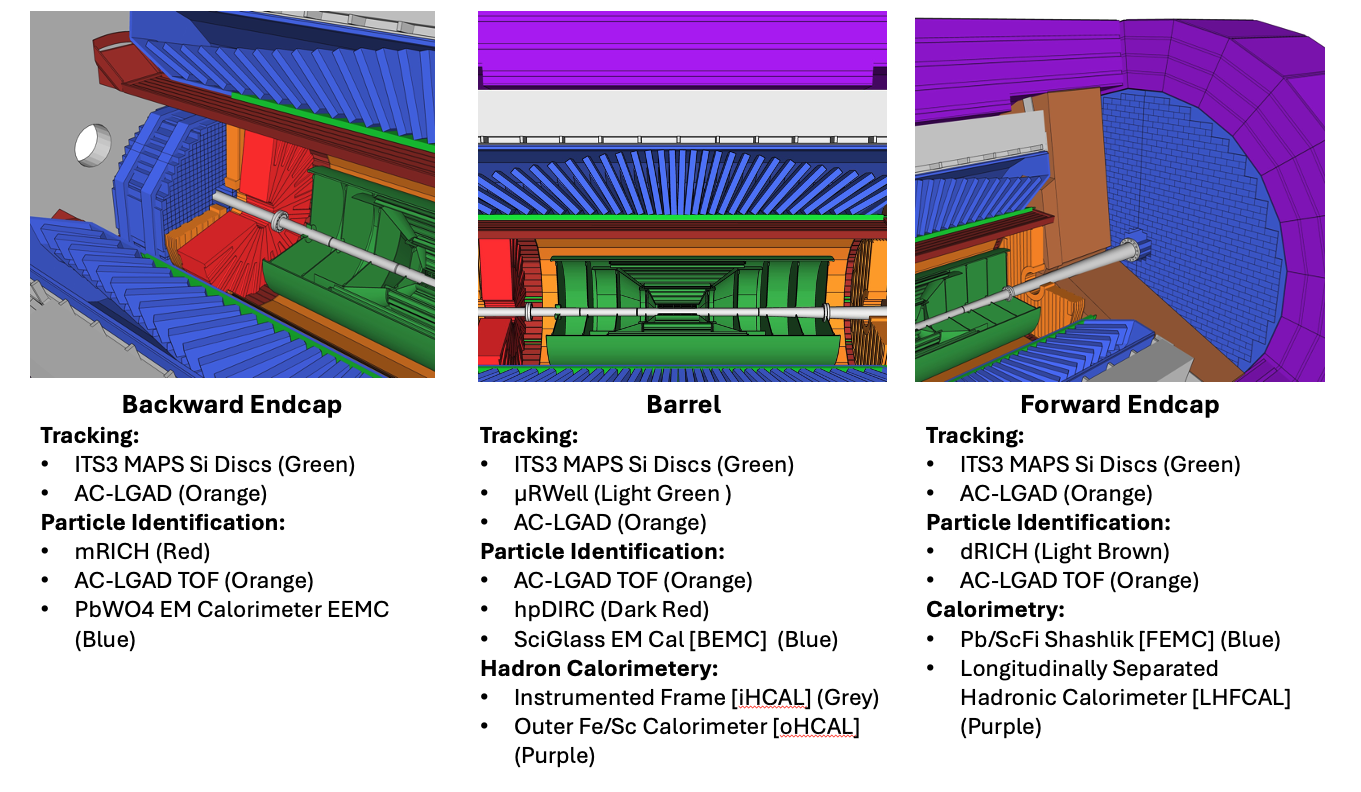}
        \caption{Principal components of the ECCE central detector: backward/electron endcap (left), barrel (center), and forward/hadron endcap (right).}
\label{fig:ecce-sectional-diagram}
\end{figure*}


The ECCE detector consists of three major components: the central detector (shown in Fig.~\ref{fig:ecce-sectional-diagram-1.1}) along with far-forward and far-backward system located tens of meters of the central detector within the accelerator ring.

The ECCE central detector has a cylindrical geometry based on the BaBar/sPHENIX superconducting solenoid, and has three primary subdivisions: the barrel, the forward endcap, and the backward endcap.
Henceforth ``forward" is defined as the hadron beam direction and backwards the electron beam direction. We will use electron or backward, and hadron or forward, interchangeably when describing the endcaps.

\input{sections/ecce_det_barrel_req}

\input{sections/ecce_det_fffb_req}

Table~\ref{tab:det-requirements} lists the physics requirements in the ECCE central detector, the technical challenges associated with its realization, and the ECCE solutions that achieve the stated goals.  Comments about future upgrade paths are also provided.
Table~\ref{tab:det-ff-fb-requirements} presents similar information for the far-forward and far-backwards regions. 
Figure~\ref{fig:ecce-sectional-diagram-1.1} shows the ECCE central detector and lists its key components and the technology selected for each sub-system.

Here, we provide general technical details on all eight of the systems and the components that make up the ECCE detector: magnet, barrel, backward endcap, forward endcap, far-forward detectors, far-backward detectors, electrons/DAQ, and computing.

\begin{description}

\item[Magnet] ECCE intends to reuse the BaBar super-conducting solenoid that is
currently installed in the sPHENIX experiment and will be
available after its conclusion. Its reuse for the EIC was the subject of an engineering study and risk analysis in 2020~\cite{BaBarJLABStudy} whose main conclusion was that the “magnet should be suitable for prolonged use as part of the detector system for the EIC project.”
Additional performance assessment will be
conducted during an sPHENIX long-duration high field test (at 1.4~T)
planned in 2022.
This test, followed by the first full duration run of
sPHENIX in early 2023, will validate the feasibility of its reuse in ECCE. 
Preparing the solenoid for reuse will involve proactive maintenance and several minor
modifications. We also plan to carry through a replacement magnet engineering and design assessment as risk mitigation, as described in Section~\ref{sec:risk}. 

\item[Barrel/Central Detector]
The ECCE central barrel region comprises the following seven subsystems:
\begin{description}
\item[]Silicon Tracker:
 The silicon barrel detector consists of five layers: three vertex layers close to the beam pipe and two middle layers  providing the central track sagitta measurements. All layers use the ITS-3 sensors with pixel pitch at 10~$\mu$m with an average material budget of 5\% of a radiation length, X$_0$, per layer.
\item[]$\mu$Rwell Tracker: The Si tracker is supplemented by two $\mu$Rwell layers, cost effective micro-pattern gas detectors, between the Si sagitta layer and the hpDIRC, and a single outer barrel $\mu$RWell layer between the DIRC and BECAL.
\item[]AC-LGAD TOF: A time of flight layer is placed just before the hpDIRC to provide a precise TOF measurement as well as an additional tracking point. See electron endcap discussion for details.
\item[]hpDIRC: 
The high-performance DIRC provides particle identification with three standard deviations separation for $\pi/K$ up to 6~GeV/$c$, $e/\pi$ up to 1.2~GeV/$c$, and $K/p$ up to 12~GeV/$c$.
\item[]BECAL:
The barrel ECAL (BECAL) is a homogeneous projective electromagnetic calorimeter built out of 8960 clear scintillating glass (SciGlass) towers, arranged in 70 rings in the $\eta$ direction, with 128 towers per ring along $\phi$.
The SciGlass towers have a front face of 4~cm$ \times $4~cm and are 55~cm deep including $\sim$10cm readout, providing 16~radiation lengths and better than 4\%/$\sqrt(E)$ + (1--2)\% energy resolution. This resolution surpasses the EIC Yellow Report requirement to complement the tracking system and ease electron identification and $\pi/e$ rejection, with an eye to the future high-luminosity EIC science needs.
The towers are slightly tapered to be nearly projective to the interaction point.
\item[]oHCAL/iHCAL:
The ECCE outer barrel hadronic calorimeter (oHCAL) is integrated into the barrel flux return for the ECCE solenoid magnet and has been built and will be optimized by the sPHENIX experiment.  It consists of 32 sectors of 1020 magnet steel, with an inner and outer radius of 1.9~m and 2.6~m respectively. Each sector is 6.3~m long and weighs 14~tons. The barrel inner HCAL (iHCAL) is a hadronic calorimeter that is integrated into the BECAL support frame. Its design consists of 32~sectors of stainless steel, with an inner radius of 135~cm and an outer radius of 138.5~cm.  
\end{description}

\item[Backward Endcap]
The ECCE backward endcap region for electron detection comprises the following five subsystems:
\begin{description}
\item[]Tracker:
The silicon electron endcap detector consists of four disks which provide precise measurements of charged tracks (especially electron tracks) in the backward pseudorapidity region. The technology for the silicon disks is the ITS-3 silicon sensor with pixel pitch at 10~$\mu$m. The detector mechanical structure design will be informed by the EIC eRD111 studies. In addition, the AC-LGAD TOF detectors described below will provide an additional high-precision tracking point after the disks at a large distance from the interaction point.
\item[]mRICH:
The design goal of the modular RICH (mRICH) is to 
achieve $3\sigma$ K/pi separation in the momentum range from 2 to 10~GeV/$c$.  Excellent $e/pi$ separation for momenta below 2~GeV/$c$ is expected.
In addition, the RICH detectors contribute to $e/\pi$ identification.   When combined with an EM calorimeter, the mRICH and hpDIRC will provide excellent suppression of the low-momentum charged-pion backgrounds, which limits the ability to measure the scattered electron in kinematics where it loses most of its energy. 
\item[]AC-LGAD TOF: A time of flight measurement using AC-LGAD technology will be used for PID in the momentum range below the Cherenkov detectors thresholds. These detectors also provide a high-precision tracking point.
\item[]EEMC:
 The Electron Endcap EM Calorimeter (EEMC) is a high-resolution electromagnetic calorimeter that is capable of providing precision scattered electron and final-state photon detection in the region $-3.7 < \eta < -1.5$.
 The detector is comprised of
 2~cm x 2~cm x 20~cm \PbWOiv crystals which provide 20 radiation lengths. The overall design concept is the same as in the EIC Yellow Report~\cite{AbdulKhalek:2021gbh}.
 \item[]Iron Flux Return:
 As an active electron endcap hadron calorimeter provides no substantial benefits to the scientific program, the iron flux return will be passive.
 We note that adequate space remains available for a possible upgrade path towards high-luminosity running allowing the measurement of the jet distribution in the low-$x$, high-$Q^2$ region.
\end{description}

\item[Forward Endcap]
The ECCE forward endcap region for hadron detection comprises five subsystems:
\begin{description}
\item[]Tracker:
The silicon hadron endcap detector consists of five disks, which provide precisely measured space points for charged particle tracking in the forward pseudorapidity region. This detector will improve the determination of the decay vertex of weakly decaying particles and measure the majority of the charged particles in asymmetric $e + p$ and $e + A$ collisions. The technology for the silicon disks is ITS-3 silicon sensor with pixel pitch of 10~{\textmu}m. The detector mechanical structure design will be informed by the EIC eRD111 studies. An AC-LGAD TOF detector placed in front of the dRICH will provide an additional high-precision tracking point.
\item[]AC-LGAD TOF:  time of flight measurement using
AC-LGAD technology will be used for PID in the
momentum range below the Cherenkov detectors
thresholds. These detectors also provide a high-
precision tracking point.
\item[]dRICH:
The dual-radiator Ring Imaging Cherenkov (dRICH) detector is designed to provide continuous hadron identification in the (outgoing) ion-side with 3$\sigma$ or more for $\pi/K$ from $\sim$0.7~GeV/$c$ to $\sim$50~GeV/$c$, and for $e/\pi$ from a approximately 200~MeV/$c$ up to $\sim$15~GeV/$c$.
\item[]FEMC:
The forward ECal (FEMC) will be a Pb-Scintillator shashlik calorimeter. Its towers have an active depth of 37.5cm with an additional 5cm readout space. Each tower consists out of 66 layers of alternating 1~cm$\times$1~cm$\times$0.16~cm Pb and 1~cm$\times$1~cm$\times$0.4~cm scintillator.
\item[]LFHCAL:
The forward HCal (LFHCAL) is a steel-tungsten-scintillator calorimeter. Its towers have an active depth of 1.4~m with an additional space for the readout of about 20-30~cm depending on their radial position. Each tower consists out of 120~layers of alternating 5cm x 5cm x 1.6cm steel and 5cm x 5cm x 0.4cm scintillator material and 20 layers of alternating tungsten and scintillator material of the same size. In each scintillator a loop of wavelength shifting fiber is embedded. Ten consecutive fibers in a tower are read-out by a single silicon photo multiplier, leading to 7~samples per tower, with the last 10 layers acting as tailcatcher.
\end{description}
\item[Far-forward detectors]
 The auxiliary detectors consist of a set of trackers and calorimeters that are closely integrated with the beam-line elements. The detector are designed to measure very forward particles to high precision with a high rejection of beam-related background. The far forward detection systems consist of the following components:
\begin{description}
\item[]B0 spectrometer:
The B0 spectrometer measures charged particles and photons at forward ($\eta > 3$) angles to facilitate studies of exclusive processes and general process characterization. This subsystem is designed for reconstructing charged particles with angles of $5.5 < \theta < 20.0$~mrad, and also large angle protons from nuclear breakup. The B0 detector is embedded in the first dipole magnet after the interaction point (B0pf). It consists of four layers of AC-LGAD tracking planes followed by an array of \PbWOiv crystals for the photon detection. The \PbWOiv array consists of 250 crystals, each 10 cm long with a surface area of 2x2 cm$^2$.
\item[]Zero-Degree Calorimeter:
The ZDC consists of a single unit with four different calorimeter layers. 
\begin{itemize}
            \item \PbWOiv Crystal calorimeter: This is a silicon pixel layer plus a layer of \PbWOiv crystals intended to measure low energy photons.  
            In front of the crystal layer, a silicon pixel layer is attached. 
            \item W/Si sampling calorimeter: This is an ALICE FoCal-E style calorimeter~\cite{ALICE:2020mso} and consists of alternating tungsten plates and silicon sensor planes. It is meant to measure the residual photon energy escaping the \PbWOiv Crystals in the shower development of high-energy photons and neutrons. 
            \item Pb/Si sampling calorimeter: This is a calorimeter with 3 cm-thick lead plane absorbers and active silicon pad layers, where the pad-layer design is as in the W/Si calorimeter.
            \item Pb/Sci. sampling calorimeter: This is to measure hadron shower energy and uses 3 cm thick lead plane absorbers with 2 mm-thick scintillator planes as active materials. The calorimeter is segmented as 10~cm x 10~cm on a plane and 15 layers of scintillator planes will be read together, making a tower.
        \end{itemize}
\end{description}
\item[Far-backward detectors] 
The auxiliary far-backward detectors consist of a low-$Q^2$ tagger and a luminosity monitor.
\begin{description}
\item[]Low-$Q^2$ tagger: The tagging system is made up of two detection systems which are located at different distances from the beam, each including two AC-LGAD tracking layers followed by a high-resolution \PbWOiv calorimeter.
\item[]Luminosity monitor: A forward \PbWOiv calorimeter with a passive x-ray absorber and a two-arm $e^+/e^-$ pair spectrometer measures the beam luminosity. Each includes AC-LGAD tracking layers and a high-resolution \PbWOiv calorimeter.
\end{description}

\item[Electronics/DAQ]The ECCE DAQ is a fully streaming readout (SRO) design capable of supporting high bandwidth to the Event Buffer and Data Compressor (EBDC) computers as well as high bandwidth to the data storage. A key component of this design is the Data Aggregation Module (DAM), for which we take as the current model the ATLAS FELIX board that will be used by sPHENIX in their hybrid streaming DAQ. We assume the development of a specific iteration of a FELIX-like board~\cite{Anderson_2016} as the DAM board for ECCE (also referred to as ”EIC-FELIX” in the text that follows) that will serve as a common interface for all of the subsystems. The use of a common interface reduces the number of electronics designs that need to be verified and supported in the experiment.

The general design of the ECCE data acquisition builds on the sPHENIX DAQ system, which already demonstrates almost all of the concepts of the envisioned ECCE DAQ system. However, while sPHENIX had to be a hybrid of triggered and streaming readout components, the ECCE DAQ will be built around a trigger-less Streaming Readout (SRO) concept, similar to the streaming readout systems currently in use at JLab and in the ALICE experiment at the LHC.

\item[Computing] The ECCE computing will be based on a distributed model with multiple sites for calibration, storage and analysis.  Computing resources must be sufficient to do calibrations and reconstruction in near real-time.    Disk space should be sufficient for holding up to 3 weeks to allow time for data quality checks.   Tape storage will used for long term backup of the filtered data.

\end{description}

\begin{figure*}[!thb]
\centering
    \includegraphics[width=1.8\columnwidth]{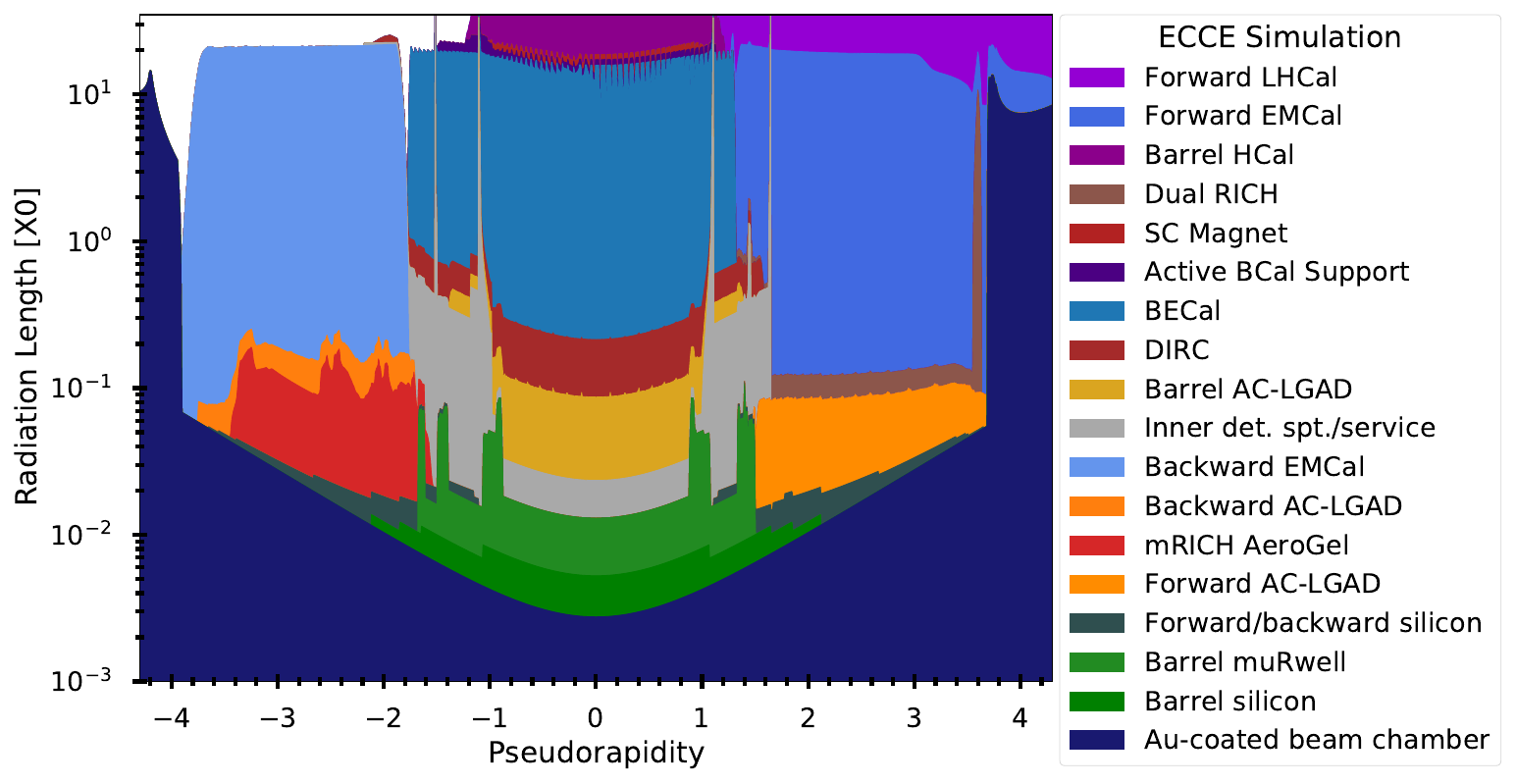}
    \caption{The stacked plot of material distribution in the ECCE detector subsystems, which is quantified as the radiation length that photons from the central interaction point observe and is averaged over azimuth.  }
    \label{fig:ECCE-materials}
\end{figure*}

Figure~\ref{fig:ECCE-materials} shows the material distribution of the ECCE central detector via a radiation length scan of the detailed ECCE \geant model. The large acceptance and low mass inner tracker (green) is hermetically enclosed by the PID detectors (red and yellow) and EM calorimetry (blue). Hadronic calorimeters further cover $\eta>-1.1$.

\section{Magnet} 

\begin{figure}[htb]
    \centering
    \includegraphics[width=0.9\columnwidth]{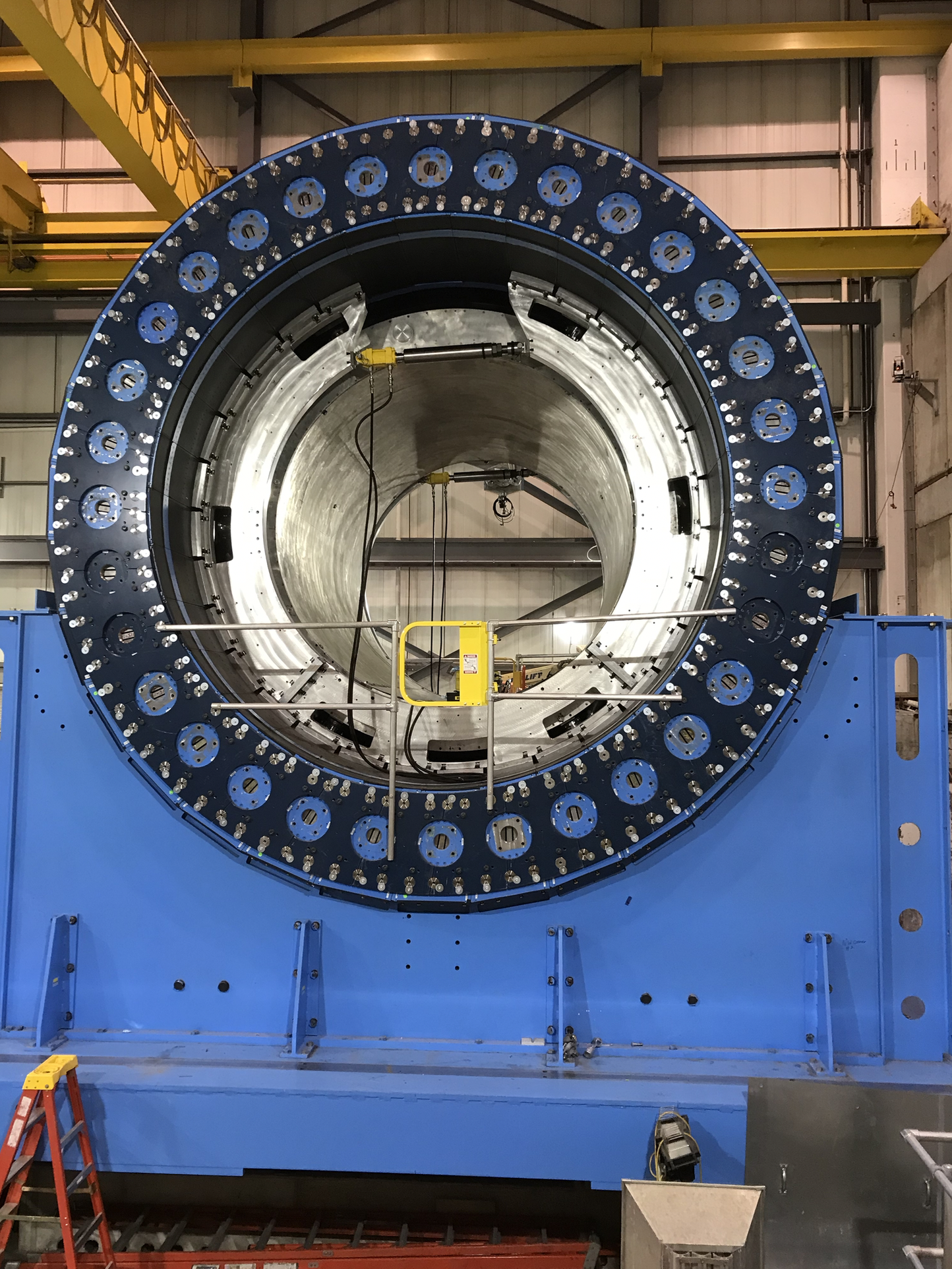}
    \caption{The BaBar solenoid in late February 2022, during installation in the sPHENIX experiment. The solenoid is surrounded by the barrel outer hadronic calorimeter and flux return. The barrel flux return (outer hadronic calorimeter) and BaBar solenoid are items planned to be reused by the ECCE experiment. The experimental cradle may also be reused.}
    \label{fig:magnet-2}
\end{figure}
\begin{table}[t]
  \caption{Design parameters of the BaBar superconducting solenoid.}
  \label{tab:magnet-1}
  \small
  \begin{tabular}[b]{ll}
    \toprule
    Central Induction &  1.5~T$^{*}$ (1.4~T in ECCE flux return) \\ 
    Conductor Peak Field &  2.3~T \\ 
    Winding structure &  Two layers, graded current density \\ 
    Uniformity in tracking region &  $\pm3\%$ \\ 
    Winding Length &  3512 mm \it{at R.T.} \\ 
    Winding mean radius &  1530 mm \it{at R.T.} \\ 
    Operating Current &  4596 A (4650 A$^{*}$)\\ 
    Inductance &  2.57 H (2.56 H$^{*}$)\\  
    Stored Energy &  27 MJ\\ 
    Total Turns &  1067 \\ 
    Total Length of Conductor &  10,300 m \\ \bottomrule
    $^{*}$ Design Value &   \\ 
  \end{tabular}
\end{table}

ECCE plans to reuse the BaBar/sPHENIX solenoid and the surrounding combined hadronic calorimetry and flux containment system for the magnet.

The magnet for the BaBar experiment at PEP-II at SLAC was manufactured by Ansaldo in 1997 and commissioned in 1998. It was transferred to BNL in 2015 for use in the sPHENIX experiment and passed an initial high-field test (up to 1.3~T) in 2018. Its main design parameters are listed in Table~\ref{tab:magnet-1}. For an EIC detector the region covered by the barrel detectors should span a  pseudo-rapidity $-1<\eta<1$, corresponding to an angle of $\sim$40 degrees. This corresponds well with the BaBar solenoid, which has a warm bore diameter of 2.84 meters and a coil length of 3.512 meters, corresponding to a 39 degree angle. 

The reuse of the BaBar solenoid for the EIC was the subject of an engineering study and risk analysis in 2020~\cite{BaBarJLABStudy}.  The main conclusion of the assessment was that the “magnet should be suitable for prolonged use as part of the detector system for the EIC project.” The report also suggested the implementation of several maintenance and improvement modifications, including new protection circuits such as voltage taps, inspection and, as needed, reinforcement of the internal mechanical support, including new strain gauges, and replacement of control instrumentation sensors. The implementation of some of these suggestions would involve opening the magnet cryostat, which could create additional risk of magnet failure. In 2021 JLab engineers revisited the risk analysis and, following extensive discussions, decided that any modifications or refurbishment that require opening the BaBar solenoid cryostat would not be worth the additional risk~\cite{ecce-note-det-2021-01}. They further noted that no such actions will be necessary if the magnet continues to operate well throughout a high-field magnet test with the sPHENIX experiment flux return (which will also be re-used for ECCE) in mid-2022 and subsequent initial sPHENIX experimental operations starting from 2023 until 2025.

Further magnet engineering studies of the ECCE detector magnet indicate that the unbalanced forces on the magnet are small, a net force of 4kN or less than 1000 lbs, because the magnetic field at the locations of the ECCE forward and backward calorimeters are small and most of the magnetic flux is returned through the barrel. These small forces should not present a substantial engineering difficulty in the proposed ECCE configuration.

The scope of the reuse of the BaBar solenoid in ECCE includes a review by a panel of experts (following initial sPHENIX running), the disconnect of the magnet in IP-8 and move to IP-6, a new valve box, and assembly and magnet mapping in IP-6. The risk mitigation strategy associated with the reuse of the BaBar solenoid, including the design of a potential replacement magnet, are discussed in Section~\ref{sec:risk}. 

\section{Tracking}
\label{sec:tracking}

\begin{figure}[!t]
\centering
    \includegraphics[width=0.49\textwidth]{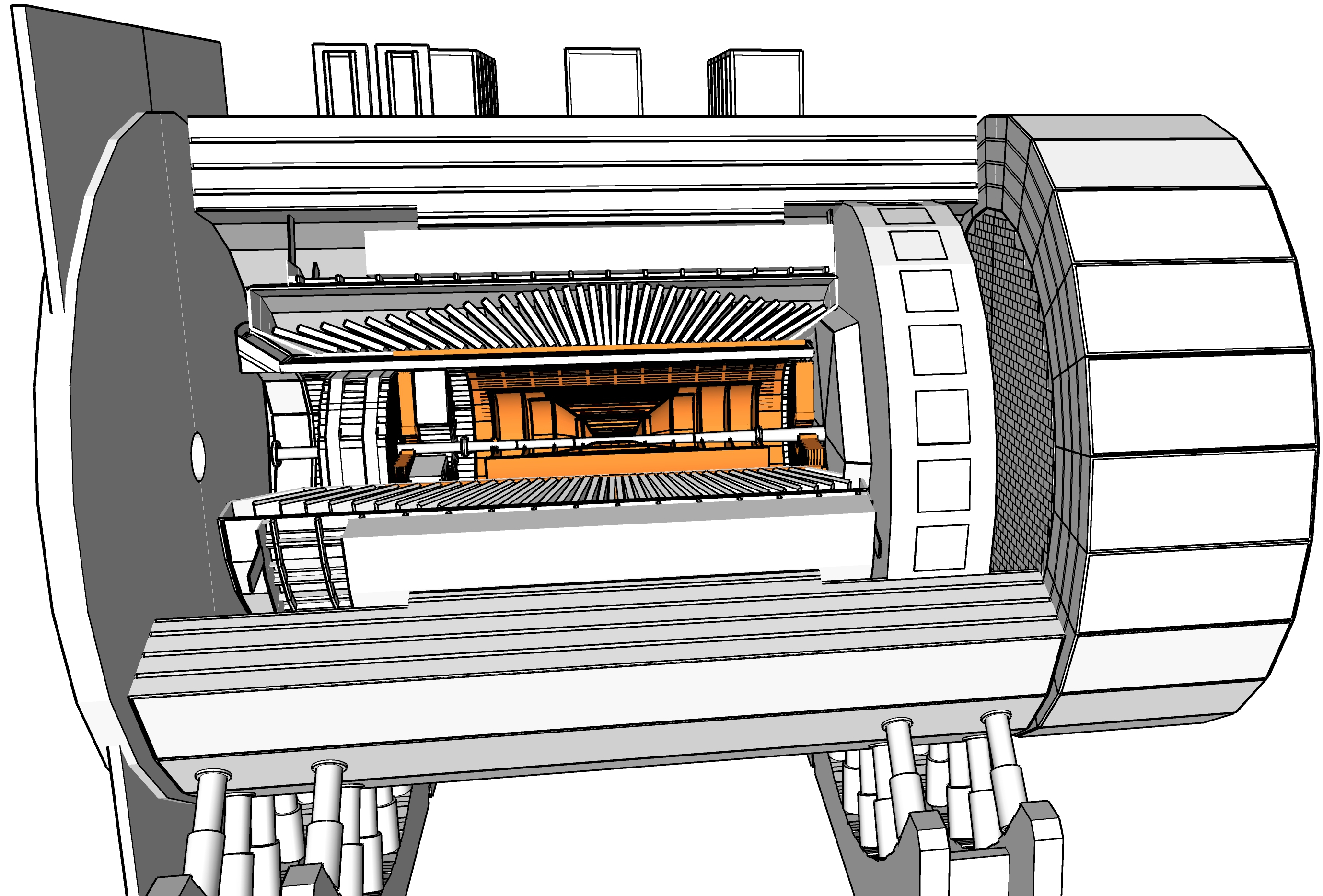}
    \includegraphics[width=0.49\textwidth]{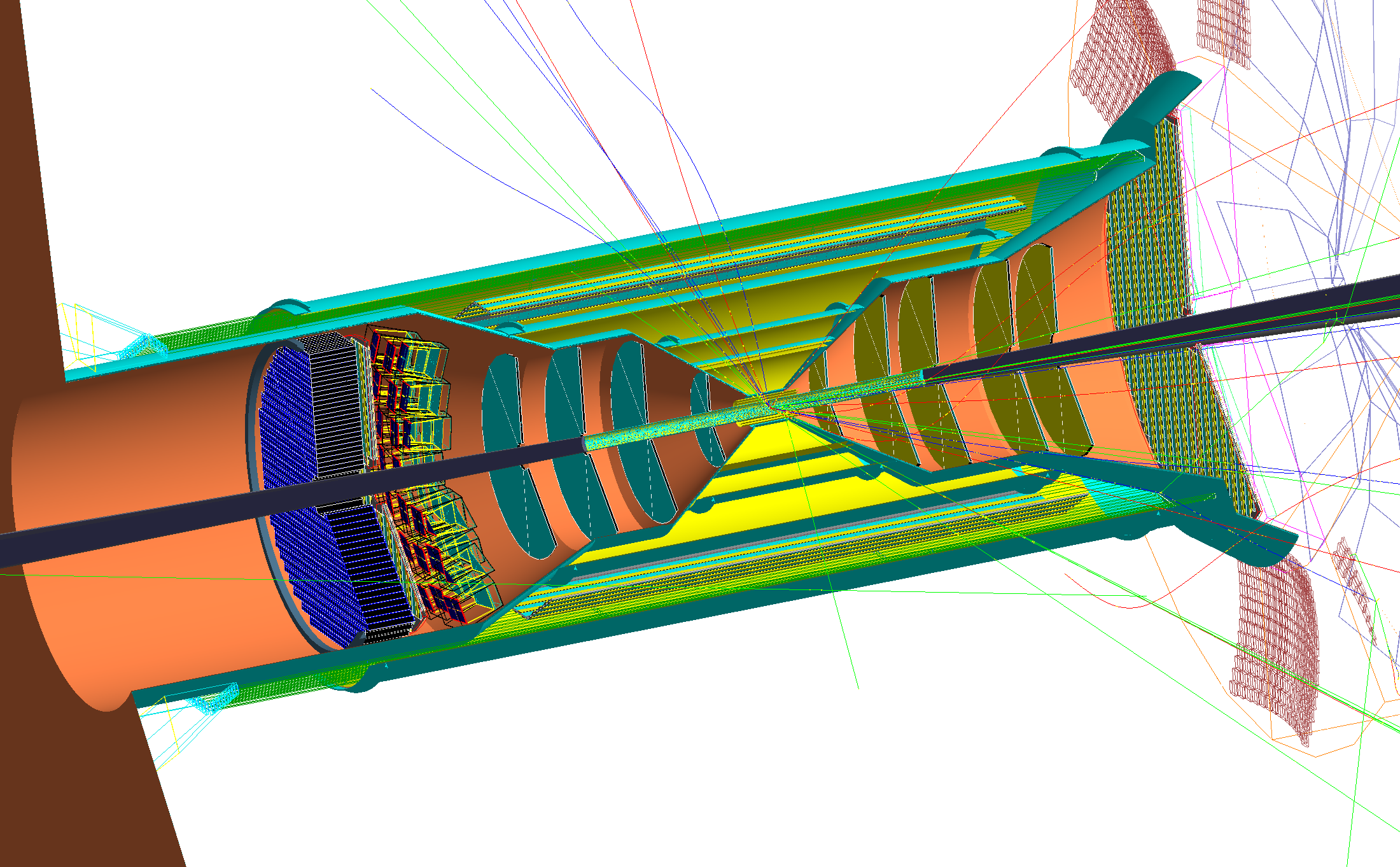}
    \caption{Tracking system of ECCE in mechanical model (top) and \geant simulation (bottom). The tracking system is tightly integrated with the PID detectors which is also shown on the right. Support and cabling for the these detectors was implemented (copper-colored cylinder-cone) to count for its material and acceptance effects.}
    \label{fig:tracking}
\end{figure}
\begin{figure*}[th!]
    \centering
    \includegraphics[width=\textwidth]{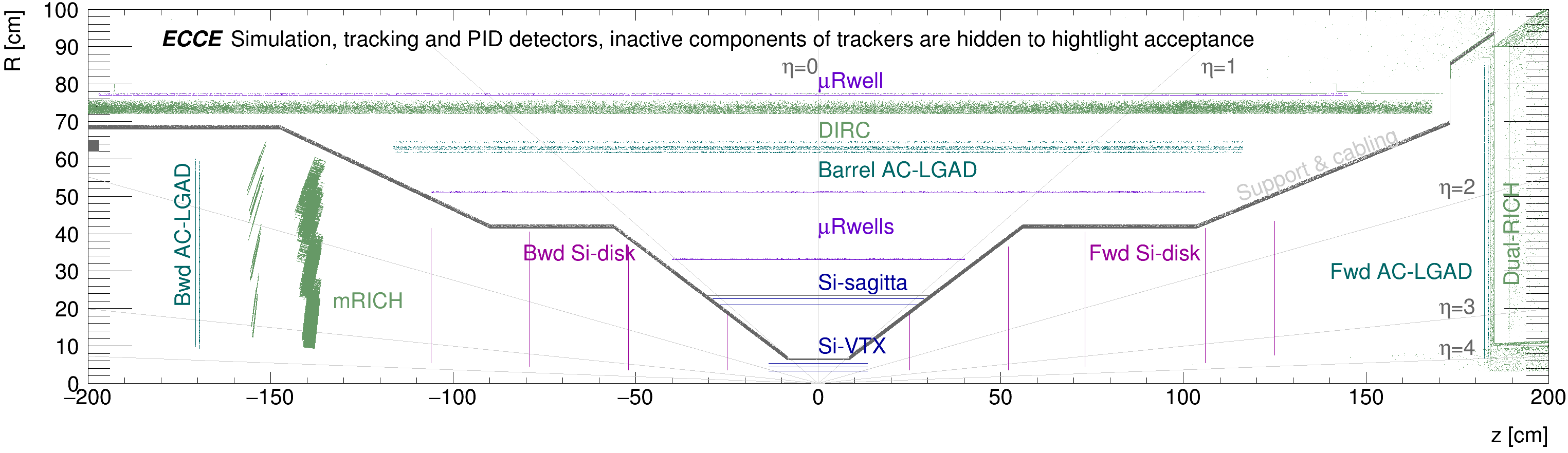}
    \caption{Schematic view of the ECCE tracker, including silicon, \mrwell, AC-LGAD, DIRC, mRICH and dRICH detector systems. {\bf{Need better version of this image.}}}
    \label{fig:ecce-tracker-upper-hemisphere}
\end{figure*}

ECCE features a hybrid tracking detector design (Figure~\ref{fig:tracking}) using three state-of-the-art technologies to achieve high precision primary and decay vertex determination, as well as, excellent tracking momentum and distance of closest approach resolution in the $|\eta| \leq 3.5$ region with full azimuth coverage~\cite{ecce-note-det-2021-03}. 
The ECCE tracking detector consists of the Monolithic Active Pixel Sensor (MAPS) based silicon vertex/tracking subsystem, the \mrwell tracking subsystem and the AC-LGAD outer tracker, which also serves as the ToF detector. 
The ECCE tracking design has been optimized assisted by Artificial Intelligence (AI) taking into account BaBar magnet coverage, integration with the other detector subsystems, and cost. 

The tracker geometry is shown in schematic form in Fig.~\ref{fig:ecce-tracker-upper-hemisphere} which displays the detector in the $R-z$ plane.
The barrel layers centered at $z=0$ have a cylindrical geometry, while the endcap layers centered at $R=0$ are disks oriented around the $z$ axis. 
The MAPS silicon detector contains 3-layer silicon vertex layers, 2-layer silicon sagitta layers, five disks in the hadron endcap and four disks in the electron endcap region. 
This silicon vertex/tracking detector provides the desired primary vertex and displaced vertex reconstruction also documented in the EIC yellow report~\cite{AbdulKhalek:2021gbh} and the tracking momentum and DCA$_{2D}$ resolutions (see Fig.~\ref{fig:tracking-momentum-performance} and Fig.~\ref{fig:tracking-dca-performance}) for heavy flavor measurements. 
For the barrel layers at large radii, which have the largest surface area, cylindrical \mrwell gas trackers are used to optimize performance at reduced overall cost. These are introduced both right outside the Si tracker and in front of the barrel EM calorimeter.
In addition, an AC-LGAD based ToF layer in each section  provides a precision space-time measurement on each track.
The tracking system is thus tightly integrated with the PID detectors.

\subsection{MAPS}

The silicon vertex and sagitta layers utilize MAPS technology, as implemented in high-precision (10~$\mu$m pixel pitch and $\sigma_{xy} = 2.9~\mu$m) low-material ($0.05\%$/layer) ALICE-ITS-3-type  sensors~\cite{ALICEITSproject:2021kcd, colella2021alice}, used in both cylindrical and disk configurations. 

The MAPS detector systems have been costed based on the TowerJazz 65nm production line. This technology is in the prototype sensor design and characterization stage. 
Recent R$\&$D on the ITS-3 has delivered a 32 by 32 pixel matrix prototype sensor using the 65nm production line that is undergoing beam test studies at CERN. 
Validation of the curved ALPIDE (ITS-3) sensor performance was obtained by early beam test results. 
The mechanical design for the silicon tracking detector, especially for the stave and disk layout and assembly, is led by the ongoing EIC R$\&$D project eRD111. 
Reduction of the material budgets for the EIC silicon tracking detector service parts is also being studied as part of the EIC eRD104 project. Alternative silicon technologies have been explored such as the Depleted MAPS (DMAPS), and progress in the MALTA DMAPS technology has been reported in~\cite{Li:2021kus}.
All these R$\&$D activities align with other major project upgrades or construction projects such as the ALICE ITS-3 upgrade. The required sensor R\&D is included in the ECCE detector R\&D plan. 

\subsection{\texorpdfstring{\mrwell}{muRwell}}
\begin{figure}[t]
\centerline{
\includegraphics[width=0.42\textwidth]{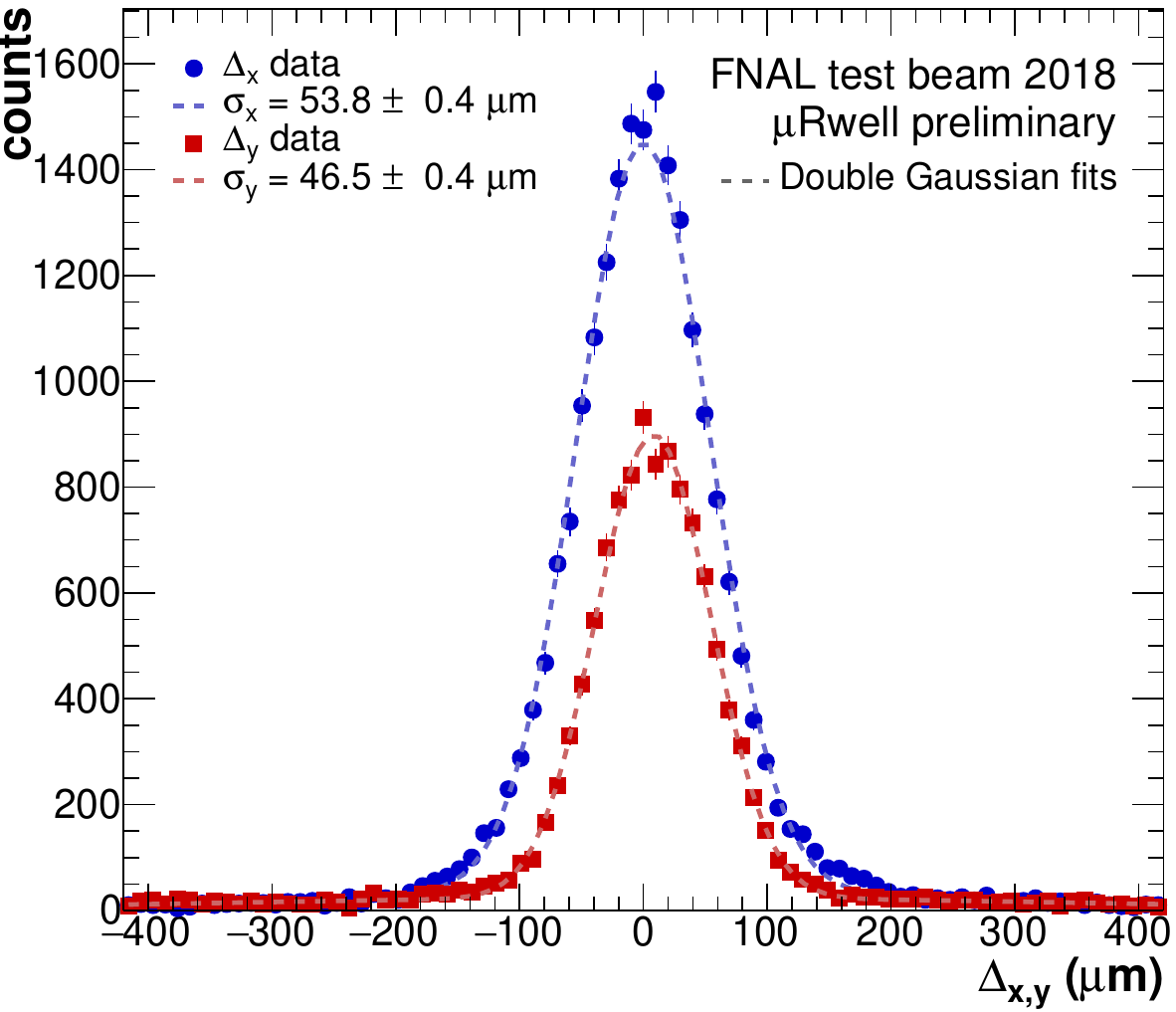}}
\caption{ Preliminary results of spatial resolution performances  of  the \mrwell prototype with 2D  X-Y strip readout layer.}
\label{fig:urwell}
\end{figure}

In ECCE \mrwell layers will form three barrel tracking layers  further out from the beam-pipe than the silicon layers. The barrel gas tracker layers include two inner barrel \mrwell layers, as well as a single outer barrel \mrwell.
The \mrwell technology is a single-stage amplification Micro Pattern Gaseous Detector (MPGD) that is a derivative of Gas Electron Multiplier (GEM) technology. It features a single kapton foil with GEM-like conical holes that are closed off at the bottom by gluing the kapton foil to a readout structure to form a microscopic {\it well} structure.  The technology shares similar performances to GEM detectors in term of rate capability, while providing a better spatial resolution than GEM. Furthermore, compared to GEMs, $\mu$RWELL has distinct advantages of flexibility, more convenient fabrication and lower production cost making it the ideal candidate for large detectors. Large area \mrwell foils have already been developed and manufactured at CERN. 
All \mrwell detectors will have 2D strip based readout. The strip pitch for all three layers will be 400~$\mu$m. Figure~\ref{fig:urwell} shows the resolution results from a \mrwell prototype detector in test beam at Fermilab (June-July 2018) as part of the EIC eRD-6 activities. 
The measurements were done using a beam hitting the detector perpendicularly, and using detailed MC simulations we estimate a 55$\mu$m resolution for a curved geometry where the particle hits the detector at an angle.

The Korean institutions in the ECCE collaboration will manufacture the \mrwell foils for the ECCE \mrwell detectors. Specifically, a Korean manufacturer (Mecaro) has demonstrated that they can produce high quality large MPGD foils for the CMS detector at the LHC, working in conjunction with member institutions of the Korean ECCE collaboration.  In addition, Chinese institutions in the ECCE collaboration have experience with the DLC resistive coating required for \mrwell detectors.

\begin{figure*}[t]
\centering
\includegraphics[width=\textwidth]{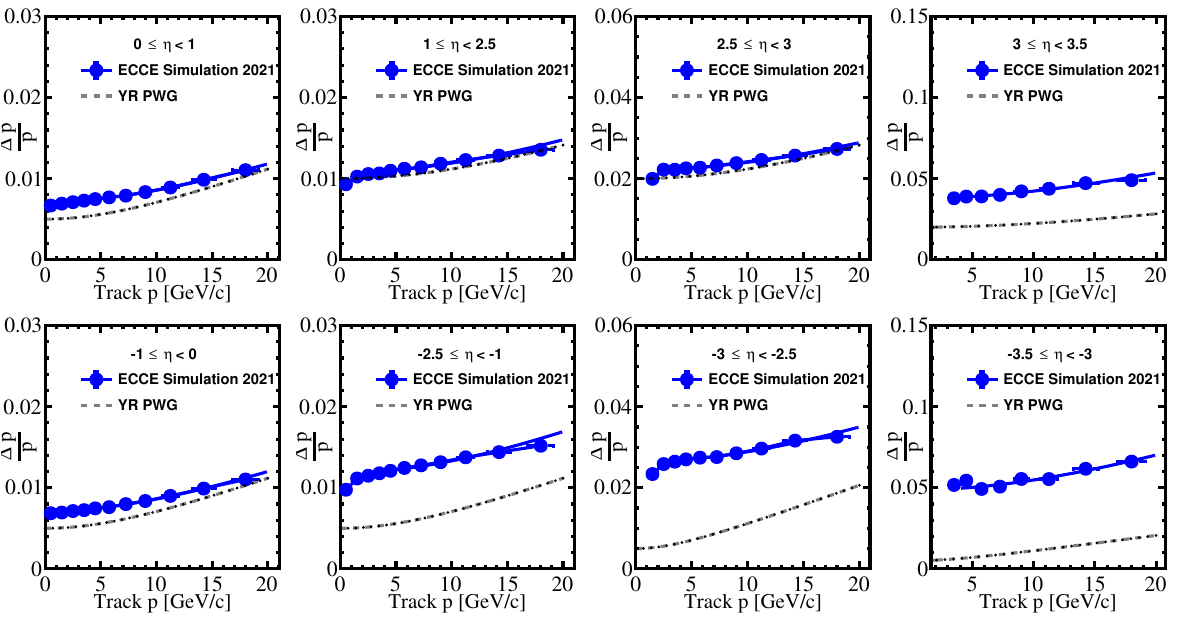}
\caption{ECCE pion track momentum resolution (data points) with the EIC Yellow Report PWG requirements for the tracker indicated by the dashed lines. Note that the ECCE performance simulations take into account materials for readout and services. The impact of these can be observed most clearly in the bins covering the barrel/barrel endcap transition regions. 
As an integrated EIC detector with all subsystems operating in a complementary way, ECCE achieves the EIC physics goals.
}
\label{fig:tracking-momentum-performance}
\end{figure*}
\begin{figure*}[t]
\centering
\includegraphics[width=\textwidth]{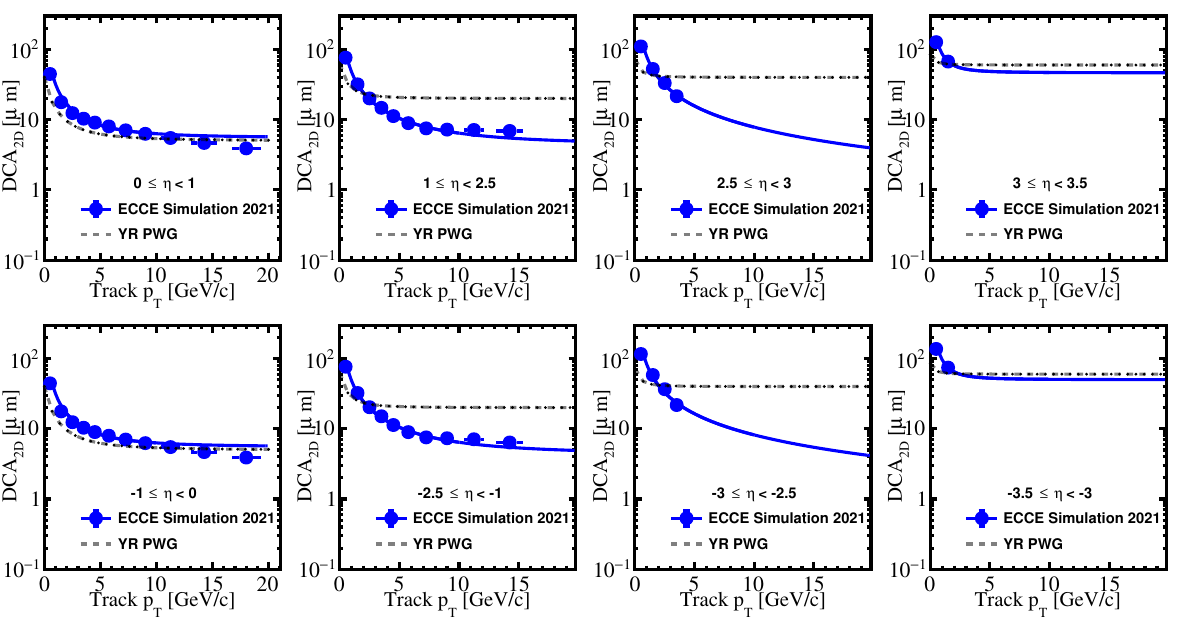}
\caption{Pion DCA$_{2D}$ resolutions (data points), which is compared to the EIC Yellow Report PWG requirement (dashed lines). The ECCE DCA resolution is consistent with Yellow Report requirements.}
\label{fig:tracking-dca-performance}
\end{figure*}

\subsection{AI optimization}
A framework for Multi-Objective Optimization (MOO) has been incorporated into the ECCE detector design simulation. AI has played a crucial role in  choosing 
the combination of technologies for the inner tracker. The choice of having ITS-3 and the \mrwell gas tracking layers, as well as the disk minimum radii were supported by AI. This has been an iterative process that evolved over time and required interplay between the ECCE teams working on Physics, Detector and Computing. 

Our approach deals with a complex optimization in a multidimensional design space driven by multiple objectives that encode the detector performance, while satisfying several mechanical constraints. 
This framework has been developed for the optimization of the inner tracker of ECCE and can in principle be extended to another sub-detector or to a system of sub-detectors, provided a viable parametrization of the detector simulation can be produced. 
Different parametrizations of the inner tracker design have been explored and most of our studies have been characterized by at least 11 parameters in the design space characterizing the location of the tracking layers in the central region and the disks in the two endcaps. 
The parametrization has been extended to include the support structure in the design optimization process and more recently to the outer tracking layers.
The different designs have been optimized with particle gun samples of pions and then studied and validated with independent data samples and physics analyses. 
At least three objective functions have been optimized simultaneously. In particular, for a 3-objective problem we utilized the momentum resolution, the polar angular resolution along with the Kalman filter efficiency of $\pi$ tracks. 
This problem has been tackled with evolutionary algorithms to assist the design during the detector proposal. A recently developed framework for MOO, {\sc pymoo}~\cite{blank2020pymoo}, has been implemented which supports algorithms like NSGA-II and NSGA-III~\cite{ishibuchi2016performance} and distributed evaluation with task scheduler like {\sc Dask}~\cite{rocklin2015dask}.
  
This approach accommodated both mechanical and geometrical constraints during the optimization process. In our studies we included at least 5 constraints (\textit{e.g.}, the outermost location as well as the difference between the outer and inner radius of a disk, or the radius of the outermost layer in the inner tracker). Overlaps in the design are excluded by a combination of constraints, ranges for the exploration of the parameters and internal checks done before and during the entire optimization process. Further details can be found in~\cite{ecce-note-comp-2021-03}.

The AI-assisted design has been used as input to multiple iterations of the ECCE tracker design, which led to the current tracker layout~\cite{ecce-note-det-2021-03} (Fig.~\ref{fig:tracking} and~\ref{fig:ecce-tracker-upper-hemisphere}), and is also contributing to the ongoing project R\&D to reduce the impact of readout and services on the tracking resolution as discussed in Section~\ref{sec:on-going-tracking}. 
  

\subsection{Expected backgrounds}

Vacuum and background estimates were done in joint working group meetings across both the ATHENA and ECCE proto-collaborations. 
A detailed simulation study was carried out to assess the collision signal and background from beam gas and synchrotron radiation on tracking detectors in the BaBar magnetic field~\cite{ecce-note-comp-2021-02}. Although the beam gas background was found to be small, the synchrotron radiation on the MAPS-based silicon trackers can be very significant and its uncertainty is large at this stage of the EIC design. A high-Z coating in the Be-section of the beam pipe (e.g. 2~$\mu$m Au coating) was shown to reduce the synchrotron hit rate in the silicon vertex tracker by four orders of magnitude resulting in a manageable hit rate~\cite{ecce-note-comp-2021-02,osti_1765663}. Therefore, all ECCE studies adopted such synchrotron shielding coating which introduces 0.06\%~$X_0$ (at $\eta=0$) of additional material to the beam pipe ($\sim 30\%$ relative increase). 

\subsection{Tracking performance}

The performance of the ECCE reference tracker design has been studied using single pions propagated through the ECCE \geant simulation framework. 
The momentum resolution is presented in Fig.~\ref{fig:tracking-momentum-performance} together with the Yellow Report requirement indicated as the dash lines. In the region, $-1<\eta<3$, the ECCE momentum resolution is consistent with Yellow Report physics requirements~\cite{AbdulKhalek:2021gbh}.   Between $1<\eta<1.5$ a substantial deviation is observed that is not obvious in Figure~\ref{fig:tracking-momentum-performance}. This difference is expected from the material for readout and services (copper-colored structure in right of Figure~\ref{fig:tracking}), whose impact is largest in this region. 
Further AI-assisted optimization in this region is on-going as discussed in Section~\ref{sec:on-going-tracking}.

In the backward region $\eta<-1.0$ and in the most forward bin the ECCE momentum resolution provided by tracking exceeds the Yellow Report requirements~\cite{AbdulKhalek:2021gbh}. However, ECCE is an integrated detector and in this region the physics performance, and in particular for $\eta<-2.5$, is achieved through excellent EM calorimetry.  Due to the limited time since the call for proposals to produce and analyze complete \geant simulations for physics performance, many of the physics studies used only tracking.  Nevertheless, these studies all show sufficient performance for the EIC physics program.  The addition of the calorimetry information will only improve these results, as shown for key physics topics.

We further note the dominant Yellow Report requirement for the momentum resolution in the backward region is driven by coherent \jpsi\  production on the nuclei, and in particular the $t$-reconstruction from the forward particles. Nonetheless, the ECCE physics studies have shown that for either a 1.4~T or even a 3.0~T field strengths the $t$-reconstruction resolution is dominated by the calorimeter.

The resolution of measurements of distance-of-closest-approach (DCA$_{2D}$), which is critical for heavy flavor measurements, is provided in Fig.~\ref{fig:tracking-dca-performance} and also compared with Yellow Report requirement. The ECCE DCA resolution is consistent with the Yellow Report, and will enable robust physics programs in heavy flavor measurements~\cite{Li:2022irb} and beyond standard model search.

\subsection{Ongoing R\&D for support structure optimization}
\label{sec:on-going-tracking}

\begin{figure}[t]
    \centering
    \includegraphics[width=\columnwidth]{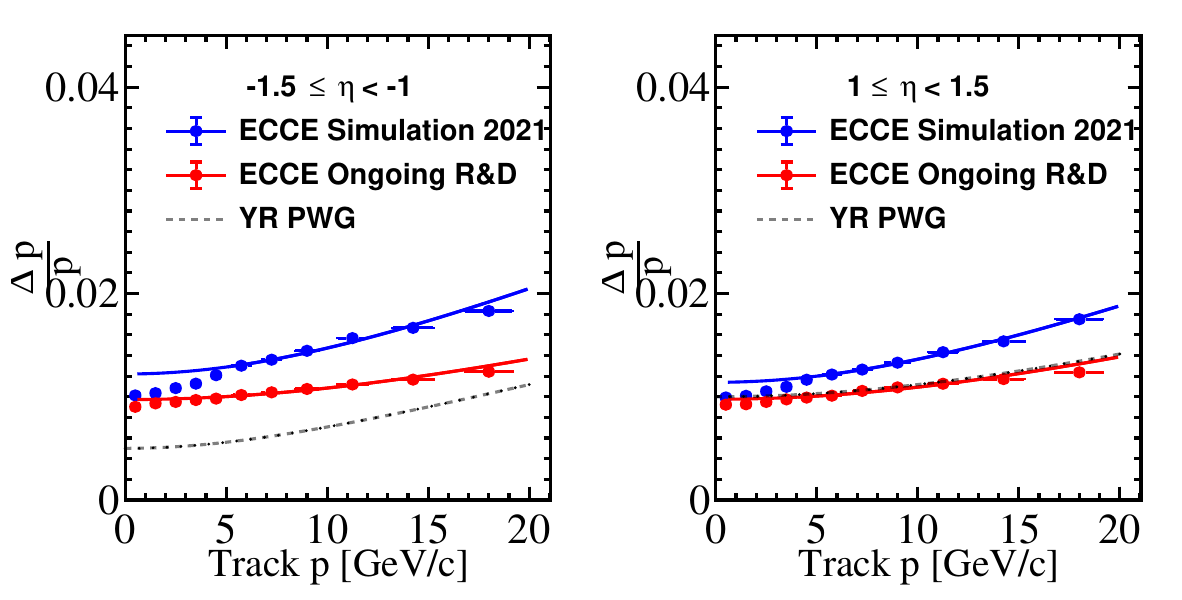}
    \caption{
 The momentum dependence of the tracker momentum resolution for the ECCE reference tracker design (ECCE Simulation, blue solid circles) and for the projective mechanical support design of the ECCE ongoing project R\&D that will continue after the proposal (red solid circles). The latter shows a reduction of the impact of readout and services on the tracking resolution. Note that the backward region (left panel) relies on the EM calorimeter, and thus a resolution larger than the EIC Yellow Report PWG requirement is acceptable.  
}
    \label{fig:improved_with_AI}
\end{figure}

Given the importance of the service structure in the tracking detector, the reduction of the impact of readout and services on tracking resolution is subject of ongoing R\&D and ECCE has made tremendous progress on this front using AI. 
The AI investigation in the ECCE framework focused on optimizing the tracker design with a projective support cone structure that reduces the amount of material a particle traverses. The design concept is illustrated in the Tracking document~\cite{ecce-note-det-2021-03} and more details on the AI based studies can be found in \cite{ecce-note-comp-2021-03}. The momentum resolutions resulting from this investigation are shown in Fig.~\ref{fig:improved_with_AI}. The largest impact is in the region between the central barrel and endcaps ($1<\eta<1.5$ and $-1.5<\eta<-1$) while the tracking momentum resolution in the central barrel as well as at large pseudo-rapidities ($|\eta|> 1.5$) is largely unaffected. 

\begin{figure*}
    \centering
    \includegraphics[width=\textwidth]{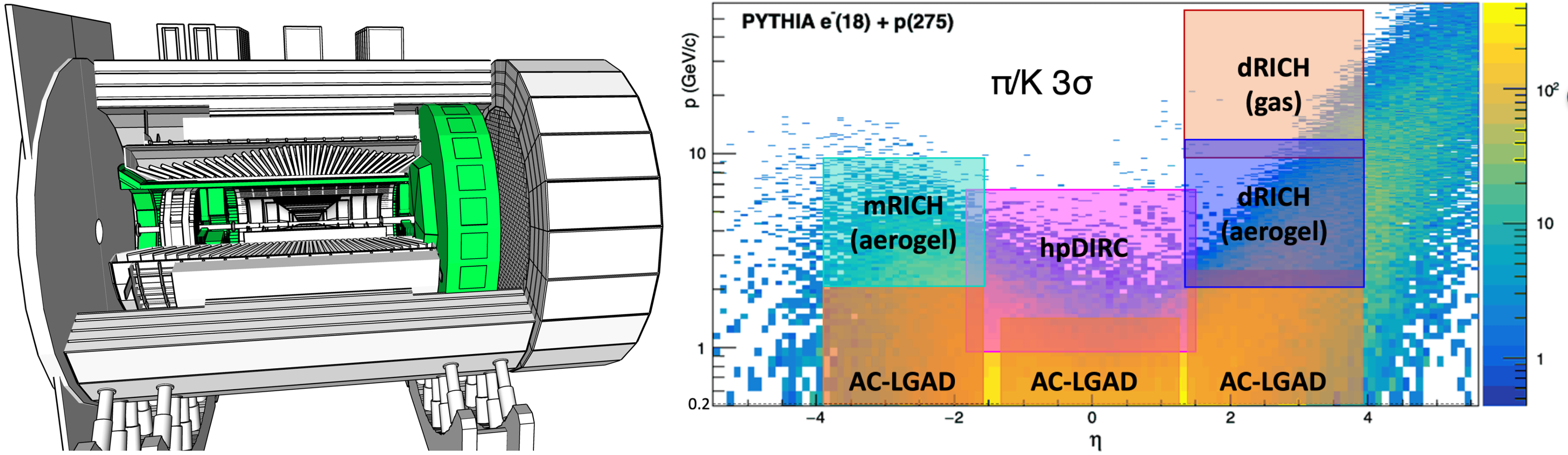}
    \caption{\label{fig:pid} Left: 3D model of the ECCE detector with the PID systems highlighted.
    Right: Expected 3 standard deviations $\pi/K$ separation coverage for the ECCE PID systems as a function of the particle momentum and pseudo-rapidity. Full coverage is achieved by making use of the veto mode of the Cherenkov detectors, complementing the TOF PID in the low momentum region. 
    }
\end{figure*}

\section{Particle Identification}
The ability to identify hadrons in the final state is a key requirement for the physics program of the EIC. 
Being able to tag the flavor of the struck quark in semi-inclusive DIS can, for instance, provide valuable information about the transverse momentum distributions~\cite{Seidl:2022dmh, Seidl:2022dhh} (and potentially orbital angular momentum) of the strange sea quark, while open charm (with subsequent decays into kaons) is important for probing the distribution of gluons in protons and nuclei. 

The choice of ECCE PID detector technologies was based on the outcome of the EIC generic R\&D program (eRD14 EIC PID Consortium and eRD29 on TOF with the LGADs technology), started in 2015, and in line with the baseline EIC detector concept in the Yellow Report~\cite{AbdulKhalek:2021gbh}. 
The longitudinally compact, modular RICH (mRICH)~\cite{Sharma:2024bjc,DelDotto:2017zbt}, the radially thin high-performance DIRC (hpDIRC)~\cite{Kalicy:2024jme}, the dual-radiator RICH (dRICH)~\cite{DelDotto:2017zbt,Rignanese:2024uhs}, and AC-LGADs based TOF, provide excellent PID over a wide momentum range for the full final state phase space~\cite{ecce-note-det-2021-04}.
The geometries of all PID detectors were optimized to fit the ECCE baseline design while maintaining the required performance. 
Figure~\ref{fig:pid}~(left) shows the four PID systems in a 3D model of the ECCE detector and (right) their $\pi/K$ separation coverage as a function of momentum and pseudo-rapidity for a sample of physics events. 
Compared to the Yellow Report reference detector, a number of key design features of the PID systems were optimized for ECCE.

\begin{figure*}[htb]
    \centering
    \includegraphics[width=1.0\textwidth]{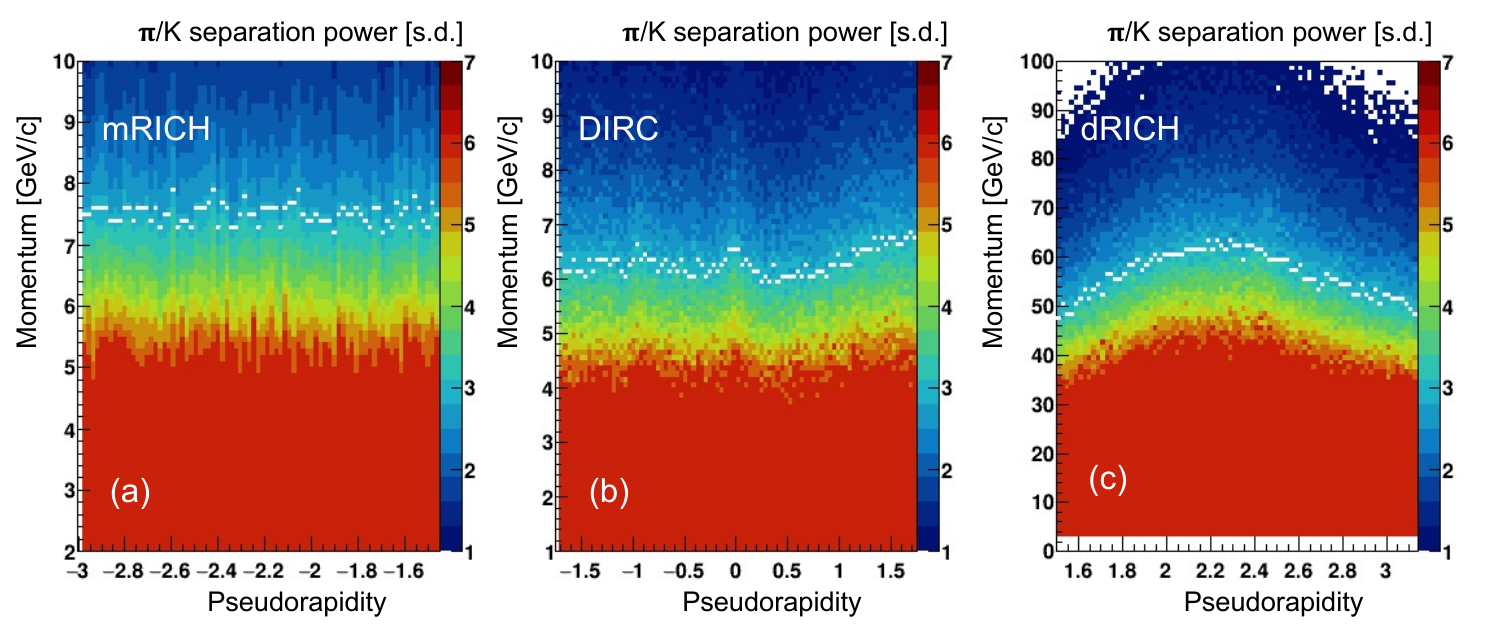}
    \caption{Parametrized $\pi/K$ separation power in ECCE as a function of particle momentum and pseudo-rapidity for mRICH (a), hpDIRC (b), and dRICH (c) based on standalone full \geant simulation and analytical calculation. The white symbol marks the maximum momentum for 3 standard deviations $\pi/K$ separation in each pseudo-rapidity bin.
    }
    \label{fig:PID_sd_1}
\end{figure*}

The expected PID performance of the three ECCE Cherenkov detectors, mRICH~\cite{Sharma:2024bjc,DelDotto:2017zbt},  hpDIRC~\cite{Kalicy:2024jme}, and  dRICH~\cite{DelDotto:2017zbt,Rignanese:2024uhs}, was obtained from standalone \geant simulation and analytical calculations, parametrized and used as input into the ECCE physics studies.
Figure~\ref{fig:PID_sd_1} shows the parametrized $\pi/K$ separation power in units of the number of standard deviations as a function of pseudo-rapidity and momentum for conservative assumptions for the tracking angular resolution.

Note that subsequent tuning of the PID detector geometries and reconstruction algorithms resulted in further improvement of the PID performance, which are not reflected in the shown parametrization.
The resulting momentum coverage for the separation of $e/\pi$, $\pi/K$, and $K/p$ with three standard deviations or more is summarized in Table~\ref{table:ECCE_PIDec} for the three ECCE Cherenkov systems.
The Cherenkov system performance is further separated into the nominal ``Ring Imaging" mode of operation, which provides positive ID of the particle type, and the so-called ``threshold mode" or ``veto mode", which uses the number of Cherenkov photons in excess of the expected background to differentiate between particle types above or below the threshold for Cherenkov light emission. The combined performance of the ECCE Cherenkov detectors meets or exceeds the ECCE PID requirements.   

\begin{table}[!t]
\centering  
\small
\caption{Summary of the PID performance of the ECCE Cherenkov systems (momentum  coverage in GeV/$c$).}
\begin{tabular}{cccccc}
\toprule
 \multirow{2}{*}{PID}  &   \multirow{2}{*}{Mode}    &  \multirow{2}{*}{mRICH}  &  \multirow{2}{*}{hpDIRC}  & \multicolumn{2}{c}{dRICH} \\
 &  &  &   & aerogel & gas\\
\midrule
$\pi/K$ & Ring Imaging & $2-9$ & $1-7$  & $2-13$ & $12-50$ \\
& Threshold  & $0.6-2$ & $0.3-1$  &  $0.7-2$ & $3.5-12$ \\
$e/\pi$ & Ring Imaging & $0.6-2.5$ & $<1.2$ & $0.6-13$ & $3.5-15$ \\
& Threshold  & $< 0.6$ &  --  &  $<0.6$ & $<3.5$ \\
\bottomrule
    \end{tabular}
    \label{table:ECCE_PIDec}
\end{table}

The Cherenkov systems provide, in addition to hadron PID, a significant contribution to the $e/\pi$ identification. 
When combined with the EM calorimeter, the mRICH and hpDIRC will provide excellent suppression of the low-momentum charged-pion backgrounds, which otherwise limit the ability of the EMCal to measure the scattered electron in kinematics where it loses most of its energy. 
The time-of-flight (TOF) system, using the AC-LGAD technology, will provide hadronic PID and electron identification in the momentum range below the thresholds of the Cherenkov detectors and provide a time resolution of 25~ps and a position resolution of about 30~$\mu$m over a nearly 4$\pi$ coverage. 

\begin{figure*}
    \centering
    \includegraphics[width=\textwidth]{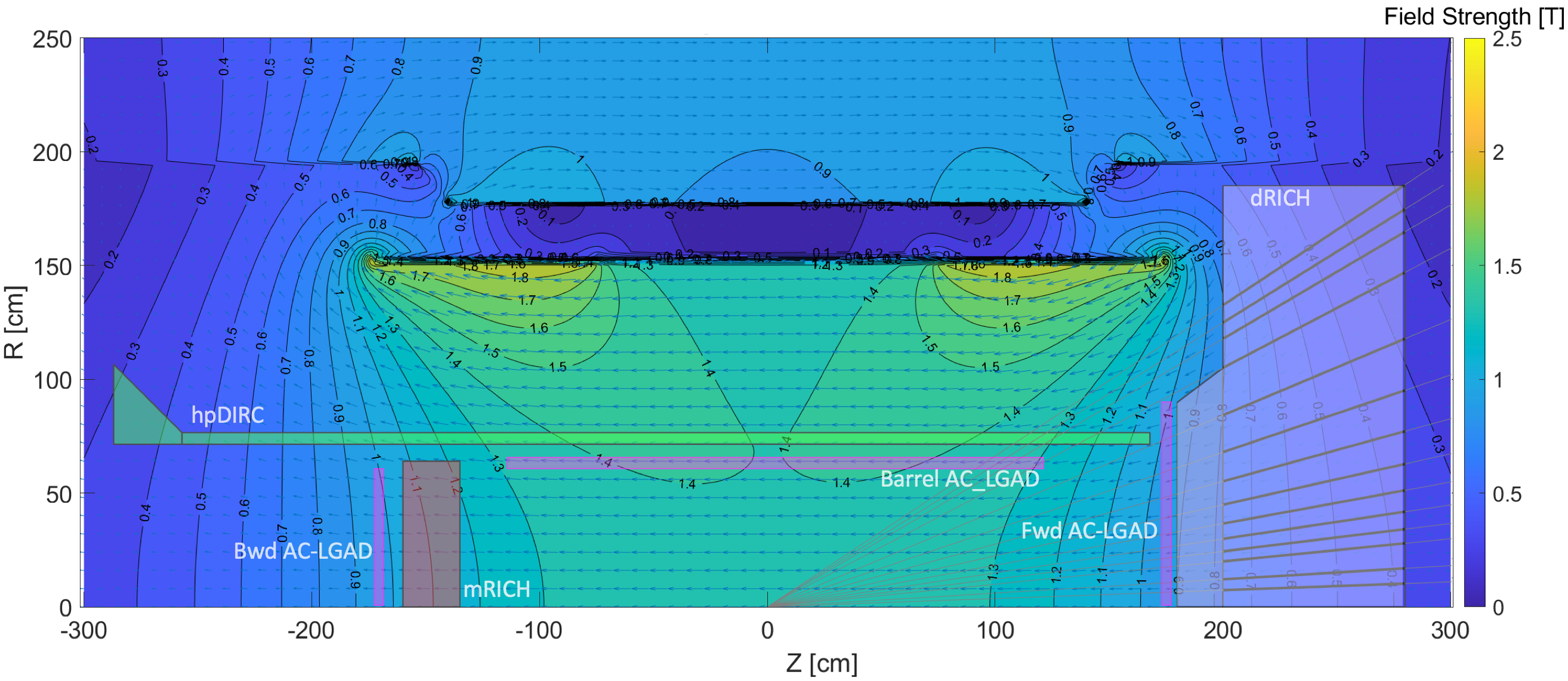}
    \caption{ECCE magnetic field map with the PID detector envelopes overlaid. }
    \label{fig:Bfield_0}
\end{figure*}

Figure~\ref{fig:Bfield_0} shows the realistic ECCE magnetic field with highlighted PID detectors envelopes. 
In the region of the hpDIRC detector plane, where the MCP-PMTs will be located, the magnetic field is at a level of 0.3--0.4~T, which provides a large safety margin in terms of the MCP-PMT field tolerance. 
Both RICH detectors in ECCE assume the use of SiPM, which are insensitive to magnetic fields of this strength, as their baseline photosensor.
Bending of the charged particle tracks in RICH detectors
can potentially have an impact on performance, but studies of the deviation of charged tracks from a straight line within the gas volume of the dRICH show a negligible impact on performance.

\subsection{mRICH}

The novel design of the modular RICH (mRICH) modules consists of four components. A block of aerogel serves as the Cherenkov radiator, immediately followed by an acrylic Fresnel lens, which focuses the ring image and acts as a UV filter. A pixelated optical sensor is located in the image plane, and flat mirrors form the sides of each mRICH module.

Several optimizations of the ECCE mRICH design were made compared to the Yellow Report reference detector:
(1) the projective array design was optimized maximizing the acceptance, removing the polar-angle dependence, and reducing the material budget; (2) the dead region between the mRICH modules is minimized using optimized thin module walls and mirrors (shorter as well) (3) an integrated mRICH array mechanical design was designed, consistent with the simulated array configuration in \geant.

\begin{figure}
    \centering
\includegraphics[width=0.45\textwidth]{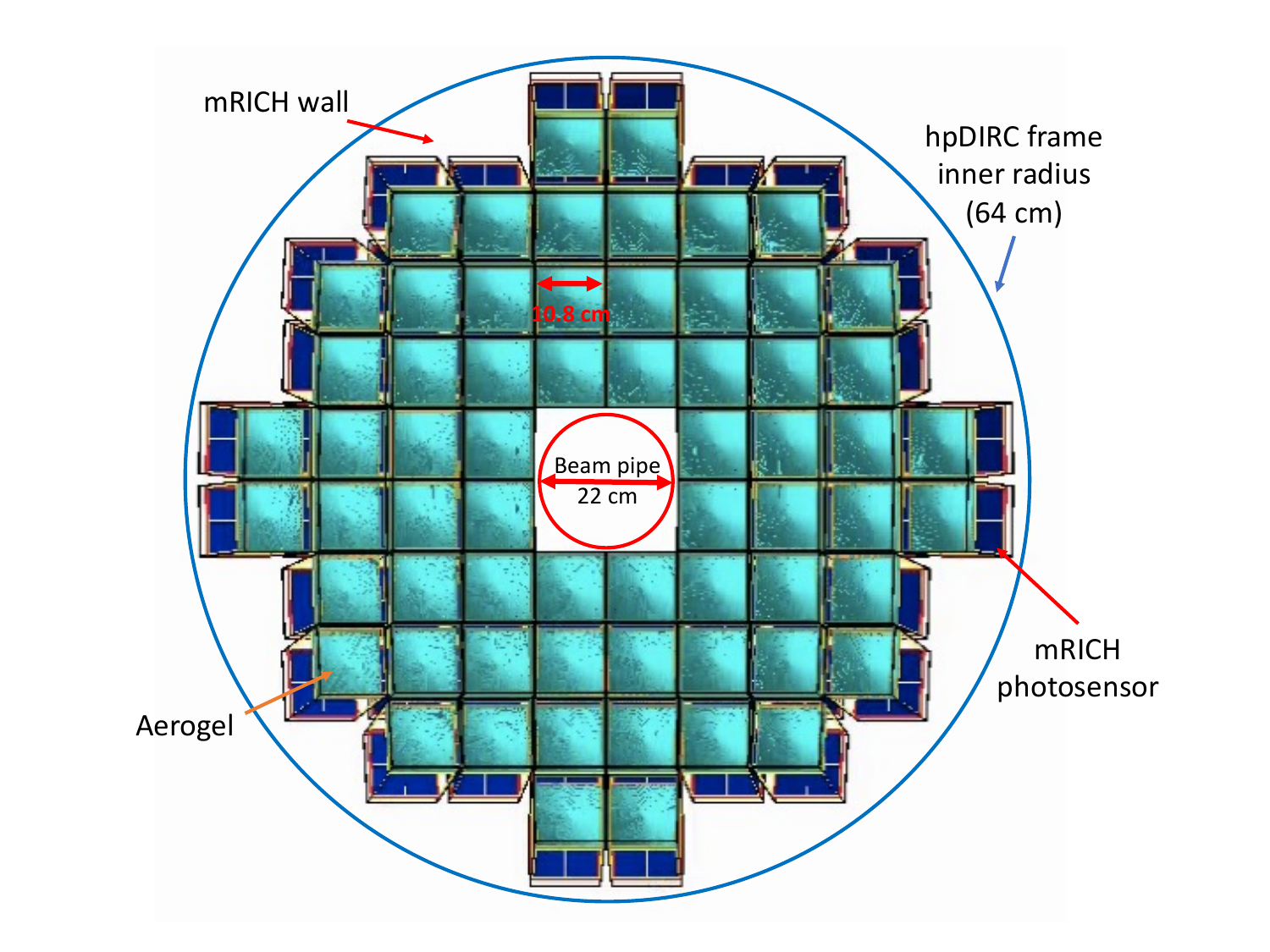}
\includegraphics[width=0.45\textwidth]{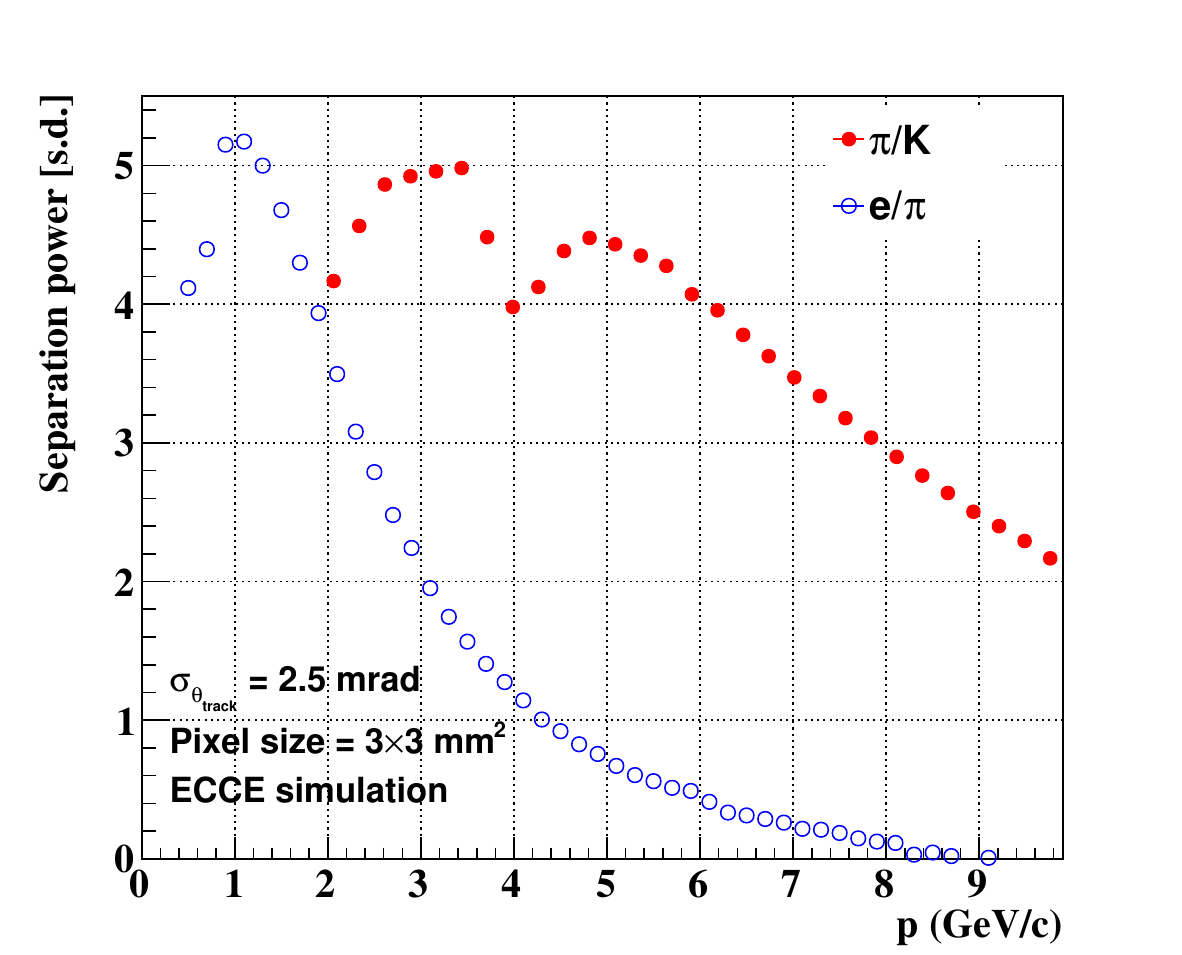}
    \caption{ Top: Front view of mRICH module array in the allocated space projected towards the IP. Bottom: The separation power of the mRICH in units of number of standard deviations as function of particle momentum from ECCE simulation.}
    \label{fig:ECCE_PID_mRICH}
\end{figure}

To study the performance of mRICH setup in ECCE, a set of tracks from the most demanding parts of the phase space were used, where the performance is expected to deteriorate, setting a lower limit on the performance and comparing it to what we see from the parametrizations. 
The study specifically focuses on the cases where the particles are incident at the surface of the aerogel closer to the outer edges with an outward angles and tracking angular resolution of 2.5~mrad. Fig.~\ref{fig:ECCE_PID_mRICH} shows the results for the $e/\pi$ and $\pi/K$ separation. 
The dips in the $\pi/K$ separation at 2 and 3.8~GeV/$c$ are due to the Cherenkov thresholds for kaons and protons in the aerogel. 
The obtained results show better performance than that used in the parametrization, shown in Fig.~\ref{fig:PID_sd_1}a, which indicates a better momentum reach once the mRICH reconstruction is further optimized.

\subsection{hpDIRC}

The radially-compact hpDIRC is based on a fast focusing DIRC design.
Thin rectangular bars, made of synthetic fused silica, serve as Cherenkov radiators and guide the photons to the readout section where they are focused by a lens and recorded by an array of pixelated photon sensors, placed on the back surface of a fused silica prism expansion volume.
Key features of the hpDIRC include three-layer spherical lenses, photosensors with small (3~mm$\times$3~mm) pixels, and fast readout electronics.

Compared to the Yellow Report reference detector, several hpDIRC design aspects were optimized for ECCE.
The expansion volume and readout were moved from the hadron side to the electron side for better detector integration and to minimize gaps in the EM calorimeter coverage. 
The bar box radius was decreased to match the EM calorimeter barrel size and the number of bar boxes, as well as the number of bars per bar box, were tuned to optimize the azimuthal coverage of the hpDIRC and to be consistent with the reuse of the BaBar DIRC bars. 
None of these changes had a significant impact on the performance of the hpDIRC.

\begin{figure*}
    \centering
    \includegraphics[width=1.0\textwidth]{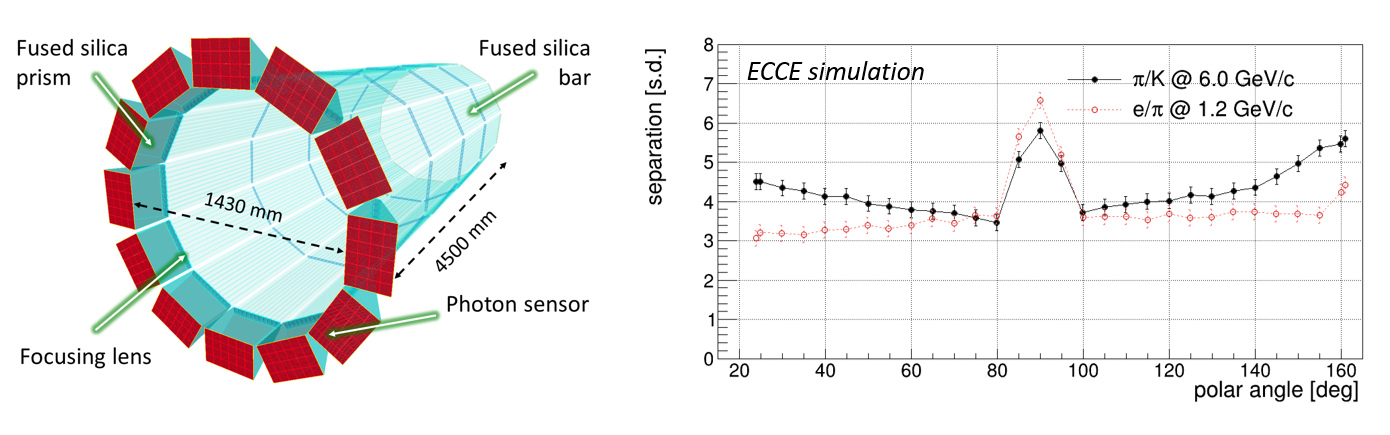}
    \caption{hpDIRC geometry (left) and expected PID performance (right) from the \geant standalone simulation. The $e/\pi$ and $\pi/K$ separation power is shown in units of number of standard deviations as a function of the particle polar angle for $e/\pi$ at 1.2~GeV/$c$ and $\pi/K$ up to 6~GeV/$c$.
    }
    \label{fig:ECCE_PID_hpDIRC}
\end{figure*}

Figure~\ref{fig:ECCE_PID_hpDIRC} shows the hpDIRC geometry as well as and the expected performance of the hpDIRC from the standalone \geant simulation studies for two particular cases. 
The black points show the separation power for charged pions and kaons as a function of the polar angle at a momentum of 6~GeV/$c$ while the red points show the same quantity for charged pions and electrons at 1.2~GeV/$c$. 
The expected particle identification performance of the hpDIRC exceeds the ECCE PID goal of three standard deviation separation power for $e/\pi$ up to 1.2~GeV/$c$ and $\pi/K$ up to 6~GeV/$c$ for the entire polar angle range.

\subsection{dRICH}

The dual-radiator Ring Imaging Cherenkov (dRICH) detector configuration for ECCE consists of six identical, transversely open sectors.
Each contains two radiators (aerogel and C$_2$F$_6$ gas), sharing the same outward focusing mirror and readout planes, which are instrumented with highly segmented photosensors (3~mm$\times$3~mm pixels), located outside of charged particle acceptance. 
The photosensor tiles are arranged on a curved surface to compensate for aberrations. 
Due to the open geometry of the dRICH sectors,
photons from a Cherenkov cone may split over two or more sectors 
Relative to the Yellow Report reference detector, the ECCE dRICH radial size was scaled down by 25\% to fit into the envelope defined by the HCAL.   It was also moved about 40~cm closer towards the IP to maintain the original acceptance.

\begin{figure*}[htb]
    \centering
    \includegraphics[width=1.6\columnwidth]{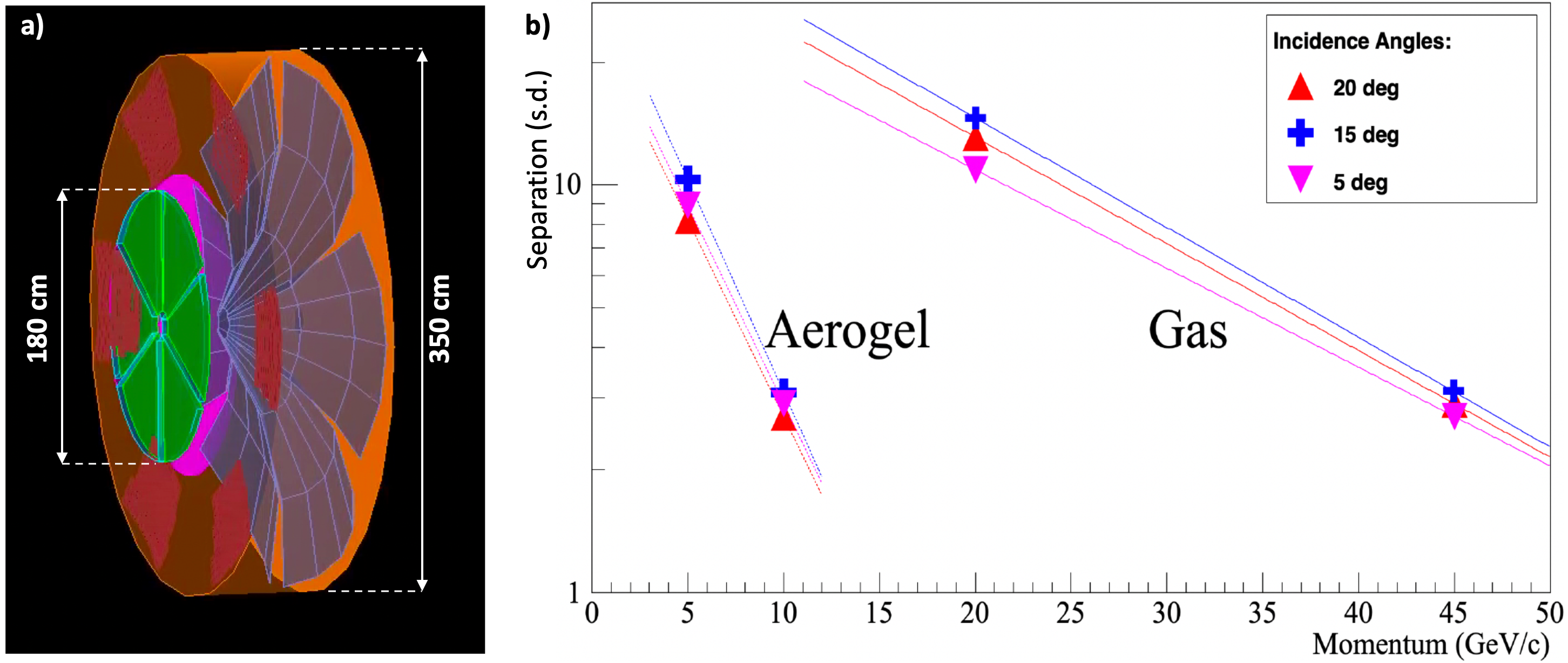}
    \caption{ dRICH geometry (a) and expected performance (b) from the ECCE \geant simulation. The $K/\pi$ separation power is shown as a function of momentum for a simplified dRICH geometry (flat detector plane). 
    }
    \label{fig:ECCE_PID_dRICH}
\end{figure*}

Figure~\ref{fig:ECCE_PID_dRICH} shows the preliminary results of the dRICH $K/\pi$ separation power for three incidence angles and selected momenta. 
The results are obtained from the full ECCE simulation framework with the realistic magnetic field map and conservative tracking resolution. 
Note that the simulated design uses a simplified flat detector plane and that the mirror curvature is not fully optimized yet.
The results are in good agreement with expectations and already reach the desired 3~standard deviations or more over almost the full required momentum range.
Further improvement of the dRICH performance is expected once the planned AI-based geometry optimization is completed.

\subsection{AC-LGAD-based TOF}

The AC-LGAD TOF technology is based on a simple p–n diode concept~\cite{Giacomini:2019kqz}, where the diode is fabricated on a thin high-resistivity p-type silicon substrate. A highly-doped p–layer (the ``gain” layer) is implanted under the n-type cathode. Application of a reverse bias voltage creates an intense electric field in this region of the sensor to start an avalanche multiplication for the electrons. The drift of the multiplied carriers through the thin substrate generates a fast signal with a time resolution of $\sim$20--30 ps. 

\begin{figure}[htb]
    \centering
    \includegraphics[width=\columnwidth]{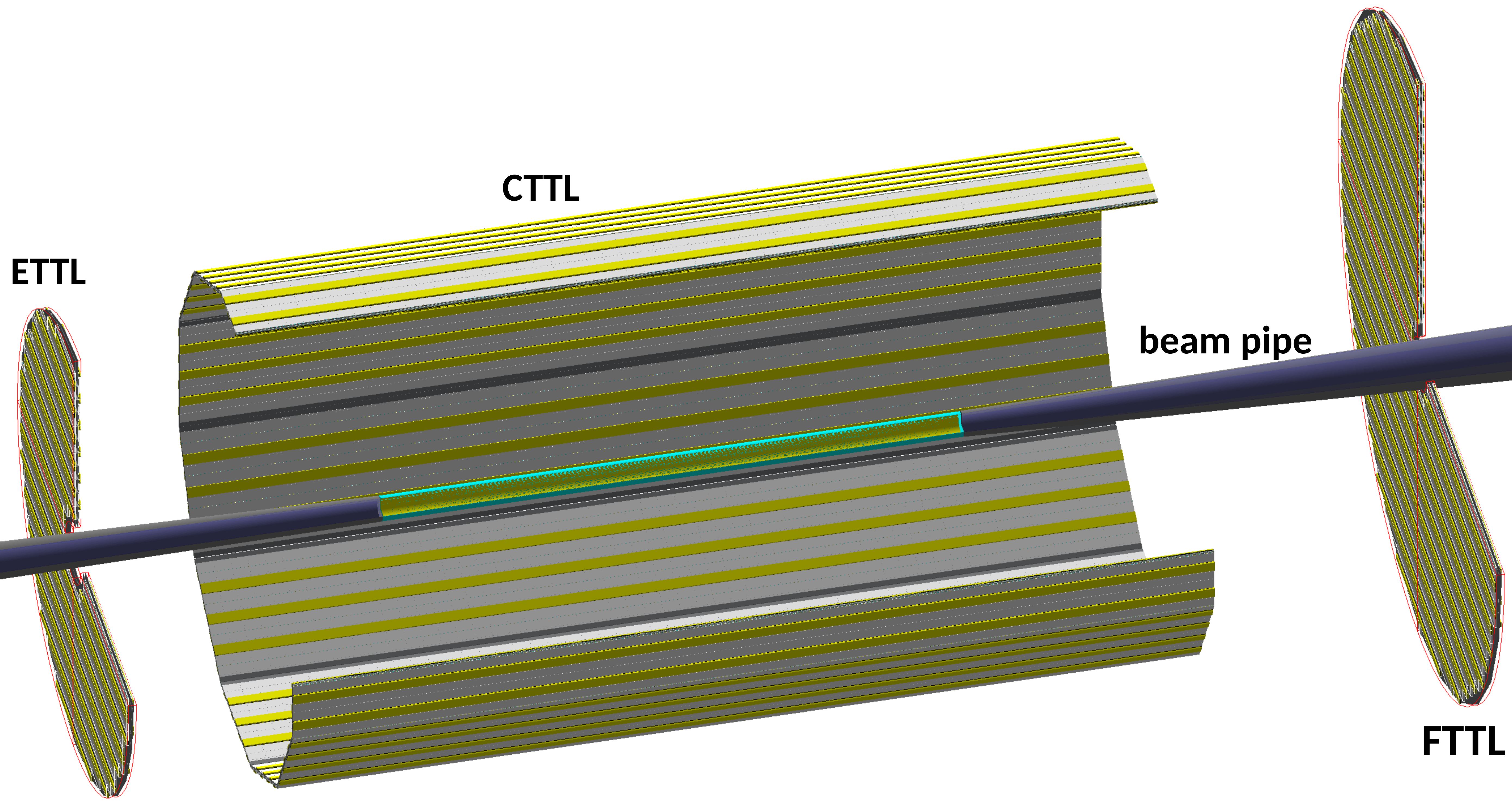} 
    \caption{A schematic view of the timing and tracking layers (TTLs) in ECCE as simulated with \geant. The different subdetectors are called ETTL (electron endcap), CTTL (barrel) and FTTL (hadron endcap).}
    \label{fig:ECCE_PID_TOFSetup}
\end{figure}

TOF layers were placed in each section of the ECCE detector and their positions were optimized to best compliment the Cherenkov detectors to cover the lowest possible particle momenta with a nearly 4$\pi$ coverage, and maximize the time (25~ps) and position (pixel granularity of 0.5$\times$2.6~mm$^{2}$) resolution.
We further plan to use the DIRC timing measurement to supplement the AC-LGAD TOF measurement. This is especially useful for the $\eta \approx -1.5$ region where a gap exists in the AC-LGAD coverage and the DIRC offers excellent TOF resolution.
Figure~\ref{fig:ECCE_PID_TOFSetup} (left) shows a visualization of the AC-LGAD geometry from the full \geant simulation. Figure~\ref{fig:ECCE_PID_TOFSetup} (right) summarizes the performance of the TOF layers in each sector of the ECCE detector for $\pi/K$, $e/\pi$, and $K/p$ separation.

\begin{figure}[htb]
    \centering
    \includegraphics[width=\columnwidth]{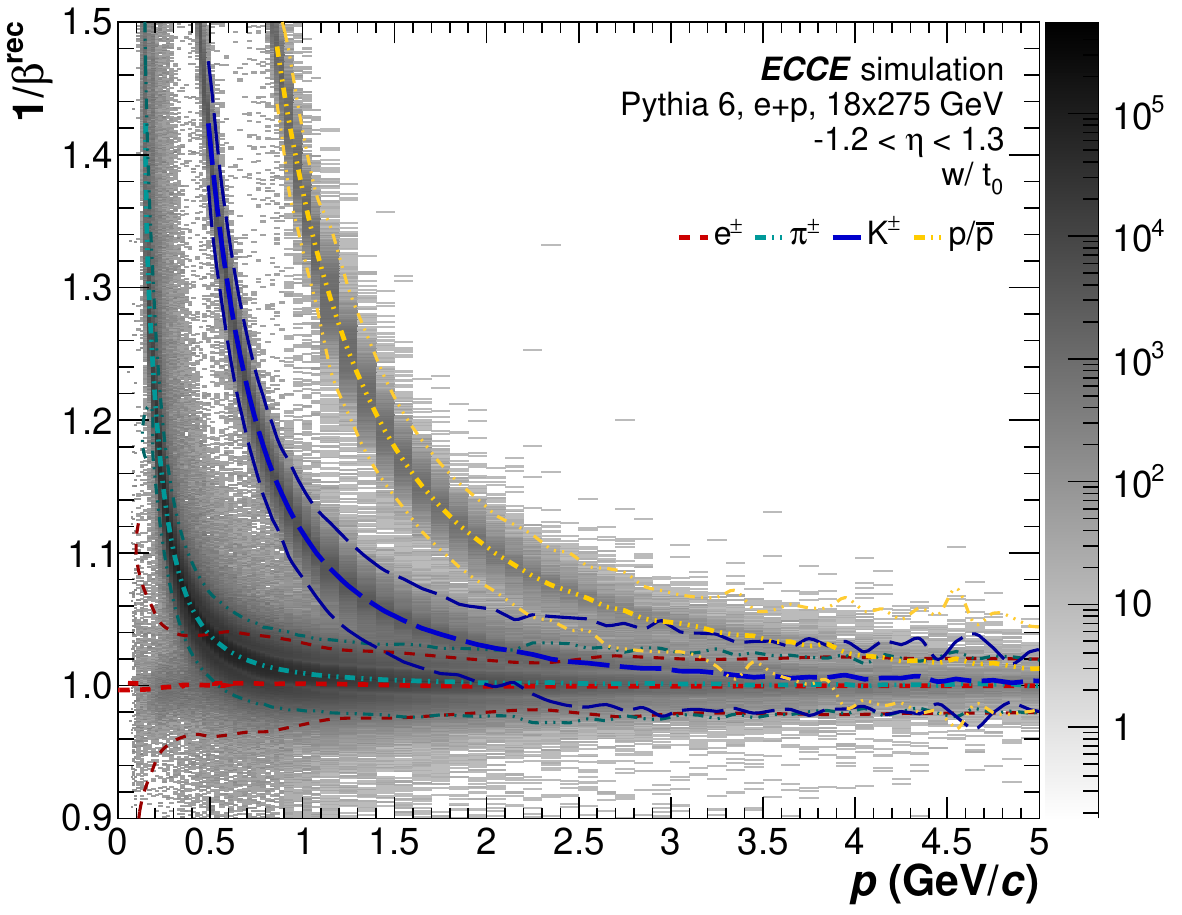} \hspace{0.012\textwidth}
    \includegraphics[width=0.9\columnwidth]{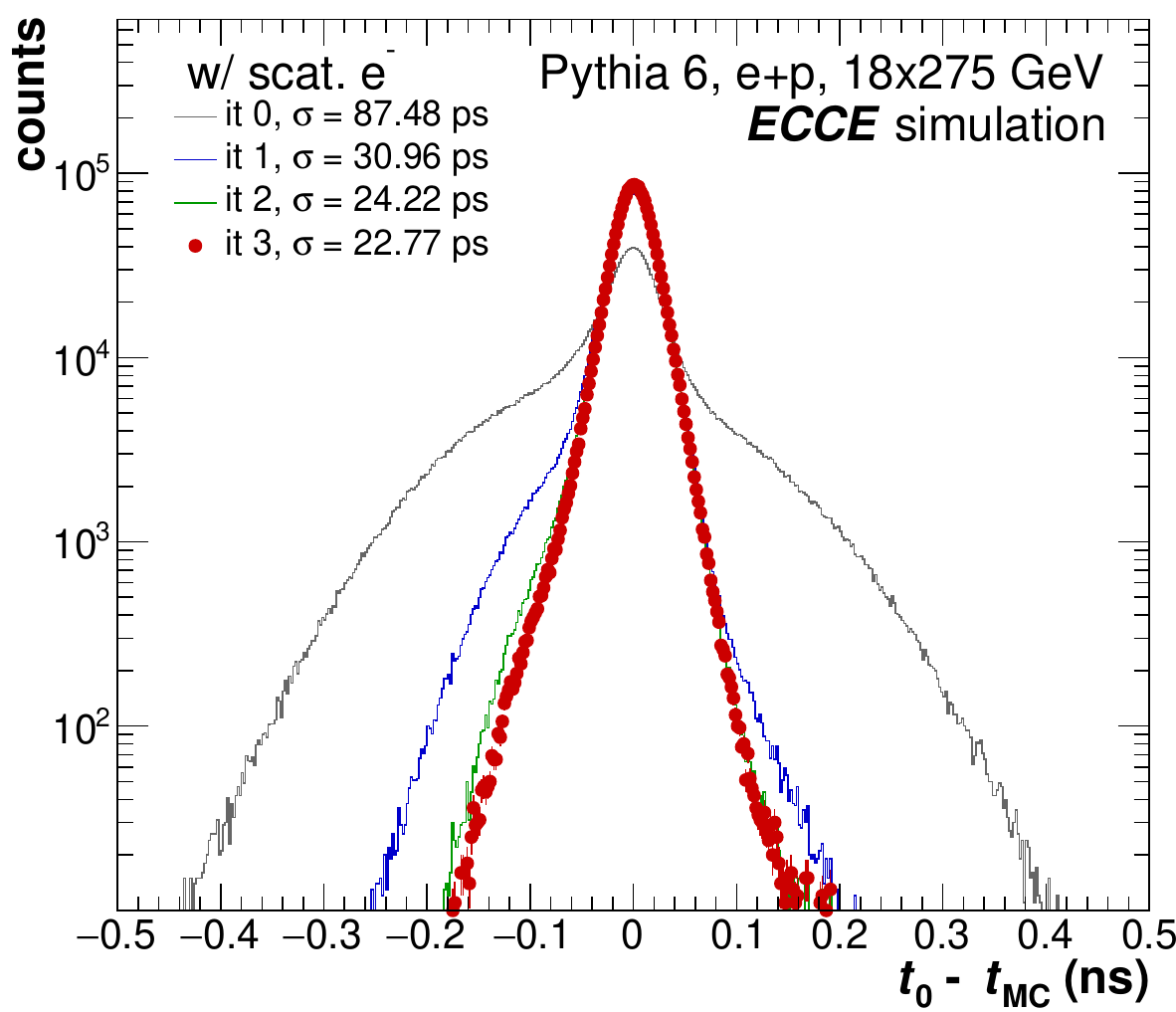}
    \caption{(left) Expected $1/\beta$ performance of the AC-LGADs TOF in the barrel as a function of particle momentum, assuming 25~ps time resolutions from full simulations including the start time estimates. (right) Expected start time ($t_0$) resolution as a function of iteration, for events where the scattered electron could be identified. 
    }
    \label{fig:ECCE_PID_TOFPerf}
\end{figure}

The PID performance in terms of 1/$\beta$ vs. $p$ for the central barrel, as a benchmark, is shown in Fig.~\ref{fig:ECCE_PID_TOFPerf} (left) for an expected timing precision of 25~ps. The long dashed lines indicate the $\pm$3$\sigma$ range around mean 1/$\beta$ values for each particle species. As shown, the $\pm$3$\sigma$ bands for pions and kaons are well separated over a momentum range of $0.1<p<1.3$~GeV/c, while proton identification is further extended to around 2.2~GeV/c. For electrons, clean separation from pions is achieved for $p<0.45$~GeV/c by at least 3$\sigma$. Similar performance studies have also been carried out for endcap TOFs. 

\begin{table}[htb]
\centering  
\caption{Summary of the Momentum coverage in GeV/c of the ECCE Time-of-Flight detector in corresponding regions.}
\begin{tabular}{cccc}
\toprule
 PID  &   ETTL    &  CTTL &  FTTL \\
\midrule
$e/\pi$ & $< 0.5$ & $< 0.45$  & $< 0.6$ \\
$\pi/K$ & $< 2.1$ & $< 1.3$  & $< 2.2$ \\
$K/p$   & $< 3.3$ & $< 2.2$  & $< 3.7$ \\
\bottomrule
    \end{tabular}
    \label{table:ECCE_MC_TOF}
\end{table}

The resolution of the start time, $t_{0}$, is self-determined by the scattered electron and final-state hadrons via an iterative fitting procedure.  The $t_{0}$ resolution was included in all performance studies and is shown in Fig.~\ref{fig:ECCE_PID_TOFPerf} (right).
In addition to providing hadronic PID, the excellent position resolution of AC-LGADs TOF system significantly improves the momentum resolution of high momentum particles in the very forward region.

\clearpage
\section{Electromagnetic and Hadronic Calorimetry}

The ECCE electromagnetic calorimeter system\cite{ecce-paper-det-2022-02} consists of three components which allow high precision electron detection and hadron suppression in the backward, barrel, and forward directions.
Hadronic calorimetry is essential for the barrel and forward endcap regions for hadron and jet reconstruction performance. Jet yields in the backward region were found to be sufficiently infrequent that hadronic calorimetry would provide little to no scientific benefit. The details for all six calorimeters envisioned for ECCE can be found in Tab.~\ref{tab:calospecs}.

\begin{figure}[htb]
  \centering
  \includegraphics[width=\linewidth]{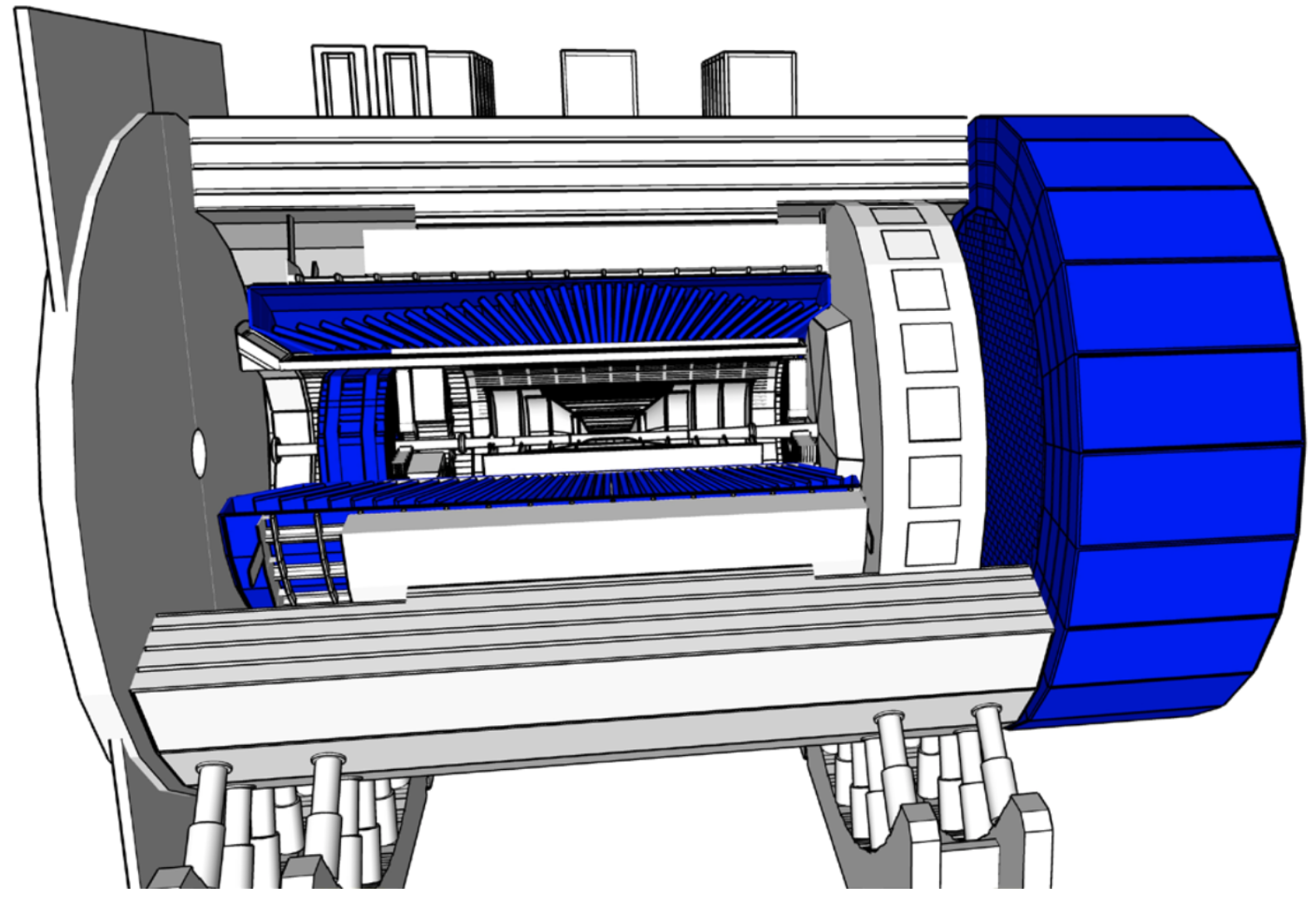} 
  \includegraphics[width=\linewidth]{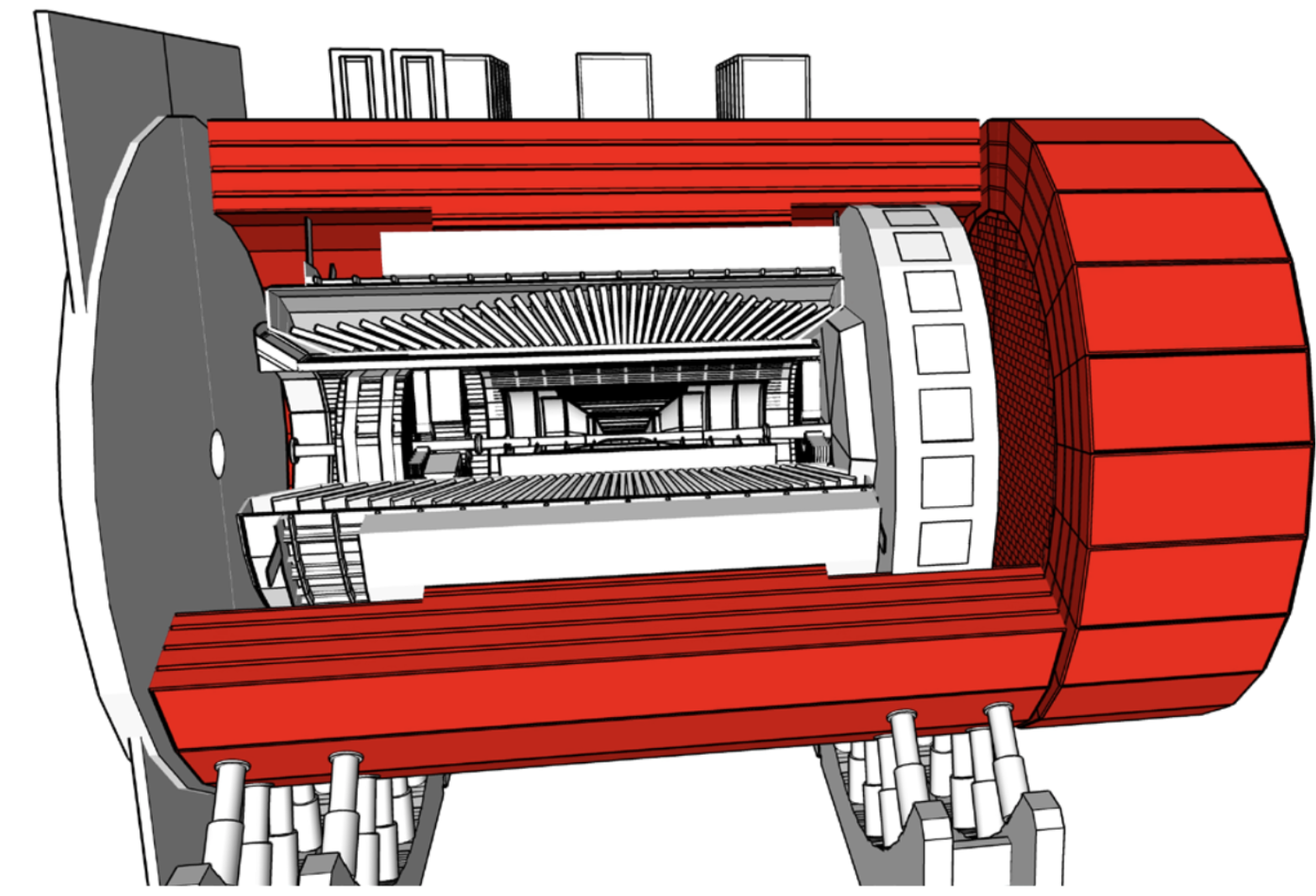} 
  \caption{The upper figure shows the locations of the ECCE electromagnetic calorimeter systems in blue while the upper figure shows the locations of the ECCE hadronic calorimeter systems in red.   As can be seen, the endcap region contains both electromagnetic and hadronic calorimetry.}
  \label{fig:SketchupCalos} 
\end{figure}
\begin{figure*}[ht]
  \begin{minipage}{\textwidth}
    \centering
    \footnotesize
    \captionof{table}{Specifications and properties for the electromagnetic and hadronic calorimeters from the Geant simulation. Note that $d_{act}$ does not include readout. The acceptance of the EEMC can be achieved with a small inner calorimeter as discussed in the text. The energy resolutions for EEMC, BEMC and OHCAL are those expected from prototype tests or experiments~\cite{AbdulKhalek:2021gbh,Buchner:1988fu,E705:1993imr,Aidala2018}.
    Further details can be found in the \cite{ecce-paper-det-2022-02}.}
    \begin{tabular}{l| ccc ccc}
    
    \toprule
    \setlength\columnsep{0.01em}
    
                                & EEMC                      & BEMC                      & FEMC                                  & IHCAL                 & OHCAL                         & LFHCAL \\ \midrule
    tower size                  & $2 x 2 x 20$ cm$^3$       & $4 x 4 x 45.5$ cm$^3$     & in: $1 x 1 x 37.5$ cm$^3$             & $\Delta\eta \sim 0.1$   & $\Delta\eta \sim 0.1$       & $5 x 5 x 140$ cm$^3$  \\
                                &                           & projective                & out: $1.6 x 1.6 x 37.5$ cm$^3$        & $\Delta\varphi\sim 0.1$ & $\Delta\varphi \sim 0.1$    &                       \\
                                &                           & projective                & out: $1.6 x 1.6 x 37.5$ cm$^3$        & $l \sim 4.5$ cm         & $l \sim 88$ cm              &                       \\
    material                    & \PbWOiv                  & SciGlass                  & Pb/Scintillator                       & Steel/                & Steel/                        & Steel/W/              \\
                                &                           &                           &                                       & Scintillator          & Scintillator                  & Scintillator          \\
    $d_{abs}$            & -                         & -                         & $1.6$ mm                              & $13$ mm               & in: $10.2$ mm                 & $16$ mm               \\
                                &                           &                           &                                       &                       & out: $14.7$ mm                &                       \\
    $d_{act}$            & $20$ cm                   & $45.5$ cm                 & $4$ mm                                & $7$ mm                & $7$ mm                        & $4$ mm                \\
    $N_{layers}$         & 1                         & 1                         & $66$                                  & $4$                   & $5$                           & $70$                  \\
    $N_{towers(channel)}$& $2876$                    & $8960$                    & $19200/34416$                         & $1728$                & $1536$                        & $9040(63280)$         \\
    $X/X_O$                     & $\sim 20$                 & $\sim 16$                 & $\sim 19$                             & $\sim 2$              & $36-48$                       & $65-72$               \\
    $R_M$                       & $2.73$ cm                 & $3.58$ cm                 & $5.18$ cm                             & $2.48$ cm             & $14.40$ cm                    & $21.11$cm             \\
    $f_{sampl}$          & $0.914$                   & $0.970$                   & $0.220$                               & $0.059$               & $0.035$                       & $0.040$               \\
    $\lambda/\lambda_0$         & $\sim 0.9$                & $\sim 1.6$                & $\sim 0.9$                            & $\sim 0.2$            & $\sim 4-5$                    & $7.6-8.2$             \\
    $\eta$ acceptance           & $-3.7 < \eta < -1.8$      & $-1.7 < \eta < 1.3$       & $1.3 < \eta < 4$                      & $1.1 < \eta < 1.1$    & $1.1 < \eta < 1.1$            & $1.1 < \eta < 4$      \\
    resolution                  &                           &                           &                                       &                       &                               &                       \\
    \hspace{0.012\textwidth} - energy     & $2/\sqrt{E} \oplus 1$ & $2.5/\sqrt{E} \oplus 1.6$ & $7.1/\sqrt{E} \oplus 0.3$             &                       & $75/\sqrt{E} \oplus 14.5$   & $33.2/\sqrt{E} \oplus 1.4$   \\
    \hspace{0.012\textwidth} - $\varphi$  & $\sim 0.03$               & $\sim 0.05$               & $\sim 0.04$                           &                       & $\sim 0.1$                    & $\sim 0.25$           \\
    \hspace{0.012\textwidth} - $\eta$     & $\sim 0.015$              & $\sim 0.018$              & $\sim 0.02$                           &                       & $\sim 0.06$                   & $\sim 0.08$           \\ \bottomrule
 \end{tabular}
 \label{tab:calospecs}
    \end{minipage}
\end{figure*}

\subsection{Electron Endcap EM Calorimeter (EEMC)}
\begin{figure*}[!ht]
\centering
    \includegraphics[width=.49\textwidth]{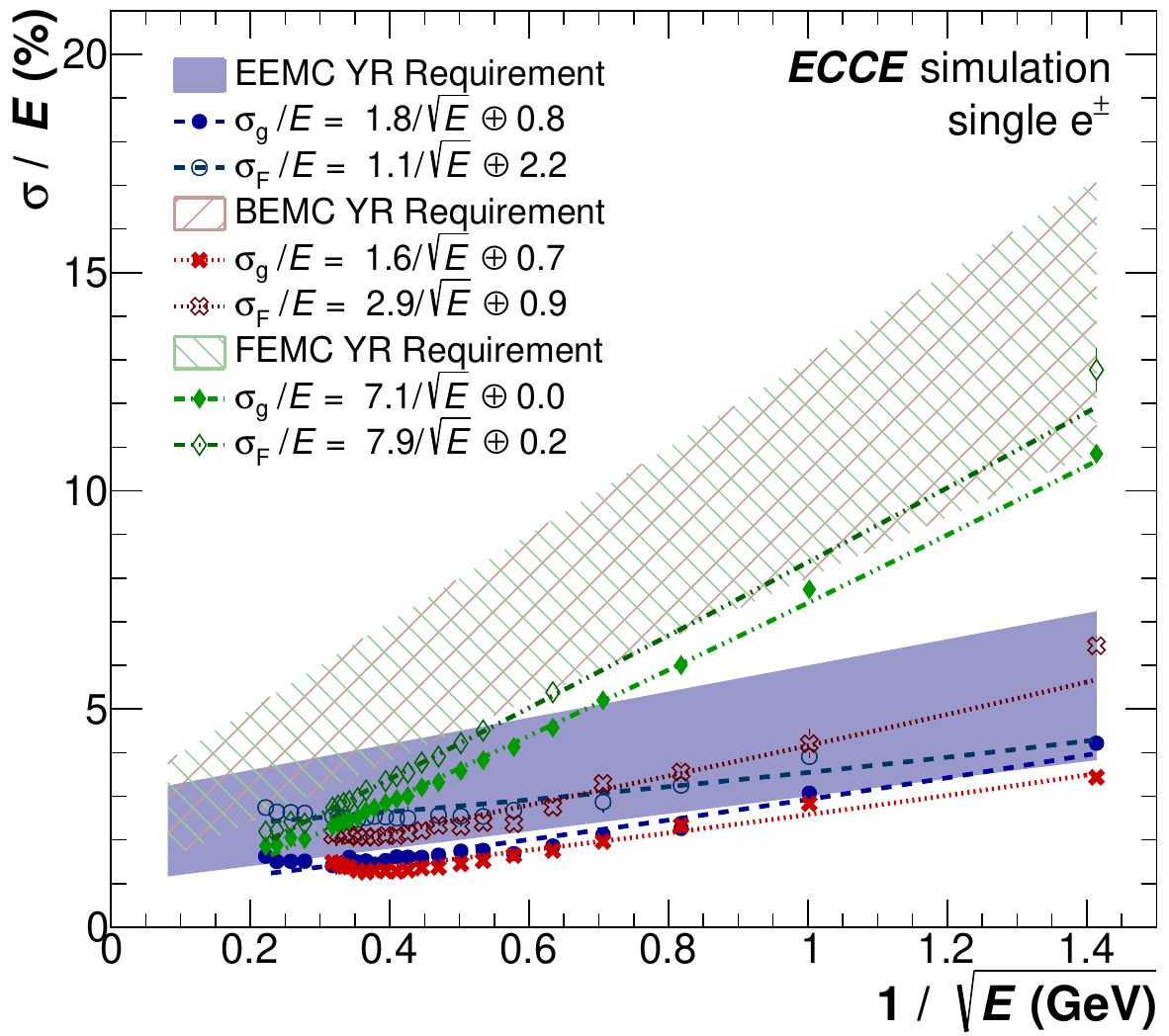}
    \includegraphics[width=.49\textwidth]{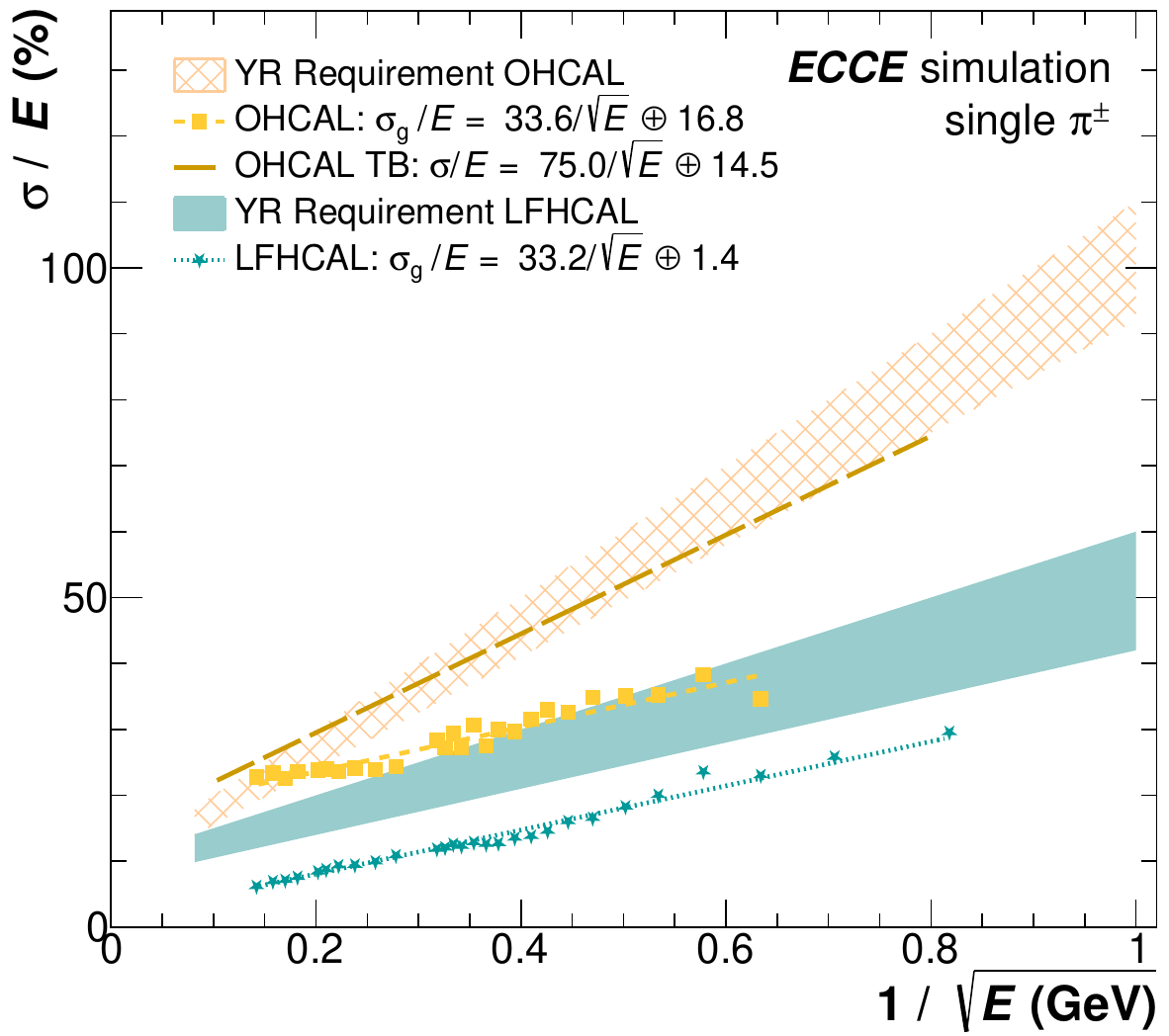}
    \caption{The electron (top) and pion (bottom) energy resolution of the electromagnetic and hadronic calorimeters, respectively, compared to the Yellow Report requirement (shaded/hashed area).
    The data points and fits indicated as $\sigma_g/E$ are based on the Gaussian width of the resolution peaks, while $\sigma_F/E$ is based on the FWHM.
      The energy resolution based on a test beam for the OHCAL is shown for comparison \cite{sPHENIX:2017lqb}.}
    \label{fig:CaloResolution}
\end{figure*}

The EEMC is a high-resolution electromagnetic calorimeter designed for precision measurements of the energy of scattered electrons and final-state photons in the electron-going region. 
The required energy resolution is driven by the need for a precise measurement of the scattered electron's energy and direction to determine the event kinematics in inclusive DIS events.

The design of the EEMC is based on an array of approximately 3000 lead tungsten crystals (\PbWOiv) of size $2\times 2\times 20$~cm$^3$ ($\sim$22$X_0$) and transverse size equal to its Moliere radius~\cite{Horn:2019beh,Horn:2015yma} readout by SiPMs yielding an expected energy resolution of 2\%/$\sqrt{E}$ + 1\%, based on prototype beam test measurements by the EEEMCAL consortium and documented in the Yellow Report~\cite{AbdulKhalek:2021gbh}.
Fig.
\ref{fig:CaloResolution} shows the EEMC 
performance in the full ECCE detector simulations, 
consistent with the 
measurements.
The corresponding particle identification power is shown in Fig.~\ref{fig:pidperf} for distinguishing electrons and pions (left) as well as separating the two photons from a neutral pion decay.

The choice of technology and overall design concept is common for all three proto-collaborations, with additional details of the development of this detector by the EEEMCal consortium summarized in the expression of interest~\cite{EEEMCAL_EOI}. The ECCE design only includes the \PbWOiv crystals due to the overall small detector radius. The EEEMCAL Consortium is planning to support one or more EIC detectors as needed and is therefore part of multiple detector proposals.

The EEMC is located inside the inner universal frame and allows to reconstruct particles with $-3.4<\eta<-1.8$. 
The lower $\eta$ boundary is constrained by the proximity to the beam pipe and integration concerns with the beam pipe flange directly in front of the EEMC \cite{EEEMCAL_integration} in combination with the crystal dimensions.
To extend the reach of the backward EEMC to a pseudorapidity of $-3.7$ one can envision a small inner calorimeter of 208 crystals and an outer calorimeter just behind it. There is sufficient longitudinal space accommodate this, but moving the outer calorimeter back could impact the acceptance in the transition region between the EEMC and the central barrel. If possible, this arrangement would allow the outer calorimeter to be removed over the beam pipe flange for maintenance, and separate removal of the small inner calorimeter in two halves. We intend to pursue this improvement to the baseline design as part of a detailed, integrated mechanical engineering design of the ECCE detector. 

The EEEMCAL team 
has begun to organize activities into mechanical design, scintillator, readout, and software/simulation among the collaborating institutions. Design activities of the mechanical support structure commenced in 2021. The design is based on models of existing detectors that the team has recently constructed, in particular the Neutral Particle Spectrometer at Jefferson Lab~\cite{Horn:2019beh}. As such, it is maturing rapidly and a document on mechanical design and integration has been completed~\cite{EEEMCAL_integration}. 

\subsection{Barrel EM Calorimeter (BEMC)}
\begin{figure*}[htb]
\centering
    \includegraphics[width=0.48\textwidth]{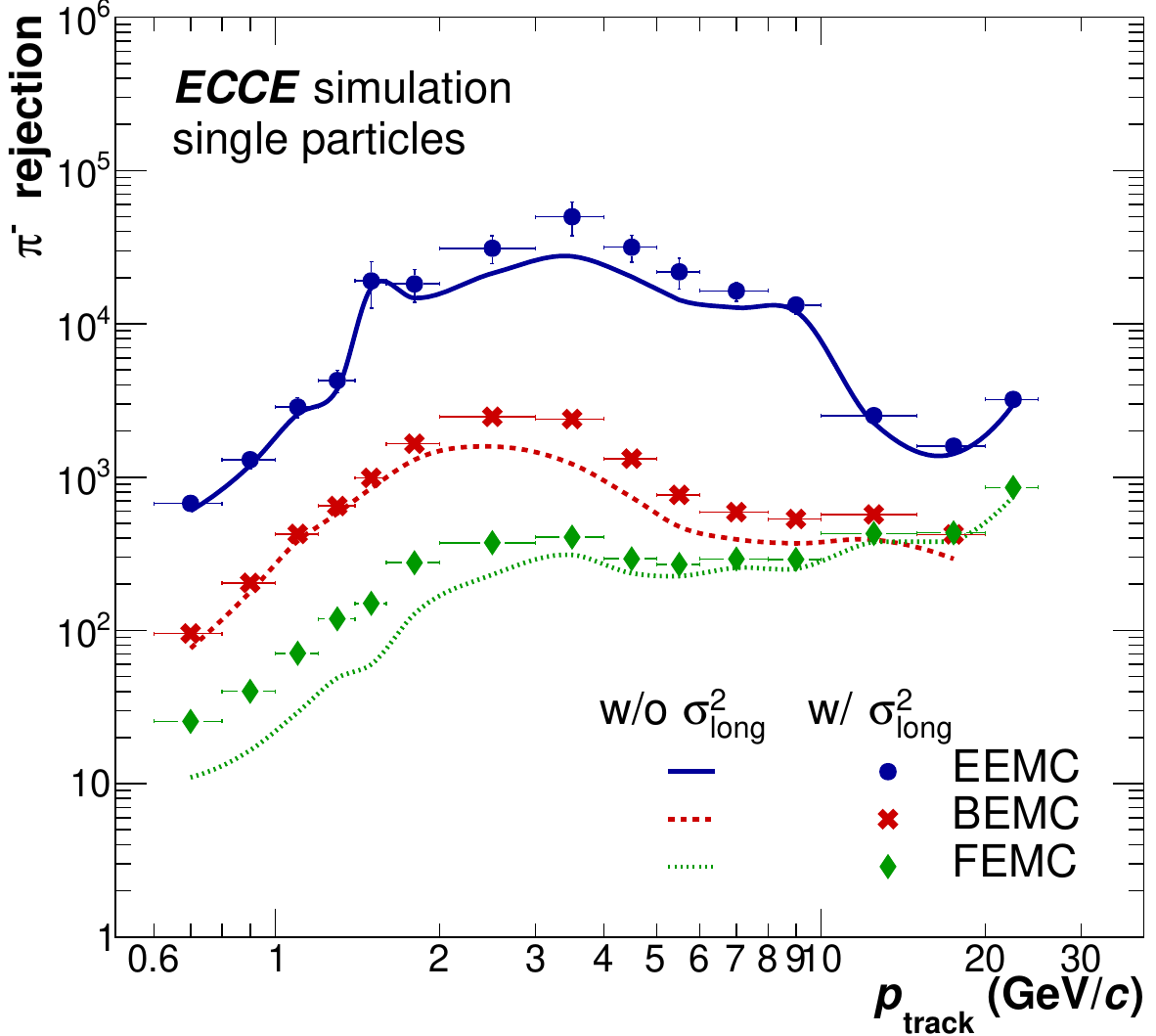}    
    \includegraphics[width=0.48\textwidth]{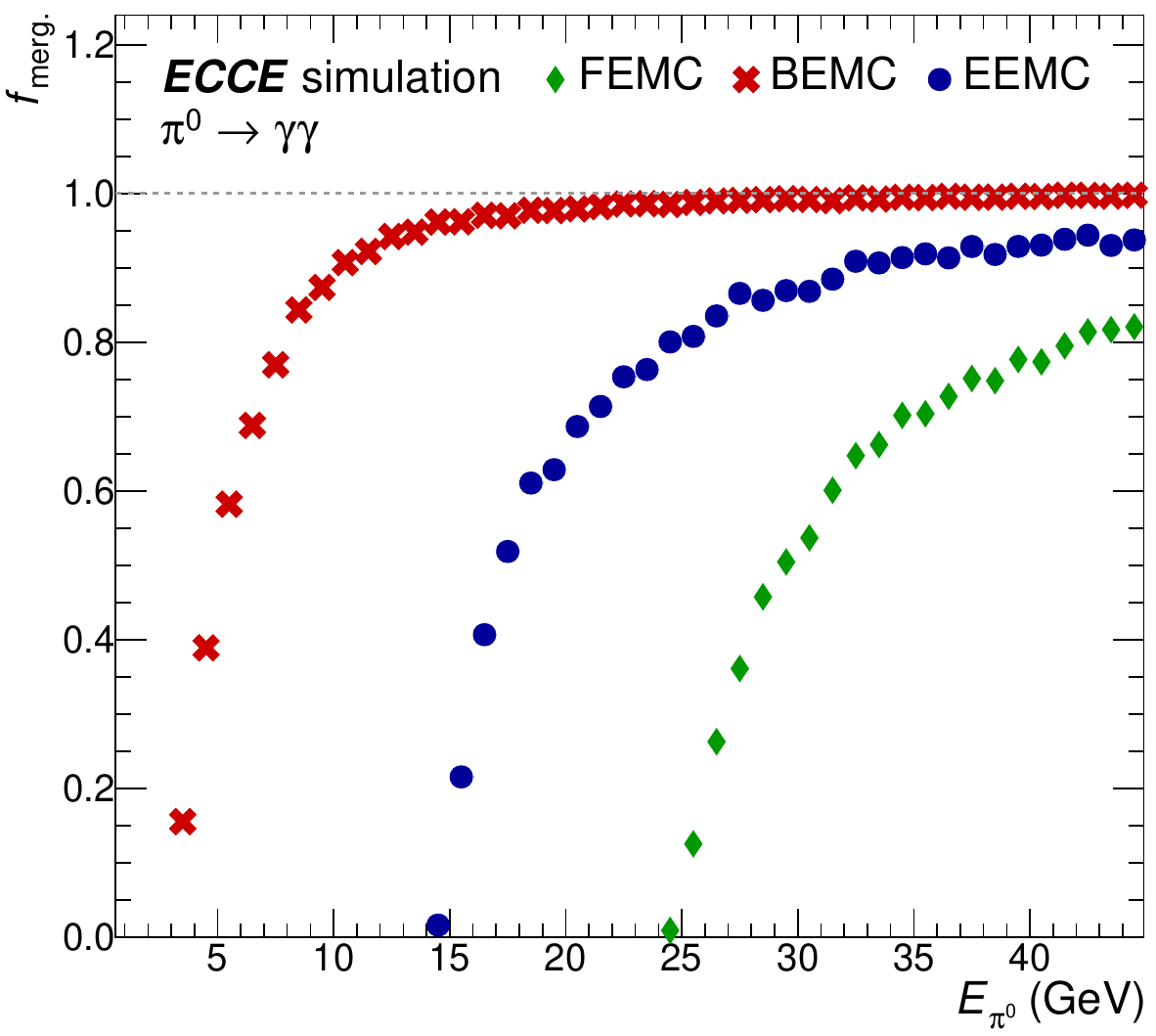}    
    \caption{\label{fig:pidperf} (top) Pion rejection factor for the different ECals with $E/p > 1-1.6\,\sigma_E/E$ and shower shape cuts applied as a function of true and reconstructed momentum. (bottom) Fraction of neutral pions for which the showers from their decay photons are merged into a single cluster and can not be reconstructed using an invariant-mass-based approach for the different electromagnetic calorimeters.}
\end{figure*}

The barrel electromagnetic calorimeter (BEMC) is a projective homogeneous calorimeter based on an inorganic scintillator material that produces the shower due to high Z components. 
This allows for a cost-effective solution that provides excellent energy resolution and sufficient $e/\pi$ rejection to achieve the EIC physics, which can be seen in Fig.~\ref{fig:pidperf}.
Further improvements are expected by determining exactly the Birk's constant and using shower shape criteria to distinguish elongated hadronic and rounder electromagnetic showers.
The reference design of the BEMC is based on an array of approximately 9000 Scintillating Glass
(SciGlass) blocks of size 4 x 4 x 45.5 cm$^3$, plus an additional 10cm of radial readout space. SciGlass has an expected energy resolution of 2.5\%/$\sqrt{E}$ + 1.6\% based on earlier measurements~\cite{Buchner:1988fu,E705:1993imr}, comparable to \PbWOiv for a significantly lower cost. The energy resolution of the BEMC is shown in red in
Fig.~\ref{fig:CaloResolution} (left) in its
optimal acceptance (-1.4 $< \eta< <$ 1.1).

\begin{figure*}[!ht]
\centering
    \includegraphics[width=0.9\textwidth]{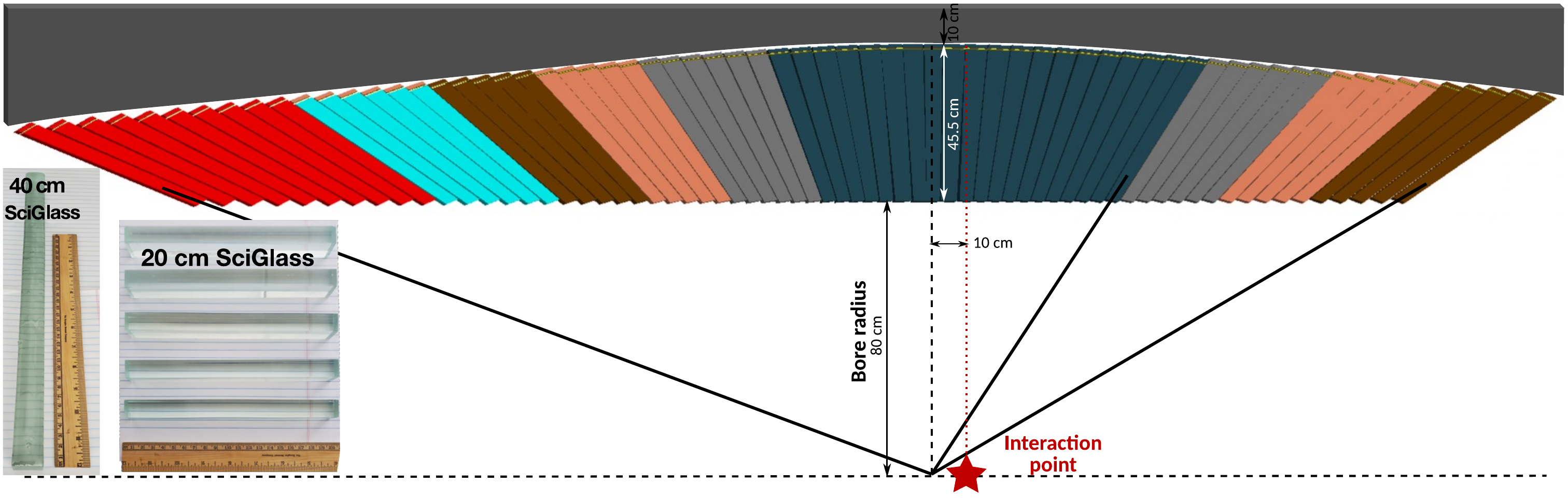}
    \caption{\label{fig:bemc} Side cut view of the barrel assembly from Geant4 illustrating the six different families of glass block sizes needed to achieve the needed projectivity. Also shown is a schematic of the support box (grey) based on the PANDA design that holds readout, cooling, and other services and mounts to the outer universal frame.}
\end{figure*}

The development of SciGlass started with the generic detector R\&D~\cite{sciglass}. During this phase the team worked in close contact with producers of SciGlass to establish robust QA protocols at all stages of production to ensure the quality needed for the EIC. The validation of large-scale SciGlass production is now continued in the ongoing project R\&D (eRD105).
An initial 40 cm SciGlass bar of high quality has been produced this Fall (see Fig.~\ref{fig:bemc} bottom right), and a prototype with nine
20-cm long SciGlass bars recently saw a successful beam test at Jefferson Lab, confirming the expected energy resolution. It is expected that multiple 45-cm long SciGlass bars will be produced in the next few months. 

Just as for the EEMC, the BEMC attaches to the outer universal frame.
Adapting the geometry of the homogeneous barrel EM calorimeter at PANDA~\cite{PANDA:2008rpr}, the BEMC towers are organized in 128 blocks by $\phi $ slice and 70 blocks in $\eta$, which will be assembled in super modules stretching the full length in $\eta$ and 8 towers in $\varphi$ for installation in the universal frame. Figure \ref{fig:bemc} (top) shows a sketch of the BEMC illustrating the at least six different families of glass blocks needed to achieve the required projectivity in $\eta$. For comparison, PANDA uses 11 different crystal types for their barrel. The optimal number of families still has to be determined, optimizing for efficient production as well as minimal leakage between towers. Also indicated is a schematic of the support box (modeled after the PANDA barrel calorimeter) for readout and other services that mounts to the outer universal frame.

The BEMC has been designed with projectivity in $\eta $ and $\phi $. This requires that the tower angular deflection depends on its location in the calorimeter. Additionally, the towers have a stronger inclination at higher absolute pseudorapidities, leading to an asymmetric tapered shape of the glass blocks, which increases with $|\eta|$. Their front face is tilted such that it is facing the interaction point shifted by $z = -10$ cm and tilted  $10 ^{\circ}$ in the azimuthal direction, to avoid channeling between the towers.  
Such a projective design delivers a more uniform performance, mainly aimed at the transition regions between
the barrel and forward and backward regions, as defined by the length to bore ratio of the magnet.
All the towers have the same length, 45.5 cm (not including $\sim$ 10cm readout),  and inner size 4 x 4 cm in the present simulation. However, the upper area sections vary from 5 to 6.6 cm  in each side depending on their location. 
\subsection{Barrel Hadron Calorimeters: OHCAL and IHCAL}
The energy resolution of reconstructed jets in the central barrel will be dominated by the track momentum resolution, as the jets in this region are relatively low momentum and the measurement of the energy in the hadronic calorimeter does not improve knowledge of the track momentum. For jet reconstruction, the primary use for a hadronic calorimeter in the central barrel will be to collect neutral hadronic energy and thus improve the overall knowledge of the Jet Energy Scale (JES). For this purpose, the Yellow Report indicates that a resolution of $(80-100)\%/\sqrt{E} \oplus (7-10)\%$ will be adequate. Therefore, we decided to reuse the sPHENIX Outer Hadronic Calorimeter (OHCAL), which instruments the barrel flux return steel of the BaBar solenoid to provide hadronic calorimetery with an energy resolution of $75\%/\sqrt{E} \oplus 14.5\%$, as measured in test beam. 
We also plan to instrument the support for the barrel electromagnetic calorimeter to provide an additional longitudinal segment of hadronic calorimetry. This will provide an Inner Hadronic Calorimeter (IHCAL) very similar
in design to the sPHENIX inner HCAL.  The inner HCAL is useful to monitor shower leakage from the barrel electromagnetic calorimeter as well as improve the calibration of the combined calorimeter system. 

The basic calorimeter concept for the IHCAL/OHCAL is a sampling calorimeter with absorber plates tilted from the radial direction.  
This design provides more uniform sampling in azimuth and gives some information about the longitudinal shower development. 
The outer HCAL uses tapered 1020 magnet steel plates which maintain a uniform gap size for the scintillating tiles. 
The inner HCAL will be made from stainless steel or aluminium  as it sits inside the magnetic field. The Inner HCAL will not require tapered plates as studies have shown that taperig the shorter inner HCAL plates is not necessary, and tapering them substantially increases the machining cost. Extruded tiles of plastic scintillator with an embedded wavelength shifting fiber are interspersed between the absorber plates and read out at the outer radius with silicon photomultipliers (SiPMs).  
A 12 degree tilt angle relative to the radius is chosen in the outer HCAL so that a radial track from the center of the interaction region traverses at least four scintillator tiles. The inner HCAL is tilted at 36 degrees, in the opposite direction compared to the outer HCAL.    
Each tile has a single SiPM, and the analog signal from each tile in a tower (five for the OHCAL, four for the IHCAL) are ganged to a single preamplifier channel to form a calorimeter tower.  
Tiles are divided in slices of pseudorapidity so that the overall segmentation is $\Delta \eta \times \Delta \phi \sim 0.1 \times 0.1$. 
The Outer HCal is longitudinally symmetric around the interaction point and requires 24 tiles along the $\eta$ direction. 
The design thus requires 12 different shapes for tiles for each longitudinal segment. The inner HCAL is extended along the backwards direction, and is comprised of 12 tiles in  the forward $\eta$ direction and 15 tiles in the backward $\eta$ direction. There are 1536 readout channels (towers) in the OHCAL and 1728 channels for the IHCAL. 

\subsection{Hadron Endcap Electromagnetic (FEMC) and Hadronic Calorimeter (LFHCAL)}
\begin{figure}[htb]
    \centering
    \includegraphics[width=0.9\columnwidth]{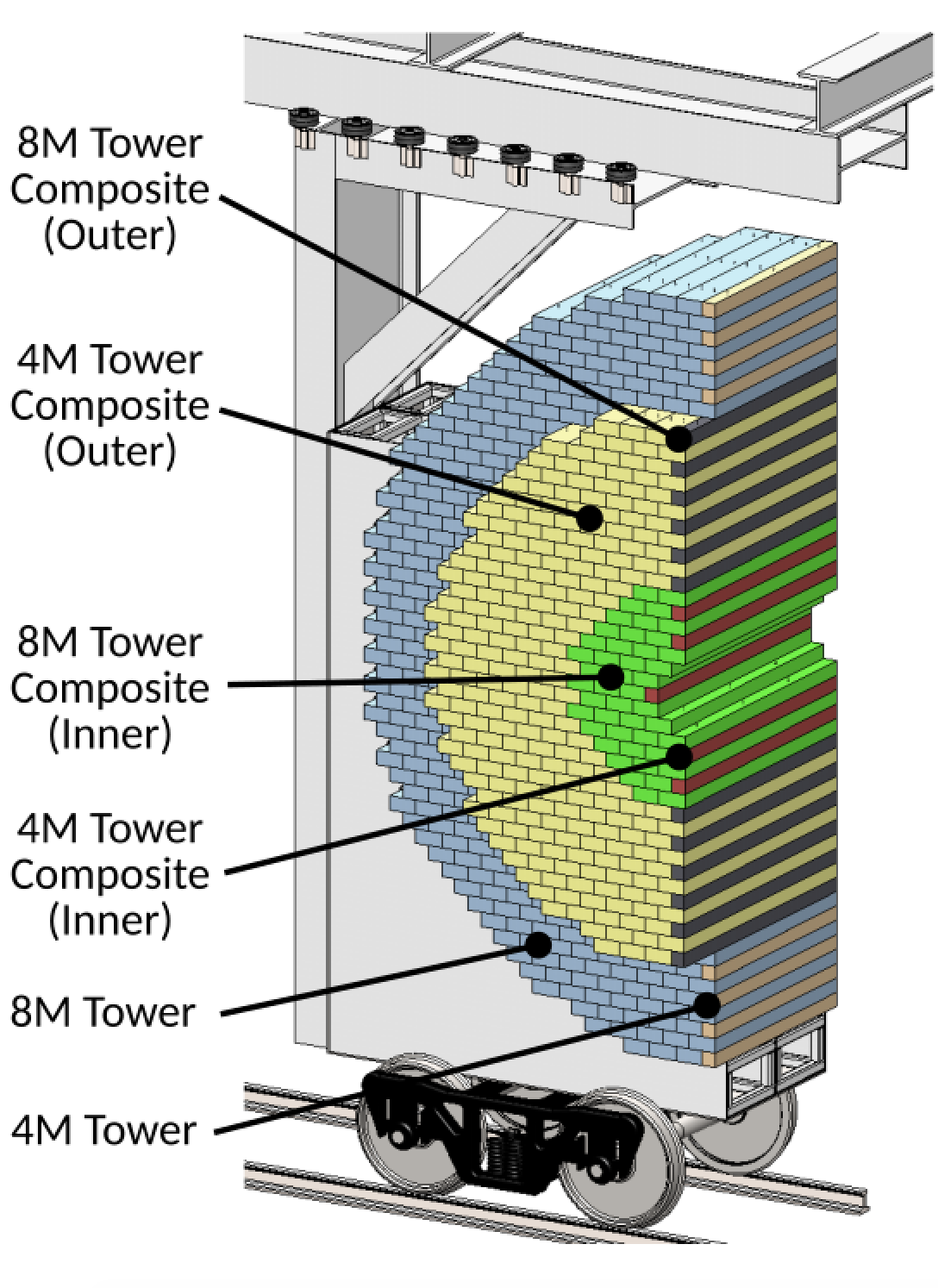}
    \caption{Detail of the  forward electromagnetic calorimeter, FEMC, showing a fully assembled half disk.   These detectors will be on rails allowing them to come together around the beamline.}
    \label{fig:fcal}
\end{figure}
\begin{figure*}[t]
    \centering
    \includegraphics[width=0.9\textwidth]{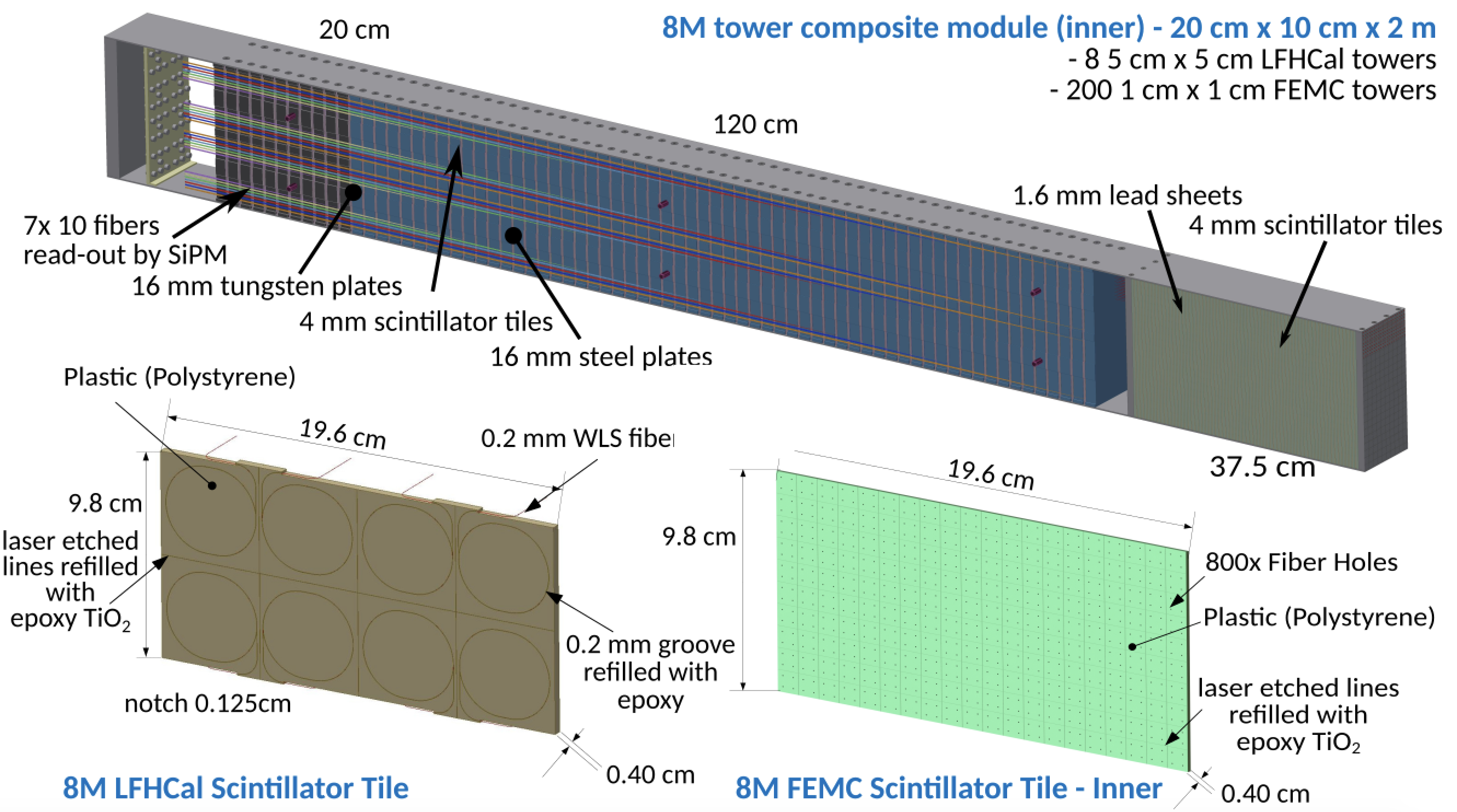}
    \caption{Details of the LFHCal design indicating  the 8-tower module design and the individual scintillator tile designs for the LFHCAL-FEMC 8M tower inner module.}
    \label{fig:lfhcal}
\end{figure*}
The desired performance in the forward region is 
governed by the jet energy resolution requirements, 
as well as very good energy resolution (35\%/$\sqrt{E}$ to reach the desired resolution in $\delta x$) for the physics processes connected to the origin of mass. Additionally, an excellent position resolution in particular within the ECal is required for PID within the jet. Within this region a higher particle density is expected than in the central barrel, supporting the need for excellent position an energy resolution in both calorimeters. Both detector systems need to be able to handle the expected energies of incoming particles up to 150 GeV. Due to the asymmetric collision system, these calorimeters are therefore focused strongly on high energetic particle shower containment while still providing good energy resolution at low energies.

We envision the forward calorimeter system as an integrated ECal and HCal, where the installation units, where appropriate, are constructed in a common casing. These so-called modules will consist of an electro-magnetic calorimeter segment in the front which is part of the forward EMCal (FEMC) followed by a hadronic calorimeter segment which is part of the longitudinally separated HCal (LFHCal). In between these segments a read-out section is foreseen for the ECal. The modules of up to four different sizes will be installed in half shells surrounding the beam pipe, which are movable on steel trolleys to give access to the inner detectors in the barrel in the hadron going direction. 
This integrated E- and HCal design reduces the dead material in the detector acceptance and allows for an easier installation in the experimental hall. 

The forward ECal (FEMC) will be a Pb-Scintillator shashlik calorimeter. It is placed after the tracking and particle identification detectors and made up of two half disks with a radius of about 1.83m as shown in Fig.~\ref{fig:fcal}. The calorimeter is based on the lead-scintillator ”shashlik” calorimeter designs already utilized for ALICE, STAR and PHENIX. However, it employs more modern techniques for the readout as well as scintillation tile separation. 
The towers were designed to be smaller than the Moliere-radius in order to allow for a further shower separation at high rapidity. 

The towers have an active depth of $37.5$~cm with and consist out of
66 layer of 0.16~cm Pb sheets and 0.4~cm scintillator material, as can
be seen in Tab.~\ref{tab:calospecs}.  Due to the high occupancy of the
detector at large pseudorapities and the collimation of the particles
in this area in physical space, the tower size will vary depending on
its radial position with respect to the beam axis. Towers which are
close to the beam pipe ($R < 0.8$~m) will have a active tower size of
1~cm$\times$1~cm$\times$37.5~cm. For the outer radii this granularity
is not necessary and thus the size is increased to
1.65~cm$\times$1.65~cm$\times$37.5~cm.  In order to collect the light
produced in the scintillator tiles, each scintillator and Pb-plate is
pierced by four 0.2~mm-wavelength shifting fibers. These fibers are
used to collect the light generated in the scintillators across all 66
layers. All four fibers are read out together by one silicon
photomultiplier (SiPM).  The FEMC is constructed with modules size of
at least 5~cm$\times$5~cm$\times$37.5~cm (1M module) up to
10~cm$\times$20~cm$\times$37.5~cm (8M modules) aligning with the
module sizes of the hadronic calorimeter. In order to separate the
light produced in different segments of the 8M-tile a gap between the
1~cm$\times$1~cm tower tiles is created by edging into the scintillator
using a laser. These 0.37~mm deep gaps are then refilled with a
mixture of epoxy and titanium-oxide in order to reduce the light cross
talk among different towers. Depending on their radial position this
leads to either 72 or 200 read-out towers in one 8M modules.

The longitudinally segmented forward HCal (LFHCAL) is a
Steel-Tungsten-Scintillator calorimeter adapted from the PSD
calorimeter for the NA61/SHINE experiment~\cite{Guber:109059}, but it
has been severely modified to meet the desired physics performance
laid out in the Yellow Report. It is made up of two half disks with a
radius of about 2.6~m.

The LFHCAL towers have an active depth of 1.4~m with an additional
space for the readout of about 20--30~cm depending on their radial
position, as seen in Table~\ref{tab:calospecs}. Each tower consists
out of 70 layers of 1.6~cm absorber and 0.4~cm scintillator
material. For the first 60 layers the absorber material is steel,
while the last 10 layers serve as tail catcher and are thus made out
of tungsten to maximize the interaction length within the available
space. The front face of the tower is 5~cm$\times$5~cm.

In each scintillator a loop of wavelength shifting fiber is embedded,
as can be seen in Fig.~\ref{fig:fcal} (middle). Ten consecutive fibers
in a tower are read out together by one Silicon photo multiplier,
leading to seven samples per tower. The towers are constructed in
units of 8-,4-, 2- and 1-tower modules to ease the construction and
reduce the dead space between the towers and the active detection
area. Similar as for the FEMC the scintillator tiles in the larger
modules are made out of one piece and then separated by a gaps
refilled with epoxy and titanium oxide to reduce light cross-talk
among the different readout towers. For the same purpose the
wavelength shifting fibers running on the sides of the towers are
grouped early on according to their readout unit and separated by thin
plastic pieces over the full length. They terminate in one common
light collector which is directly attached to a SiPM. The entire
detector will consist of 63280 readout channels grouped in 9040
read-out towers.

The majority of the calorimeter will be built out of 8-tower modules
($\sim$1091) which are stacked in the support frame using a lego-like
system for alignment and internal stability, as can be seen in
Fig.~\ref{fig:fcal}~(left). The remaining module sizes are necessary
to fill the gaps at the edges and around the beam pipe to allow for
maximum coverage. The absorber plates in the modules are held to their
frame by four screws each. To leave space for the read-out fibers, the
steel and scintillator plates are not entirely square but equipped
with 1.25~mm grooves, creating the fiber channels on the sides. These
fiber channels are covered by 0.5~mm thin steel plates for protection
after module installation and testing, in order to protect the fragile
fibers. For internal alignment we rely on the usage of 1--2~cm steel
pins in the LFHCal part which are directly anchored to the steel or
tungsten absorber plates. Afterwards the modules will be
self-supporting within the outer support frame. The steel in the
LFHCAL serves as flux return for the BaBar magnet, thus a significant
force is exerted on the calorimeter, which needs to be compensated for
by the frame and internal support structure.  The achieved energy
resolution accoding to the simulations for both calorimeters can be
found in Fig.~\ref{fig:CaloResolution}. The required resolutions can
be met in both cases and further improvements can be expected using
machine learning for the clusterization which proves challenging in
this direction. The excellent position resolution in the FEMC should
in addition allow the effective separation of electrons and pions as well
neutral pion decays, as seen in Fig.~\ref{fig:pidperf}. 

\section{Far-Forward/Far-Backward Detectors}

\begin{figure*}[!ht]
\centering
    \includegraphics[width=0.9\textwidth]{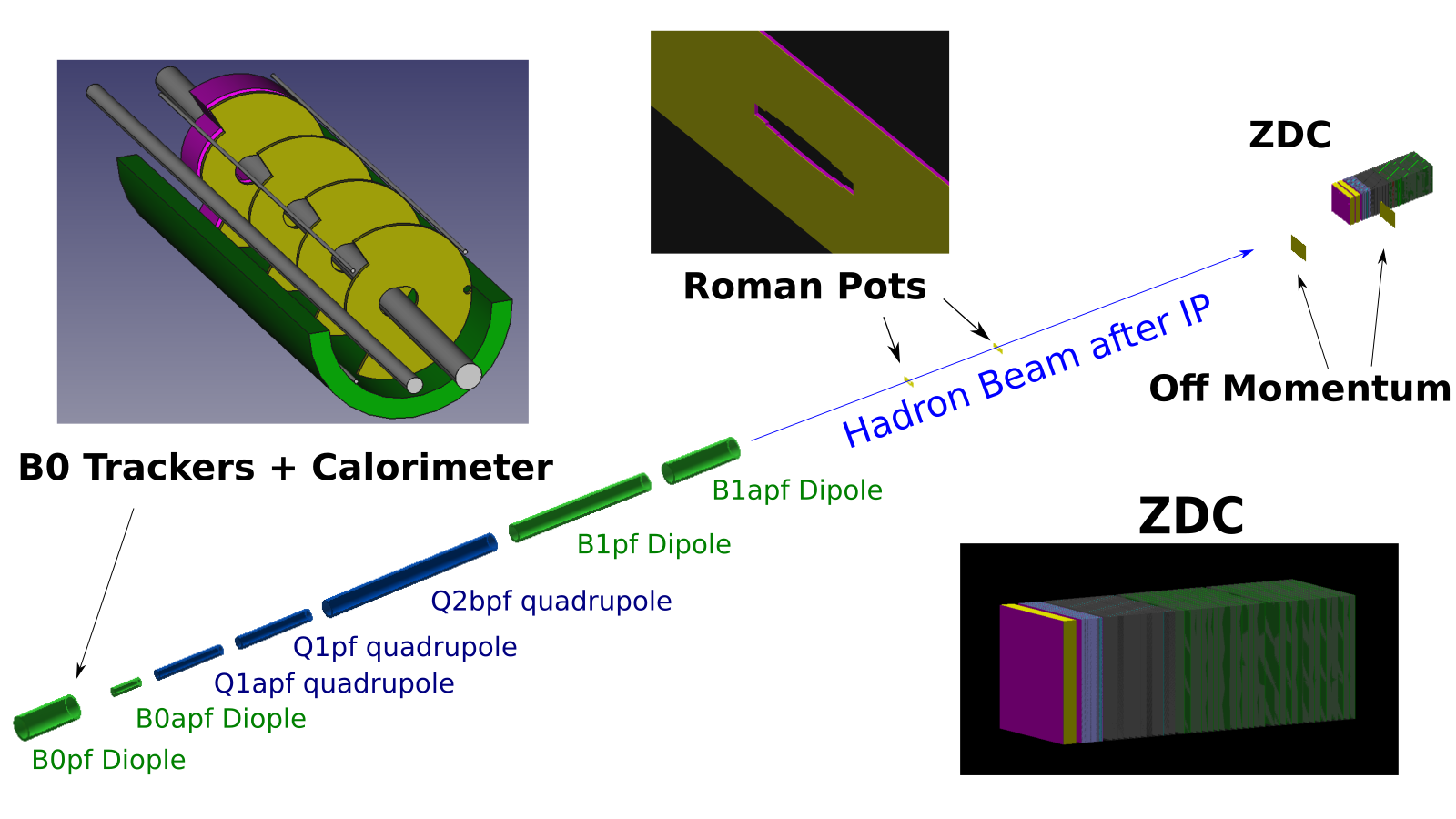}
    \caption{\label{fig:far-forward} The layout of the EIC Far-Forward region.}
\end{figure*}

A schematic of the far-forward detectors is shown in
Figure~\ref{fig:far-forward} and include the B0 spectrometer,
off-momentum trackers, Roman Pots and ZDC (see
Table~\ref{tab:FF_accep} for position and dimensions). The
far-backward region consists of two detector systems (low-$Q^2$ tagger
and luminosity monitor). All far-forward/far-backward detectors are
required for the EIC physics as described in the Yellow Report.  The
following describes their setup and performance.  For further details,
see Ref.~\cite{ecce-note-det-2021-05}.

\begin{table*}[t]
   \caption{Summary of far-forward detector locations and angular
     acceptances for charged hadrons, neutrons, photons, and light
     nuclei or nuclear fragments. In some cases, the angular
     acceptance is not uniform in $\phi$, as noted in the table. For
     the three silicon detectors (Roman Pots, Off-Momentum Detectors,
     and B0 spectrometer) a depth is not given, just the 2D size of
     the silicon plane. For the Roman Pots and Off-Momentum Detectors,
     the simulations have two silicon planes spaced 2m apart, while
     the B0 detectors have four silicon planes evenly spaced along the
     first 1.0~m length of the B0pf dipole magnet bore. The planes have
     a "hole" for the passage of the hadron beam pipe that has a
     radius of 3.2~cm. }
    \label{tab:FF_accep}
    \centering
    \resizebox{\textwidth}{!}{
    \begin{tabular}{lcccc}
    \toprule
        \textbf{Detector}\ & \textbf{(x,z) Position [m]} & \textbf{Dimensions}& \textbf{$\theta$ [mrad]} & \textbf{Notes} \\
         \midrule
         ZDC & (-0.96, 37.5) & (60cm, 60cm, 1.62m) & $\theta < $ 5.5 & $\sim$4.0 mrad at $\phi = \pi$ \\
         Roman Pots (2 stations) & (-0.83, 26.0) (-0.92, 28.0) & (30cm, 10cm) & $0.0 < \theta$ $< 5.5$ & 10$\,\sigma$ cut. \\
         Off-Momentum Detector & (-1.62, 34.5), (-1.71, 36.5) & (50cm, 35cm) & $0.0 < \theta < 5.0$ & $0.4 < x_{L} < 0.6$ \\
         B0 Trackers and Calorimeter & (x = -0.15, $5.8<$\,z\,$< 7.0$) & (32cm, 38m) & $6.0 < \theta < 22.5$ & $\sim$20\,mrad at $\phi$=0 \\
      \bottomrule
    \end{tabular}
    }
\end{table*}

\subsection{B0 Detector}
\label{sec:b0}

The B0 spectrometer is located inside B0pf dipole magnet. Its main use is to measure forward going hadrons and photons for exclusive reactions.
The B0 acceptance is defined by the B0pf magnet. Its design is challenging due to the two beam pipes (electron and hadron) that it needs to accommodate and the fact that they are not parallel to each other due to the 0.025 mrad IP6 crossing angle. Moreover, the service access to the detectors inside of the dipole is only possible from the IP side, where the distance between the beam pipes is narrowest. Following these limitations the B0 detector require using compact and efficient detection technologies. 

Our design uses four AC-LGAD tracker layers with 30 cm spacing between each layer. They will provide charged particle detection for $6<\theta<22.5$ mrad. The use of AC-LGAD sensors will allow good position and timing resolutions. The AC-LGAD sensor will have a 3.2x3.2 cm$^2$ area, with four dedicated ASIC units on each sensor. In addition, a \PbWOiv calorimeter will be positioned behind the fourth tracking layer at 683 cm from the IP. 
Using the \PbWOiv in the B0 calorimeter will increase the detection fraction of the two decayed $\gamma$s from the $u$-Channel $\pi_0$ production from 40\% to 100\%, and enable measurements of $u$-Channel DVCS
events which without it will be swamped by the $\pi^0$ events with single $\gamma$ detected.
The calorimeter is constructed from 10 cm long 2x2 cm$^2$ \PbWOiv crystals and positioned to leave 7 cm for the readout system. Both trackers and Calorimeter has oval holes in the center to accommodate the hadron beam pipe, and a cutaway in the side to accommodate the electron beam and allow installation and service of the detector system (see Fig.~\ref{fig:far-forward}).

\begin{figure}[ht]
\centering
    \includegraphics[width=0.4\textwidth]{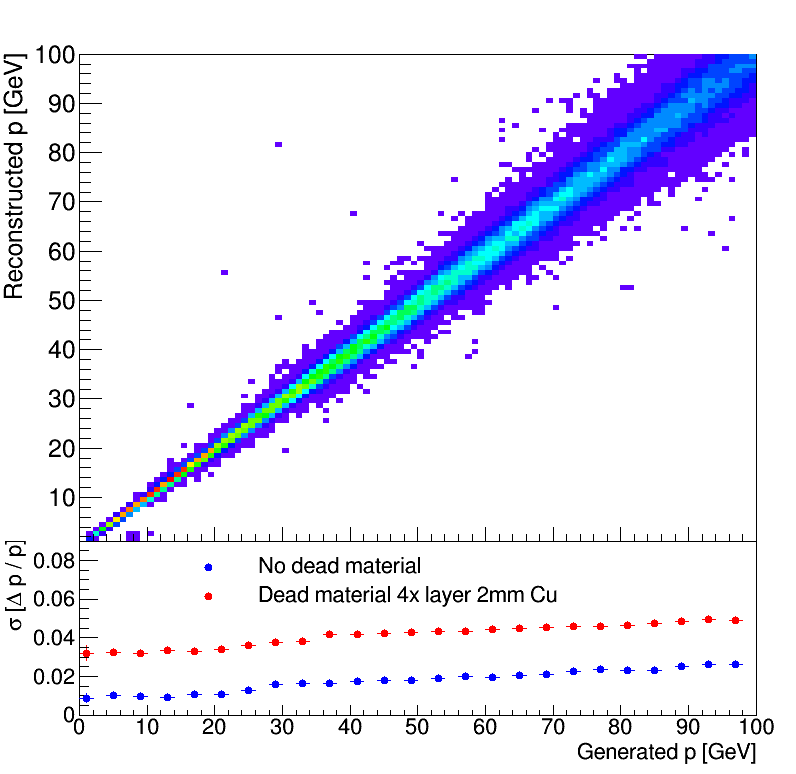}
    \includegraphics[width=0.4\textwidth]{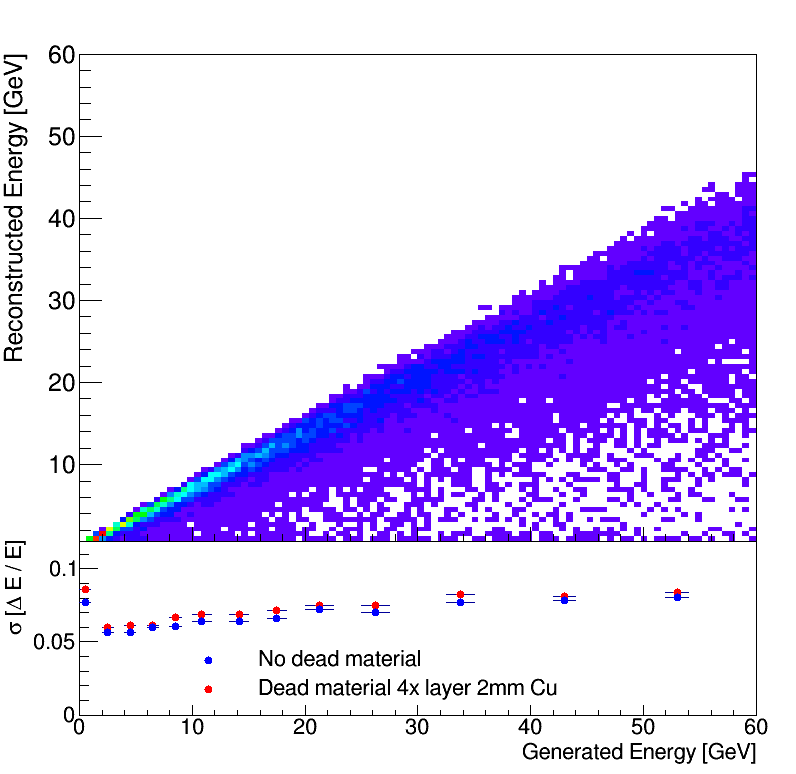}
    \caption{\label{fig:b0} (top) Reconstructed momentum and its resolution for $\mu^{-}$ tracks found in the B0 tracker; (bottom) reconstructed energy of photons and its resolution in the B0 calorimeter.}
\end{figure}

Figure~\ref{fig:b0} (left) shows the simulated momentum and its resolution $\sigma[\Delta p/ p]$ as a function of truth momentum. It is below 5\% for the studied kinematic region. The effect of the presence of dead material (2mm of Cu after each Si plane) layers on the momentum resolution is also shown and estimated to degrade the resolution by 2\% uniformly as a function of $p$.
The photon energy reconstructed in the B0 calorimeter and its resolution are shown in Fig~\ref{fig:b0} (right) for photons originating in the interaction vertex with pseudorapidity $4 < \eta < 6$ and energy $0 < E_{\gamma} < 60$~GeV. It is found to be below 7\% for the studied kinematic region. In general about 60\% of the energy is reconstructed within a 2x2 crystal grid with some dips in efficiency at low $E_{\gamma}$  and high $\eta$. 

\subsection{Roman Pots}

 Diffractive processes such as deeply virtual Compton scattering will produce protons with high energy and small $p_T$  with only a small  separation from the hadron beam.  The Roman Pots are designed to detect such particles. They 
 will consist of two double-layer 25x12 cm$^2$ AC-LGAD stations, located inside the beam line 26 and 28 m downstream the interaction point and $10\sigma$ from the main beam.
 This technology will provide the necessary position and timing resolution for a precise measurement with minimized background.
 
 The vacuum environment will require special cooling. We will use heat sinks made of  metal foam through which compressed air will flow. Such cooling systems are already in use  at the LHC.

\subsection{Off-momentum Detectors}

Off-momentum detectors  complement the Roman Pots by  measuring charged particles that have a  smaller magnetic rigidity than the main hadron beam. Such particles will be  bent outside the beam pipe. The 
detectors consist of tracking planes based on  AC-LGAD sensors. 

Good timing resolution on the order of 10~ns facilitates the rejection of pileup and beam related background, since particles that do not come directly from the interaction point will have a different flight path than the particles of interest. Such techniques have been used extensively by the CMS Precision Proton Spectrometer and the ATLAS Forward Proton Group at the LHC.

\subsection{Zero Degree Calorimeter (ZDC)}
The size of the ECCE ZDC is 60 cm$\times$60 cm$\times$162 cm, and the weight is greater than 6t. As shown in Fig.~\ref{fig:far-forward}, the ZDC consists of \PbWOiv crystal layer, W/Si layer, Pb/Si layer and Pb/Scintillator layer.  

The estimated energy resolution for high energy photons is well below the required value. For the low energy photons, estimated resolution for 100~MeV photons using 5\% smearing reaches 20\%, which is is still acceptable. 
The neutron energy resolution is consistent with and even smaller than the Yellow Report required value of $50\%/\sqrt{E}+5\%$. For 40 GeV and 20 GeV photons, the position resolution is estimated as 1.1~mm and 1.5~mm respectively. On the crystal layer, the cluster finding efficiency is $>95\%$ for both 20 GeV photons and 100 MeV photons with the seed energy requirement of 15~MeV for the clustering. 

\begin{figure}[!t]
\centering
    \includegraphics[width=0.4\textwidth]{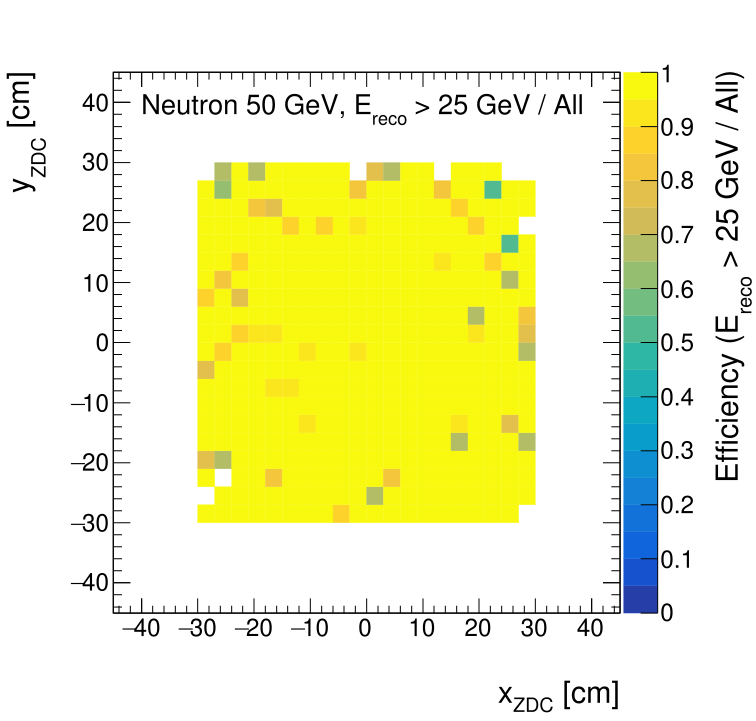}
    \includegraphics[width=0.4\textwidth]{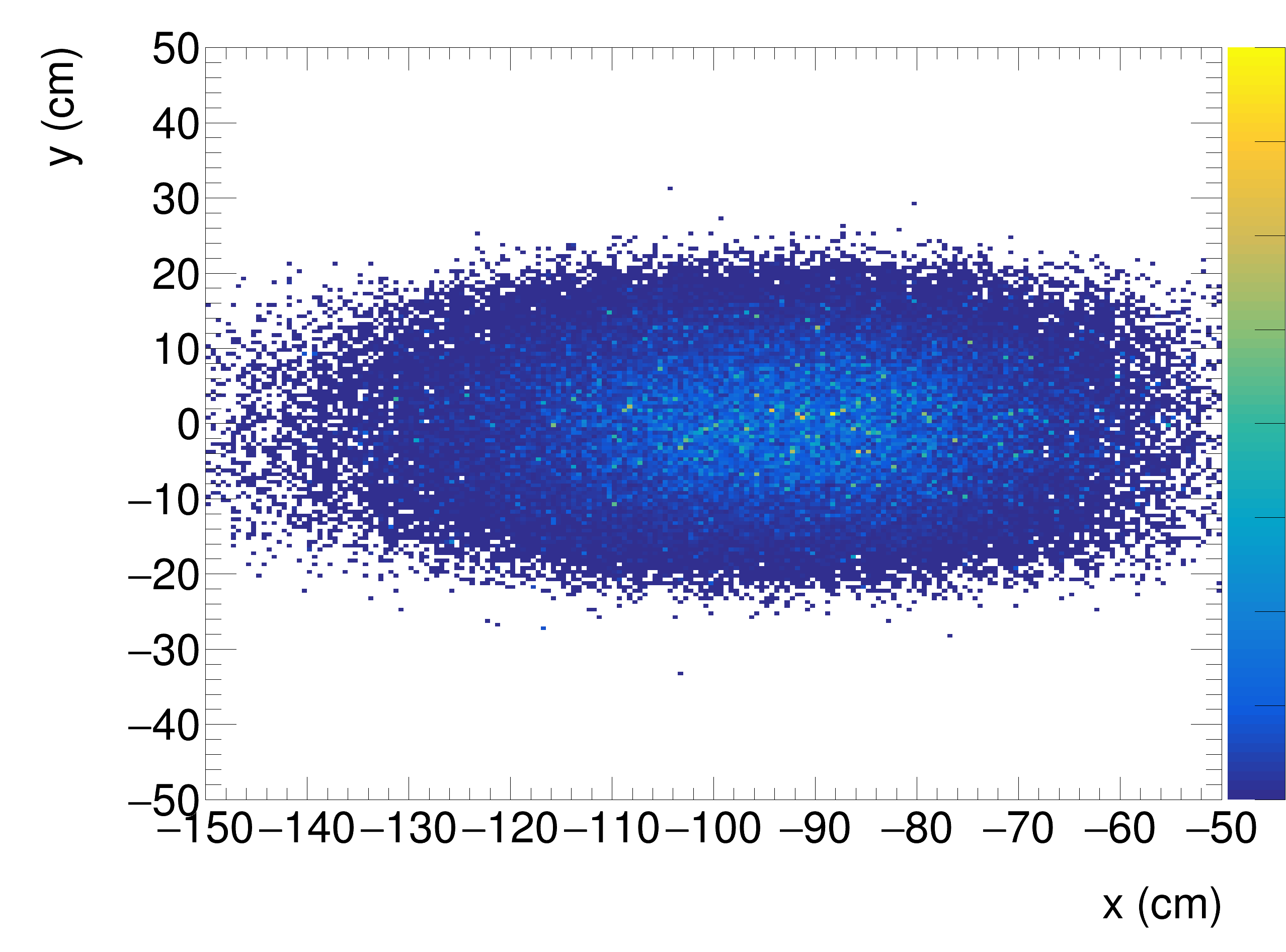}
    \includegraphics[width=0.4\textwidth]{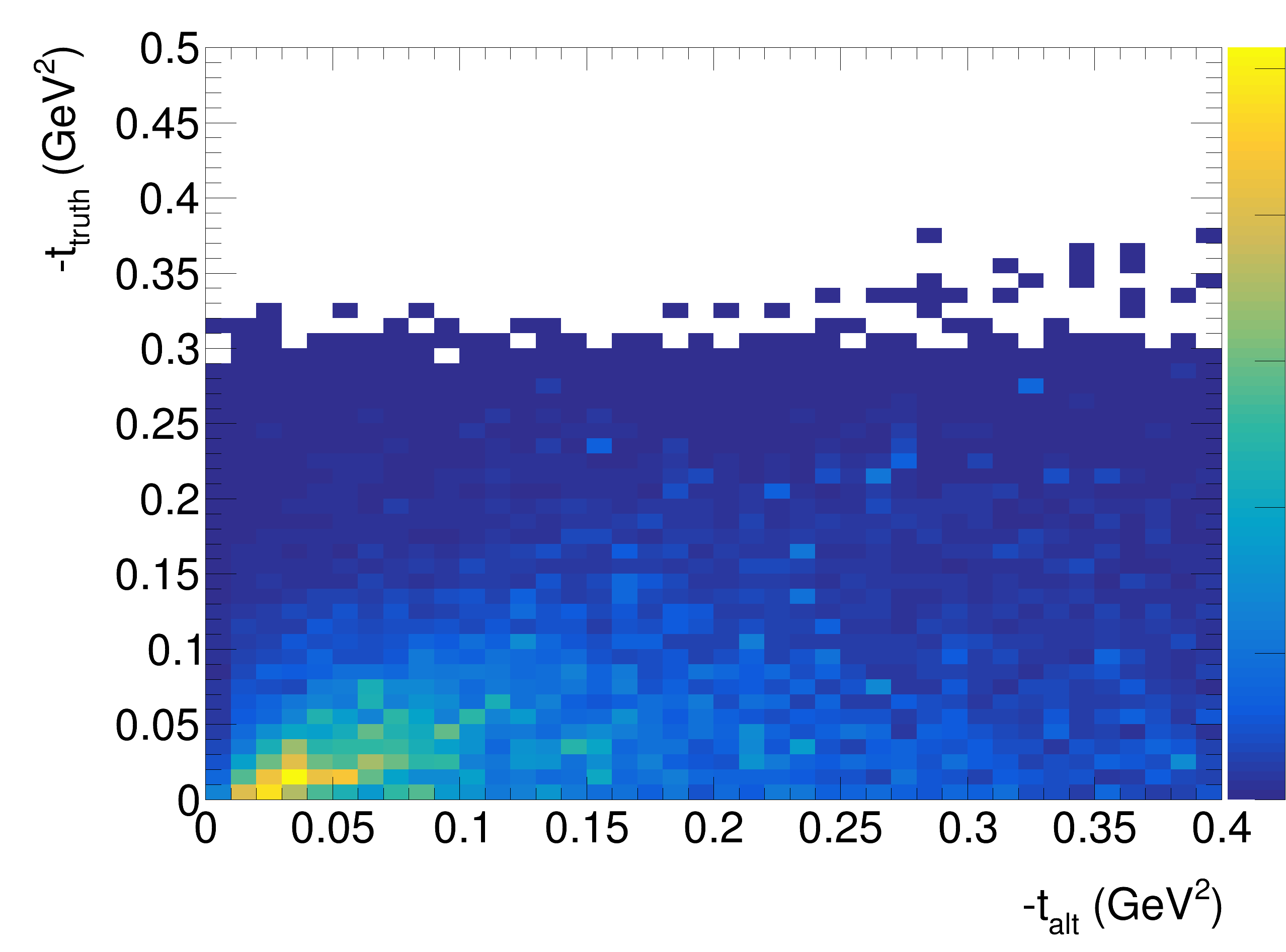}
    \caption{\label{fig:zdc} (top) ZDC detection efficiency for neutrons in its local coordinate system. (middle) Detection distribution of neutron hits in the ZDC for meson structure function processes without the beam pipe blocking contribution. $z$-axis reflects the normalized yield. (bottom) Reconstructed $t$ versus true $t$, where $t$ is reconstructed as from the baryon information, $t_{alt}=(p_p-p_n)^2$, which is reliable with a resolution of $<0.025$ GeV$^2$.}
\end{figure}

While the ZDC is used for a variety of measurements in ECCE, we evaluate its performance here using simulations of meson structure function
measurements that represent a key performance driver for this detector.
In these reactions, neutrons from the Sullivan process carry 80-98\% of the proton beam momentum and are detected at far-forward angles in the ZDC. The detection fraction for neutrons ($t$ resolution) is 59\% (0.019 GeV$^2$) at the lowest, 5 on 41, and 100\% (0.005-0.007 GeV$^2$) at the higher energy combinations.
Due to the large size and high inherent ZDC detection efficiency (Fig.~\ref{fig:zdc}~(left)), the ECCE detection efficiency for these events is quite high, $\sim 80\%$, and nearly independent of $Q^2$.  A density plot of event distribution is shown in the left panel of Fig.~\ref{fig:zdc}.  The detection efficiency is highest for events with small $-t<0.15$ GeV$^2$, which are needed for measurements such as the pion form factor, and decreases rapidly with $-t$.  The $t$-range of optimal acceptance is dictated by the size of the ZDC, as the energetic neutrons from higher $-t$ events are emitted at an angle larger than the ZDC acceptance.

We further find the ZDC to offer excellent reconstruction of $t$. Compared with the $t$ reconstruction from the measurement of the $\pi^{+}$ and $e'$ tracks, the ZDC's baryon measurement is significantly more reliable, in agreement with EIC Yellow Report studies. Due to the excellent position resolution of the ZDC, the neutron track momentum is reconstructed to within $1\%$ of the "true" momentum.
With this information, $t$ is reconstructed from the neutron track in a manner that reproduces the true value very closely, see Fig.~\ref{fig:zdc}~(right). Such a reliable reconstruction of $t$ is essential for many processes such as the pion form factor measurement, where the rapid fall off of the cross section needs to be measured to confirm the dominance of the Sullivan mechanism. The high quality ZDC proposed by ECCE is clearly of paramount importance to the feasibility of such measurements.

\subsection{\texorpdfstring{Low-$Q^2$ Tagger}{Low-Q2 Tagger}}

The low $Q^2$-tagger will facilitate measurement of reactions with small cross sections, e.g. timelike Compton scattering. Measuring the scatted electron will allow the $s$ dependence to be measured as well as giving some measure of the production four momentum transfer, or $t$. When coupled with proton detection in the far forward region there will be the possibility of applying exclusivity cuts. 

The low-$Q^2$ Tagger consists of two stations, located 24~m and 37~m from the interaction point. Each station includes a double layered AC-LGAD tracker, followed by a \PbWOiv electromagnetic calorimeter. The detectors surface areas are 40.5~cm$\times$40.5~cm at 24~m and 30~cm$\times$21~cm at 37~m and their calorimeters both use 20~cm long 2~cm$\times$2~cm \PbWOiv crystals.

The tracking planes enable the determination of the electron scattering angle, that in turn facilitate a precise determination of $Q^2$. The calorimeter provides an energy measurement to complement the tracking and provide additional shower shape information to confirm that the particle really is an electron.

\subsection{Luminosity Monitors}

For the luminosity measurements, an accuracy of the order of 1\% is required, or relative luminosity determination exceeding $10^{-4}$ precision. The latter is driven by the size of the asymmetries we want to measure. This requirement drives the utilization of several complementary approaches for both relative and absolute measurements of the luminosity, allowing us to understand and constrain the beam-size effects, synchrotron radiation, as well as systematic uncertainties. The approach we will follow is based on existing experience from HERA. The absolute luminosity is determined by correlating the total energy in the calorimeter with the number of photons. The low-$Q^2$ tagger can also provide key information on the relative luminosities and thus impose further constraints on the luminosity determination.

The luminosity monitor will be located along the photon zero-degree line in the far backward region and will  measure bremsstrahlung photons. 
 It uses both a  dedicated calorimeter to measure direct photons,
 and two spectrometer arms to measure 
 $e^+e^-$ pairs from conversions. 
The direct photon calorimeter will have a size of 16~cm$\times$16~cm  and will use 20cm long 2x2 cm$^2$ \PbWOiv crystals. The $e^+$ and $e^-$ from photon conversions  will be deflected above and below the main photon beam by a small dipole magnet before entering the spectrometer arms. Each arm includes two  8$\times$16~cm$^2$ AC-LGAD tracking layers followed by a \PbWOiv calorimeter with a matching surface area (also made of 20cm long 2x2 cm$^2$ crystals). The tracking planes in the $e^+/e^-$ arms will allow reconstructing the gamma spot to help understand and constraint beam-size effects.
 



\section{Electronics and Data Acquisition}
\label{sec:elec_daq}


The general design of the ECCE data acquisition builds on the sPHENIX
DAQ system and many of the JLAB streaming readout systems under
test~\cite{ecce-note-det-2021-05}. These systems already incorporate and demonstrate almost all
concepts of the envisioned ECCE DAQ system.  The ECCE DAQ system will
be built around a trigger-less Streaming Readout (SRO) concept from
the start. 

\begin{figure*}[th]
  \begin{center}
    \includegraphics[width=1.6\columnwidth]{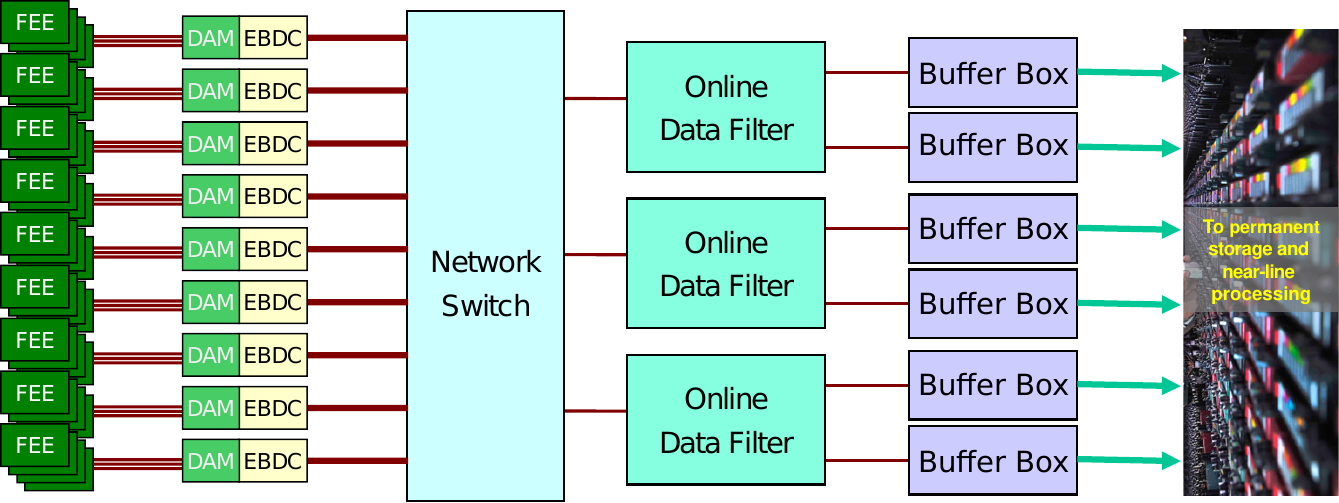}
  \end{center}
  \caption{\label{fig:DAQ_Overview} The schematic view of the ECCE
    Data Acquisition system. With the detector systems connecting to
    FEE cards from the left, the digitized data are sent to ``Data
    Aggregation Modules'' (DAM) that filter and package the data. The
    ``Event Buffer and Data Compressor'' (EBDC) nodes perform another
    filter, noise suppression, and clustering step on the scope of the
    connected detector channels, and align the hits by timing
    value. The data are then sent to processing nodes that perform a
    filtering/triggering step on the entire detector view. Data from
    selected crossings then get stored temporarily on large file
    servers (``Buffer Boxes'') before being sent to long-term storage
    at the computing center.}
\end{figure*}

As detailed in the Yellow Report~\cite{AbdulKhalek:2021gbh}, the Streaming
Readout concept has proven superior to a classic triggered scheme
in several ways. Modern readout technologies often do not follow a
strict ``event'' paradigm in the sense that data from collider
crossing $n$ are already arriving from one front-end, while other
parts can still be transmitting data from trigger $n$-1, $n$-2, or
earlier crossings. In streaming mode, there is no need to wait for the
completion of the data transmission from a given crossing, as the data
parts are later re-assembled by their embedded clock information. This
usually leads to a higher data throughput when compared to a triggered mode.

The other advantage is that classic trigger setups are always limited
in their selection power because the amount of data they can sample to
arrive at a trigger decision is generally much more restricted than in
streaming mode, where the software- or firmware-based selection
algorithms have, at least in principle, access to the data from all
detector components. The processing power to increase the quality of
the event selection has become cheaper every year, and this trend is
expected to continue.

In a trigger-less data acquisition scheme, each channel with a signal
exceeding a threshold is transferred after being labeled with a
time-stamp, irregardless of the status of the other channels. The
resulting data is often a waveform, or a list of fired pixel-type
detector elements, or some combination of both. Subsequent processing
layers reduce the amount of information by categorizing the
information by time, so that eventually the detector information of
one bunch crossing is together in one place. While traversing the
various processing layers, data get filtered and packaged, and
waveform processing and clustering algorithms are applied that further
reduce the amount of data to a few key properties.

The progression of processing layers is schematically shown in
Fig.~\ref{fig:DAQ_Overview}. With the connections from the detector,
typically fibers, coming from the left, detector-specific Front-End
Electronics (FEE) cards digitize the signals, and send digital data on
to the ``Data Aggregation Module'' (DAM). An current example of such a
DAM is the ATLAS FELIX card~\cite{Anderson_2016}.

The DAM plays a central role as it provides a common detector
interface for the expected large variety of detector readout
technologies that are found upstream of the DAM. While the DAM still
needs to run detector-specific firmware to receive and package the
data, it provides common hardware and common APIs for the subsequent
data handling, and greatly reduces the software development efforts. 

The ``Event Buffer and Data Compressor'' (EBDC) nodes, the offline
data filter, and the file servers (``Buffer Boxes'') shown in
Fig~\ref{fig:DAQ_Overview} are Linux PCs that form the next layers of
the processing chain.

The Front End Electronics including ASICs will need to be compatible
with the streaming readout DAQ system plan. FEE will need to support
continuous sampling modes and not require an external trigger to
convert detector signals because this will introduce large unwanted
DAQ deadtime. Full waveform sampling for high occupancy detectors with
zero suppression and feature extraction (time \& charge) will be
needed for a flexible streaming readout system. 

\begin{table}[t!]
\caption {\label{tab:PID_Count}PID Detector ASICs and channel counts. The ASICs for the time-of-flight detectors are currently under development in eRD112} 
\centering
\small
  \begin{tabular}{lcr}
    \toprule
    \textbf{Detector}     & \textbf{ASIC}   & \textbf{Channels}    \\
    \midrule
    hpDIRC & High Density SoC  &  69,632  \\
    CTTL    & eRD112             &
                                                                      8.6M       \\ \midrule
    mRICH  & High Density SoC  & 65.5K    \\
    ETTL    & eRD112             & 0.92M        \\ \midrule
    dRICH          & MAROC3             & 5.4K      \\
    FTTL & eRD112             & 1.84M     \\ \midrule
          
    Roman Pots          & eRD112              &  524.3K     \\
    B0 Detector              & eRD112              & 2.6M  \\
    Off-Momentum Detectors              & eRD112              & 1.8M    \\ \midrule
    Low-$Q^2$ Tagger          & eRD112             &   4.6M   \\
    Luminosity Monitor              & eRD112             & 268.4K     \\
    \bottomrule
  \end{tabular}
\end{table}

ASIC devices have been carefully evaluated for each of the ECCE
experiment detector systems and are listed for the PID detectors in
Table~\ref{tab:PID_Count}. High channel counts for the hpDIRC and mRICH detectors have
based their readout on the High Density System-on-a-Chip (HDSoC) ASIC
that is commercially produced by Nalu Scientific. The HDSoC has 64
channels and a very high bandwidth sampling ADC for waveform capture
and feature extraction modes. This ASIC will support the streaming
readout model. The dRICH detector is planning to use the MAROC3 ASIC
which is a 64-channel device that interfaces directly to a 64 pixel
maPMT device. Supporting electronics to configure the MAROC3 and
provide streaming data has been in use at Jefferson Lab for the CLAS12
RICH detector for several years and is a mature technology and the
MAROC3 device is now commercially available. The 64-channel SAMPA amplifier and digitizer ASIC is strongly
considered for the \mrwell tracking detectors and is a very good example of an
ASIC that will operate within the requirements of a streaming readout
front end. 

AC-Low Gain
Avalanche Diodes (AC-LGAD) sensors planned for the Time-Of-Flight PID
detector system, where the channel counts are very dense, as well as the far-forward detectors.
Development of front-end electronics, particularly ASIC chips, for AC-LGAD readout is part of the eRD112 project for targeted EIC detector R\&D.
The strategy is to base designs on the ATLAS ALTIROC (130~nm) and 
CMS ETROC (65~nm) designs as a starting point, and reduce 
the pixel granularity and timing jitter to meet the EIC 
requirements. 
Specifically, the IJCLab (Orsay)/ OMEGA (IN2P3-\'Ecole 
Polytechnique) group on the eRD112 team is a main 
developer of the ATLAS ALTIROC, and will play the lead 
role at the initial stage of ASIC development. A 
preliminary 130~nm ASIC design with a pitch size of 
0.5~mm$\times$0.5~mm has been achieved as a stepping stone, that meets the requirements set by the EIC Roman Pot, B0 detector, and Off-Momentum detector. Future development will focus on further improving the timing jitter and scaling up to meet the requirements of the large-scale TOF system.


The calorimeter readout in ECCE will make use of a common digitizer
design for all calorimeter systems. The development will start with
the existing 64-channel, 14-bit ADCs running at six times the RHIC
bunch crossing frequency of just below 10~MHz, at about 60~MHz
designed for the sPHENIX calorimeters. ECCE will have a common
digitizer design for all calorimeters, although the form factors may
differ depending on the detector implementation. It is likely that the
sampling frequency will be higher based on the detector requirements.
The ECCE calorimeter subsystem includes a very high channel count,
however no custom ASIC development is considered because the existing
sPHENIX 64-channel 14-bit ADC design is proven and reduces the number
of separate electronics designs that need to be developed, veriﬁed,
and maintained throughout the lifetime of the experiment.

\section{Computing plan}
\label{sec:computing}

\begin{figure*}[t]
 \centering
   \includegraphics[width=\linewidth]{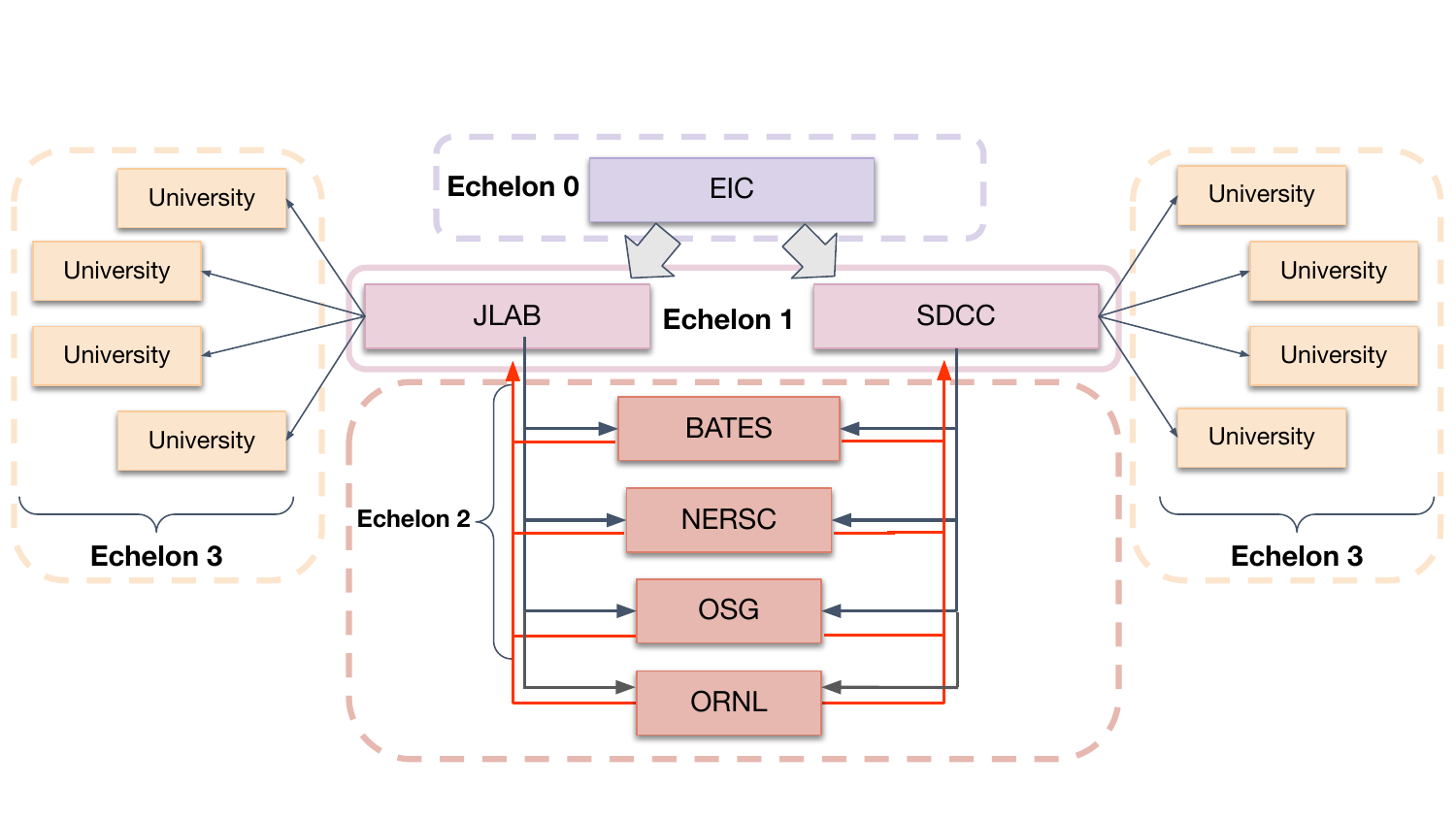}
  \caption[Federated Computing Butterfly model.]{\label{fig:computing-federated_offsite_butterfly} EIC Butterfly model of federated offsite computing~\cite{ecce-note-comp-2021-01-nim}. In this model, nearly all storage is contained in echelon 1 while large portions of the raw data processing is delegated to multiple HTC/HPC facilities. }
\end{figure*}

The ECCE consortium plans to deploy a federated computing model for the EIC, where multiple facilities are used. A similar strategy has been successfully deployed by the LHC in the form of the Worldwide LHC Computing Grid (WLCG)~\cite{SHIERS2007219}. ECCE has developed a tiered ``Butterfly'' model for EIC computing as shown in Figure~\ref{fig:computing-federated_offsite_butterfly}~\cite{ecce-note-comp-2021-01-nim}. In this model, both compute and storage resources are distributed with data storage focused at the Echelon 1 sites. This means access to data by users will be performed by connecting Echelon 3 sites directly to Echelon 1 sites. The Echelon 1 sites will themselves provide significant compute capability, but will also farm out large campaigns to Echelon 2 sites, taking advantage of the diverse computing resources available at collaborating institutions.

We have adopted a fixed-latency offline computing model where both the final calibration and reconstruction of raw data occur within 2-3 weeks of acquisition~\cite{ecce-note-comp-2021-01-nim} with resource requirements shown in Table~\ref{tab:computing-integrated_luminosity_by_year}. During this period, raw data will be buffered on disk at all of the Echelon 1 sites, along with permanent archival copies on tapes. Final calibration will be performed semi-automatically including accumulating sufficient data for tracker alignment and energy scale calibration of the calorimeters. 
The ECCE computing team is also pioneering the application of state-of-the-art AI/ML algorithms in detector optimization \cite{cisbani2020ai, ecce-note-comp-2021-03}, simulation, and PID~\cite{fanelli2020deeprich}, as well as real-time reconstruction in streaming readout~\cite{ameli2021streaming, fastML}, data reduction~\cite{huang2021efficient}, and signal processing~\cite{Miryala_2022}. AI/ML will continue to play an integral and essential role in the ECCE online and offline computing.
After calibration, data processing  will be released to multiple sites including HTC facilities at both Echelon 1 and 2 sites as in Fig.~\ref{fig:computing-federated_offsite_butterfly}. We expect that the produced simulation sample will focus on 10\% of the EIC collision cross-section that is directly relevant for the signal and background of the core ECCE physics program. These events will be simulated to $O(10)$~times the statistics in real data to constrain systematic uncertainty from the simulated sample to be much smaller than the data statistical uncertainty. The projected simulation resources are equivalent to the needs shown in the data reconstruction as in Table~\ref{tab:computing-integrated_luminosity_by_year}.

During the development of this proposal, a detailed detector model was simulated and reconstructed taking advantage of years of ongoing development and validation with the Fun4All-EIC/sPHENIX software~\cite{ecce-note-comp-2021-02, sPHENIXSoftwareFramework}. 
Fun4All was determined to be the best software stack for the ECCE proposal studies, for expediency, reliability and its familiarity within the software team.
Software is constantly evolving and choices will be re-evaluated in the coming months to ensure that over the next decade the ECCE software will incorporate the most advanced framework and packages with the aim of delivering a high performance, user-friendly, and reliable software stack. For example, the inclusion of AI as a tool to optimize detector design~\cite{cisbani2020ai} has been utilized within the ECCE software stack as described in Ref.~\cite{ecce-note-comp-2021-03}. Another example includes the integration of A Common Tracking Software (ACTS) package~\cite{Ai:2021ghi} as highlighted in Ref.~\cite{Osborn:2021zlr}, and used in preliminary ECCE tracking pattern recognition and efficiency studies. 

\begin{table*}[tb!]
  \caption{Estimate of raw data storage and compute needs for first
    three years of ECCE, assuming ramp up to full luminosity by year
    3~\cite{ecce-note-comp-2021-01-nim} }
    \centering
    \begin{tabular}{lccc}
      \toprule
      & \multicolumn{3}{c}{\textbf{ECCE Runs}} \\
      & year-1 & year-2 & year-3 \\
      \midrule
      Luminosity & $10^{33}\mathrm{cm}^{-2}\mathrm{s}^{-1}$ & $2\times 10^{33}\mathrm{cm}^{-2}\mathrm{s}^{-1}$ & $10^{34}\mathrm{cm}^{-2}\mathrm{s}^{-1}$ \\
      Weeks of Running        & 10                    & 20                      & 30                    \\
      Operational efficiency    & 40\%                  & 50\%                    & 60\%                  \\
      Disk (temporary)  &  1.2~PB & 3.0~PB & 18.1~PB \\
      Disk (permanent)    & 0.4~PB & 2.4~PB &	20.6~PB \\
      Data Rate to Storage    & 6.7~Gbps               & 16.7~Gbps                & 100~Gbps               \\
      Raw Data Storage (no duplicates) & 4~PB          & 20~PB                    & 181~PB                 \\
      Recon process time/core	& 5.4~s/ev	& 5.4~s/ev	& 5.4~s/ev \\
      Streaming-unpacked event size	& 33kB	& 33kB & 33kB \\
      Number of events produced &	121 billion	& 605 billion & 5,443 billion \\
      Recon Storage          & 0.4~PB                  & 2~PB                    & 18~PB                   \\
      CPU-core hours (recon+calib)	& 191M~core-hours	& 953M~core-hours &	8,573M~core-hours \\
      2020-cores needed to process in 30 weeks	& 38k &	189k &	1,701k \\
      \bottomrule
    \end{tabular}
    \label{tab:computing-integrated_luminosity_by_year}
\end{table*}

\section{Infrastructure/Integration}

The interaction region has an overall length of 9.5m. The ECCE detector extends from -4.5m to 5.0m around the origin. A total of half a meter of space between the end caps and the first interaction region magnets is reserved for vacuum pumps, valves, etc. The ECCE detector has an outer radius of 2.67 meters, which fits into the constraint given by the Rapid Cycling Synchrotron (RCS) located at 3.35m. To achieve the necessary alignment of the magnet with the electron direction the detector is rotated by 8 mrad in the horizontal plane. 

The central detector features service gaps for routing out cables and services. For example, service gaps between the central barrel and the forward calorimeter assembly and the backward flux return are envisioned, as indicated in the Sketchup mechanical model on the cover page. Additional space between the inner detectors and hpDIRC, and barrel EMCal and cryostat allow for routing cables out towards the service gaps. The beam pipe diameter increases in radius from the interaction point to the end caps\footnote{this is necessary to allow the cone of proton/neutron and nuclear breakup particles to pass through}, and thus includes several sections divided by flanges. This has to be taken into account for detector installation and servicing. For example, the diameter of the beam pipe flange at the location of the EEMC determines the configuration of the first layer of \PbWOiv. The beam pipe would need to be disassembled for the EEMC to be inserted/extracted from its nominal position. To maximize the EEMC acceptance and allowing for easy access the ECCE detector includes an option to separate out the inner EEMC. Taking into account the beam pipe diameter, the outer endcap detectors like the forward calorimeter assembly are foreseen to follow a clam shell design.

\section{Technology Selection, Risk and R\&D}
\label{sec:risk}

While the ECCE detector design seeks to minimize risk through strategic re-use and the selection of mature, yet state-of-the-art detector technologies, there are nevertheless risks associated with some ECCE detector technology choices. Our strategy has been to clearly identify these risks and develop an appropriate mitigation strategy, either through developing alternatives should the risks be realized or eliminating risk through an aggressive R\&D program.  We have developed an extensive risk log for the ECCE proposal that includes risk impact, likelihood and mitigation strategy for a wide array of technical and cost \& schedule risks. 

A list of specific risks related to the ECCE technology selection includes: 

\begin{itemize}
    \item {\bf BaBar Solenoid:} As a mitigation against the schedule risk posed by a potential problem with the BaBar solenoid developing during sPHENIX running, we plan to proceed with the initial engineering and design for a replacement magnet.   This work will be carried out by CEA/Saclay in close collaboration with Jefferson Lab.  A final decision to proceed with the BaBar solenoid or produce a new magnet will be taken in mid-2023 after the performance of the BaBar solenoid during the first year of sPHENIX running is reviewed by a panel of experts. The risk-mitigation decision tree is shown in Figure~\ref{fig:magnet-3}. Assuming a five-year construction for a new magnet, consistent with the duration of new SC magnets recently built as part of the Jefferson Lab 12-GeV Upgrade project, the ECCE schedule for detector construction and assembly would remain consistent with an early CD-4A date if procurement of a replacement magnet is determined to be necessary. 
\begin{figure*}[htb]
    \centering
    \includegraphics[width=0.9\textwidth]{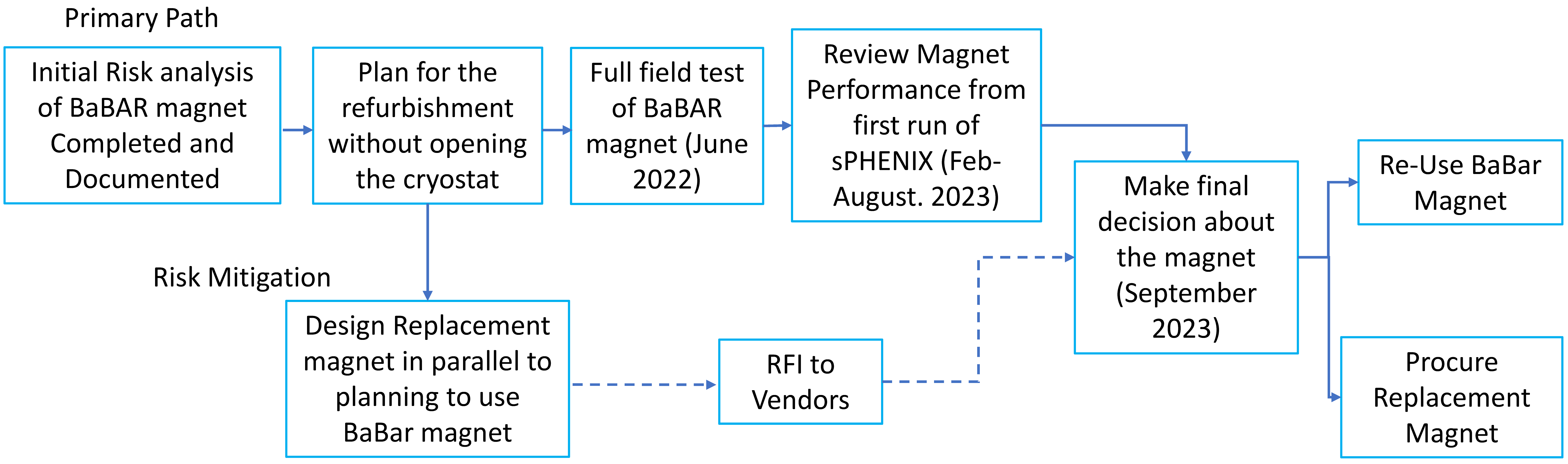}
    \caption{Decision tree for the risk mitigation strategy associated with the reuse of the BaBar solenoid.}
    \label{fig:magnet-3}
\end{figure*}    
    
    \item {\bf SciGlass Calorimetry: } The use of SciGlass for electromagnetic calorimetry in the ECCE barrel offers a low-cost solution to large area electromagnetic calorimetry with excellent energy resolution.  The performance of SciGlass has been demonstrated in short (20 cm) bars. 
    The performance validation of longer blocks is part of the ongoing EIC project R\&D (eRD105) and the demonstration of large scale commercial production with high quality and uniformity is part of an ongoing Phase2 SBIR/STTR.
    The ECCE strategy to address 
    the risk associated with SciGlass, if it is realized, is two-fold: if SciGlass cannot be produced on-schedule in sufficient quantities for ECCE needs, one option would be to refurbish half of the existing sPHENIX W/SciFi calorimeter to cover half of the ECCE acceptance, reducing the overall need for SciGlass.  The refurbished sPHENIX calorimeter could meet required energy resolution in the forward ($\eta>0$) acceptance, albeit with lower performance compared with SciGlass. SciGlass would still be used at the backwards direction ($\eta<0$) where optimal energy resolution is required.  If SciGlass were unavailable in sufficient quantity for the backwards region as well, the remaining half of the ECCE acceptance could be covered with PbGl towers at additional expense.     
    
    \item {\bf Cylindrical $\mu$RWell Tracking:} The ECCE experiment utilizes $\mu$RWell tracking layers in the central barrel as a low-mass, cost-effective means to provide the additional tracking points required to achieve the required momentum resolution.  While cylindrical $\mu$RWell detector should be technically possible, it remains to be demonstrated that they can provide stable operation at the required 55$\mu$m resolution in a magnetic field. ECCE plans an aggressive R\&D program, working with our international partners, to demonstrate the performance of cylindrical $\mu$RWell detectors and address any technical challenges that may arise. 
    
    \item {\bf AC-LGADs: } ECCE plans AC-LGAD sensors for TOF not only in the forward and backwards region but in the central barrel as well. Cylindrical detectors based on LGAD sensors have not been previously demonstrated, and AC-LGAD sensors require additional R\&D to demonstrate and characterize their performance and suitability for use in both the TOF and Roman Pot detectors in ECCE. To mitigate this risk, ECCE plans a comprehensive R\&D for AC-LGAD sensor and readout development, characterization and readout to ensure the required timing resolution can be achieved.

    \item {\bf B0 Detector:} The current design of the B0 detector calls for a crystal calorimeter to be installed after the tracking stations in the B0 warm bore to enable studies of physics processes that require $\gamma$ energy measurement such as u-channel DVCS. The installation, integration and maintenance of this detector present severe mechanical challenges due to the tight constraints in the magnet bore that will require detailed mechanical designs. If it is determined that installation of a crystal calorimeter is not feasible we will be forced to accept the loss of scope and install only the tracking planes. 

\end{itemize}

In addition to detailing risks in the ECCE risk registry, we also document potential risk opportunities. We list a few representative examples here, additional information is available in the ECCE risk registry and opportunity log, both of which are available in the ECCE supplemental material.

\begin{itemize}
    \item {\bf Reduction of the number of hpDIRC sensors:} R\&D performed for the PANDA DIRC suggests that the sensor coverage can be reduced by up to 30\% without significant impact on the PID performance. A positive outcome of the simulation study and validation in test beam would allow ECCE to take advantage of this opportunity.
    \item {\bf Improved ITS3 sensor yields:} Si tracker costs could be reduced if ITS3 sensor yield is higher than anticipated. We intend to take advantage of knowledge gained from ALICE ITS3 production, as well as with the foundry to optimize sensor yields. 
    \item {\bf hpDIRC lightguide shape:} Currently three options are being considered for the lighguide section of the bar box, which couples the narrow radiator bars to the lenses and prism. Use of one wide plate per bar box would be the most cost efficient. We intend to perform a simulation study and a test experiment with particle beams to validate this potentially cost-saving and performance-enhancing hpDIRC option for ECCE.
\end{itemize}

%

\section{Upgrades}

The ECCE baseline detector can be augmented with additional upgrades that either enhance or expand the existing physics reach:

\begin{itemize}
    \item { \bf Dual-Readout Calorimetry:} The addition of a dual-readout calorimeter, replacing the FEMC and LFHCAL in the forward region would provide a significant improvement in energy resolution for hadrons in the forward region. Because the tracking momentum resolution worsens with increasing momentum while the calorimeter energy resolution improves with increasing energy, the association of tracks with high-resolution clusters in the forward calorimeters can be used to improve the knowledge of high momentum tracks (the so-called "particle-flow" approach). With a dual-readout calorimeter, the cross-over point between the tracking and calorimeter resolution would be pushed lower, enabling this improvement for a larger fraction of the tracks detected in the forward arm. Adding such improved capabilities to ECCE would improve measurements of SIDIS hadrons, TMD measurements with jets, and the ability to reconstruct event kinematics using the hadronic remnants.  The Korean HEP community is very interested in deploying dual-readout calorimetry in ECCE as they develop the technology for future high-energy facilities. 

    \item {\bf Muon Chambers:} The addition of muon chambers to the ECCE baseline would enable the improved detection and tagging of semi-leptonic decays of heavy flavor. ECCE collaborators in Israel have expressed an interest in providing this upgrade as an in-kind contribution to ECCE. The ability to use muons for such processes as DVCS and DVMP removes an ambiguity between the produced leptons in the electron channel and the scattered electron. Such an upgrade can enhance the ability of ECCE to produce the science in the EIC white paper and NAS report. 


    
    \item {\bf Hadron Arm High-Rapidity Tracking Layer:} The addition of a small, high rapidity AC-LGAD layer ($3.0<\eta<3.5$) in front of the forward electromagnetic calorimeter could improve track momentum resolution for very high momentum ($p_{T}>20$ GeV/c) charged tracks. It would also allow the detection of hadrons that enter the forward calorimeters from outside the acceptance of the inner tracker. This would be very beneficial for the deconvolution of overlapping clusters in the forward calorimeters as a necessary component to implementing a particle flow algorithm for the reconstruction of forward jets.  

    \item {\bf Backwards Hadronic Calorimeter:} While the ECCE baseline does not include a backwards hadronic calorimeter in the electron-going region, the addition of such a calorimeter could contribute to the reconstruction of event kinematics by the double-angle of Jacquet-Blondel methods at high-y, and contribute to electron identification in the backwards region. Such a calorimeter could be based on the STAR FCS Fe/Sc hadronic calorimeter, with partial re-use of the existing STAR additional modules and new modules constructed to complete the acceptance. We have studied this extensively within ECCE, and a hadronic calorimeter in the backwards region is not required to pursue the science program in the EIC white paper or NAS report and therefore does not justify the substantial expense required at this time. However, it is possible as the EIC program matures and the EIC luminosity increases we may revisit this with a simple upgrade.

\end{itemize}

%% file: sections/ecce_det_barrel_req.tex
\begin{table*}[t]
  \tiny 
  \renewcommand{\arraystretch}{1}
  \caption{Key detector requirements for ECCE central detector, with associated challenges, and a brief description of the ECCE approach to address each issue.}
  \resizebox{\textwidth}{!} {%
  {\RaggedRight
    \begin{tabular}{p{0.3\columnwidth}p{0.3\columnwidth}p{0.3\columnwidth}p{0.3\columnwidth}}
      \toprule
      {\bf Topic} & {\bf Challenge} & {\bf ECCE solution} & {\bf Comment} \\ \midrule

Hermetic $e^-$ coverage & Leave no gaps in $e^-$ EMcal coverage  while also folding in PID/hpDIRC readout needs & hpDIRC readout in backward region; Moved EEMC inward as much as possible;  Extend BEMC longitudinally &  Good coverage for negative rapidity; performance verified with full simulations \\ \midrule

Momentum resolution in forward/backward regions at high $\eta$ & Achieve Yellow Report requirements~\cite{AbdulKhalek:2021gbh} with a realistic tracker including support materials in the BaBar solenoid&   Five ITS3 Si disks forward and four disks backward; Additional AC-LGAD tracking before (after) dRICH (mRICH) & Used AI optimization. Upgrade option: AC-LGAD ring in forward region behind dRICH for $\eta = 3$--3.5 \\ \midrule
      
Backward Particle Identification & Constrained space to maximize EMCal coverage & AC-LGAD TOF for low-momentum; mRICH for hadron PID & mRICH is a space-efficient solution \\ \midrule

      Backward $e^-$ PID, $\pi^-$ suppression up to $10^{-4}$ & Highest precision  EM calorimetry &  Use all \PbWOiv &   Can separate out EMCal to reach beyond $\eta = -3.4$ \\ \midrule

Barrel PID – $e/\pi$ separation up to $10^{-2}$--$10^{-4}$, down to 0.2~GeV/$c$ &
  Need good EMcal  resolution; need  additional $e/\pi$ below  2~GeV/c &
   55~cm long SciGlass towers  for high precision EMcal;  hpDIRC for $\pi$ veto  down to $p = 0.3$~GeV/c;  AC-LGAD TOF for $p \leq 0.4$~GeV. &
  Leave 4~cm for \mrwell  between hpDIRC and  EMCal to seed PID performance   of hpDIRC and improve tacking  resolution \\ \midrule
  
Barrel PID – $\pi/K/p$  separation down to 0.2~GeV/c &
  hpDIRC only covers  down to 0.6~GeV/c &
  AC-LGAD TOF for  0.2 \textless p \textless 0.6~GeV/c  &
  \mrwell directly after hpDIRC   to increase performance. \\ \midrule
  
Barrel Tracking resolution &
   Achieve Yellow Report requirements  with a realistic tracker  including support materials  in the BaBar solenoid  &
   Three ITS3 Si vertex and  two Si sagitta layers followed  by two \mrwell, AC-LGAD, and  far outer \mrwell layer; &
  
  Used AI optimization of  tracker and support system  layout
   \\ \midrule

Forward Hadronic  calorimetry &
   Jet energy resolution \textless $50\%/\sqrt{E}$ &
  Longitudinally separated  calorimeter to meet needs  in high-$\eta$ region. &
  Upgrade Option: Dual Calorimeter (or only central in region of highest need) \\ \midrule
  
Forward Particle  Identification &
  Constrained space in  forward region &
  AC-LGAD TOF for low-  momentum; dRICH for high- momentum (C$_4$F$_{10}$ based) &
  Seed dRICH ring finder with  AC-LGAD before dRICH;  Employ recirculation and gas  recovery systems for  environmentally unfriendly  gas use. \\ \bottomrule
  
\end{tabular}%
}
}
\label{tab:det-requirements}
\end{table*}

%% file: sections/ecce_det_fffb_req.tex
\begin{table*}[ht]
  \tiny 
  \renewcommand{\arraystretch}{1}
  \caption{ECCE Detector Far-Forward/Far-Backward requirements}
  \resizebox{\textwidth}{!} {%
  {\RaggedRight
  \begin{tabular}{p{0.3\columnwidth}p{0.3\columnwidth}p{0.3\columnwidth}p{0.3\columnwidth}}
    \toprule
  {\bf Topic} &
  {\bf Challenge} &
  {\bf ECCE solution} &
  {\bf Comment} \\ \midrule
    Far-Backward – Low-$Q^2$ Tagger & Measure low-$Q^2$ photo-production with as minimal a $Q^2$-gap as possible. & Spectrometer with AC-LGAD tracking and \PbWOiv calorimetry   & \\ \midrule
  
    Far-Backward – Luminosity Detector & $e$-ion collision luminosity to better than 1\% and relative Luminosity for spin asymmetries to $10^{-4}$ & Zero Degree Calorimeter with x-ray absorber and $e^+/e^-$ pair spectrometer with AC-LGAD tracking and \PbWOiv calorimetry & two complementary detection systems \\ \midrule
  
Far-Forward – B0 Spectrometer & $\eta > 4$ charged particle tracking and $\gamma$ measurement & Four Si trackers with 10~cm \PbWOiv calorimeter & \\ \midrule
  
Far-Forward – Off-momentum Detectors & forward particles ($\Delta$,  $\Lambda$, $\Sigma$, etc) decay product measurement &  AC-LGAD detectors & Sensors on one side detect $p$, on other side $p^-$ from $\Lambda$ decay; sensors outside beam pipe \\ \midrule
  
Far-Forward – Roman Pots & Detect low-$p_T$ forward-going particles &  AC-LGAD detectors& fast timing ($\sim$35~ps) removes vertex smearing effects from crab rotation; $10\sigma$ from beam \\ \midrule

Far-Forward – Zero-degree Calorimeter &  Measure forward-going neutrons $\gamma$ and heavy-ion fission product &  FOCAL-type calorimeter with high-precision EM and Hadron Calorimetry & Upgrade option: AC-LGAD layer to capture very high rapidity charged tracks \\ \bottomrule
  
\end{tabular}%
}
}
\label{tab:det-ff-fb-requirements}
\end{table*}

%% file: sections/summary.tex
\section{Summary}
\label{summary}

In summary, the ECCE detector has been designed to address the full scope of the EIC physics program as presented in the EIC white paper [3] and the NAS report. ECCE can be built within the budget envelope set out by the EIC project while simultaneously managing cost and schedule risks.
This detector proposal has been reviewed and has been selected to be the basis for the project detector for the future collider.

%% file: sections/acknowledgements.tex
\section{Acknowledgements}
\label{acknowledgements}

We thank the EIC Silicon Consortium for cost estimate methodologies concerning silicon tracking systems, technical discussions, and comments.  We acknowledge the important prior work of projects eRD16, eRD18, and eRD25 concerning research and development of MAPS silicon tracking technologies.

We thank the EIC LGAD Consortium for technical discussions and acknowledge the prior work of project eRD112.

We acknowledge support from the Office of Nuclear Physics in the Office of Science in the Department of Energy, the National Science Foundation, and the Los Alamos National Laboratory Laboratory Directed Research and Development (LDRD) 20200022DR.

This research used resources of the Compute and Data Environment for Science (CADES) at the Oak Ridge National Laboratory, which is supported by the Office of Science of the U.S. Department of Energy under Contract No. DE-AC05-00OR22725.    
The work of AANL group are supported by the Science Committee of RA, in
 the frames of the research project \# 21AG-1C028.
And we gratefully acknowledge that support of Brookhaven National Lab and the Thomas Jefferson National Accelerator Facility which are operated under contracts DE-SC0012704 and DE-AC05-06OR23177 respectively.